\documentclass[aps,floats]{revtex4-2}
\usepackage{amsmath,amssymb}
\usepackage{graphicx,epsfig}
\usepackage[greek,english]{babel}
\usepackage{bbold}

\usepackage{mathtools}
\usepackage{makecell,tabularx}
\setcellgapes{1pt}
\makegapedcells

\makeatletter\AtBeginDocument{\let\@elt\relax}\makeatother
 
\DeclareMathOperator{\erf}{erf}
\DeclareMathOperator{\erfc}{erfc}

\begin{document}
\bibliographystyle {plain}

\pdfoutput=1
\def\oppropto{\mathop{\propto}} 
\def\opsimeq{\mathop{\simeq}}
\def\opoverderline{\mathop{\overline}}
\def\operarrow{\mathop{\longrightarrow}}
\def\opsim{\mathop{\sim}}

\def\opmin{\mathop{\min}} 
\def\opmax{\mathop{\max}} 
\def\oplim{\mathop{\lim}}

\title{ Conditioning diffusion processes with respect to the local time at the origin } 


\author{Alain Mazzolo}
\affiliation{Universit\'e Paris-Saclay, CEA, Service d'\'Etudes des R\'eacteurs et de Math\'ematiques Appliqu\'ees, 91191, Gif-sur-Yvette, France}

\author{C\'ecile Monthus}
\affiliation{Universit\'e Paris-Saclay, CNRS, CEA, Institut de Physique Th\'eorique, 91191 Gif-sur-Yvette, France}


\begin{abstract}
When the unconditioned process is a diffusion process $X(t)$ of drift $\mu(x)$ and of diffusion coefficient $D=1/2$, the local time $A(t)= \int_{0}^{t} d\tau \delta(X(\tau))  $ at the origin $x=0$ is one of the most important time-additive observable. We construct various conditioned processes $[X^*(t),A^*(t)]$ involving the local time $A^*(T)$ at the time horizon $T$. When the horizon $T$ is finite, we consider the conditioning towards the final position $X^*(T)$ and towards the final local time $A^*(T)$, as well as the conditioning towards the final local time $A^*(T)$ alone without any condition on the final position $X^*(T)$. In the limit of the infinite time horizon $T \to +\infty$, we consider the conditioning towards the finite asymptotic local time $A_{\infty}^*<+\infty$, as well as the conditioning towards the intensive local time $a^* $ corresponding to the extensive behavior $A_T \simeq T a^*$, that can be compared with the appropriate 'canonical conditioning' based on the generating function of the local time in the regime of large deviations. This general construction is then applied to generate various constrained stochastic trajectories for three unconditioned diffusions with different recurrence/transience properties : (i) the simplest example of transient diffusion corresponds to the uniform strictly positive drift $\mu(x)=\mu>0$; (ii) the simplest example of diffusion converging towards an equilibrium is given by the drift $\mu(x)=- \mu \, {\rm sgn}( x)$ of parameter $\mu>0$; (iii) the simplest example of recurrent diffusion that does not converge towards an equilibrium is the Brownian motion without drift $\mu=0$.

\end{abstract}

\maketitle

\section{ Introduction }

\subsection{ Conditioning diffusion processes with respect to time-additive observables of the stochastic trajectories}

Since its introduction by Doob \cite{refDoob,refbookDoob},
the conditioning of stochastic processes (see the mathematical books \cite{refbookKarlin,refbookRogers,Borodin}
and the physics recent review \cite{refMajumdarOrland}) have found many applications
in various fields like ecology \cite{refHorne}, finance \cite{refBrody} or
nuclear engineering \cite{refMulatier,refbookPazsit}. 
Among the different conditioned diffusions that have been constructed 
besides the basic example of the Brownian Bridge,
one can cite the Brownian excursion \cite{refMajumdarExcursion,refChung}, the Brownian meander \cite{refMajumdarMeander},
the taboo processes \cite{refKnight,refPinsky,refKorzeniowski,refGarbaczewski,refAdorisio,refAlainTaboo},
or non-intersecting Brownian bridges \cite{grela}.
 Let us also mention the conditioning in the presence of killing rates \cite{refbookKarlin,Karlin1982,Karlin1983,Frydman,Steinsaltz,kolb,Evans2019,tryphon_killing,us_DoobKilling}
 or when the killing occurs only via an absorbing boundary condition
 \cite{refBaudoin,refMultiEnds,us_DoobFirstPassage,us_DoobFirstEncounter}.
Note that stochastic bridges have been studied for many other Markov processes, 
including various diffusions processes \cite{henri,refSzavits,delarue},
discrete-time random walks and L\'evy flights \cite{refGarbaczewski_Levy,bruyne_discrete,Aguilar}, 
continuous-time Markov jump processes \cite{Aguilar},
run-and-tumble trajectories \cite{bruyne_run}, or
processes with resetting \cite{refdeBruyne2022}.

A recent important generalization concerns the conditioning with respect to global dynamical constraints
involving time-additive observables of the stochastic trajectories. In particular, the conditioning 
on the area has been studied via various methods
for Brownian processes or bridges \cite{refMazzoloJstat} and for Ornstein-Uhlenbeck bridges \cite{Alain_OU}.
The conditioning on the area and on other time-additive observables has been then analyzed 
both for the Brownian motion and for discrete-time random walks \cite{refdeBruyne2021}.
This approach has been generalized \cite{c_microcanonical} to various types of discrete-time or continuous-time Markov processes, while the time-additive observable
can involve both the time spent in each configuration and the increments of the Markov process.
This general reformulation of the 'microcanonical conditioning', where the time-additive observable is constrained
to reach a given value after the finite time window $T$, allows to make the link \cite{c_microcanonical} 
with the 'canonical conditioning' based on generating functions of additive observables
that has been much studied recently in the field of dynamical large deviations
of Markov processes over a large time-window $T$ \cite{peliti,derrida-lecture,tailleur,sollich_review,lazarescu_companion,lazarescu_generic,jack_review,vivien_thesis,lecomte_chaotic,lecomte_thermo,lecomte_formalism,lecomte_glass,kristina1,kristina2,jack_ensemble,simon1,simon2,simon3,Gunter1,Gunter2,Gunter3,Gunter4,chetrite_canonical,chetrite_conditioned,chetrite_optimal,chetrite_HDR,touchette_circle,touchette_langevin,touchette_occ,touchette_occupation,derrida-conditioned,derrida-ring,bertin-conditioned,garrahan_lecture,Vivo,chemical,touchette-reflected,touchette-reflectedbis,c_lyapunov,previousquantum2.5doob,quantum2.5doob,quantum2.5dooblong,c_ruelle,lapolla,chabane,chabane_thesis}.
The equivalence between the 'microcanonical conditioning'
and the canonical conditioning' at the level of the large deviations for large time $T$
is explained in detail in the two complementary papers \cite{chetrite_conditioned,chetrite_optimal}
and in the HDR thesis \cite{chetrite_HDR}.

\subsection{ Simplest examples of time-additives for a diffusion process $X(t)$ : the occupation time and the local time}

For a one-dimensional diffusion process $X(t)$,
two basic examples of time-additive observables are :

(i) the occupation time $O_{[a,b]}(t) $ 
of the space interval $[a,b]$ during the time window $[0,t]$ 
\begin{eqnarray}
O_{[a,b]}(t)= \int_{0}^{t} d\tau \theta(a \leq X(\tau) \leq b )
\label{occupation}
\end{eqnarray}
belongs to the interval $0 \leq O_{[a,b]}(t) \leq t$.
The conditioning with respect to the occupation time of the interval $[a=0,b=+\infty[$
has been studied recently for the Brownian motion without drift \cite {refdeBruyne2021},
while the canonical conditioning with respect to the occupation time
has been analyzed for various settings \cite{touchette_occ,touchette_occupation}.

(ii) the local time $A_{x}(t) $ at the position $x$ during the time window $[0,t]$
 (see the mathematical review \cite{bjork} and references therein) 
\begin{eqnarray}
A_{x}(t)= \int_{0}^{t} d\tau \delta(X(\tau)-x) 
\label{localtimex}
\end{eqnarray}
has for physical dimension $\frac{Time}{Length}$ so that it is actually not a 'time' despite its standard name.
However, it is directly related to the occupation time as follows.
On the one hand, the local time $A_{x}(t) $ of Eq. \ref{localtimex} can be constructed from the occupation time $O_{[x-\epsilon,x+\epsilon]}(t) $
of the space interval $ [x-\epsilon,x+\epsilon]$ of size $(2 \epsilon)>0$ centered at the position $x$ in the limit $\epsilon \to 0^+$
\begin{eqnarray}
A_{x}(t)= \int_{0}^{t} d\tau \oplim_{\epsilon^+ \to 0} 
\left( \frac{\theta(x-\epsilon \leq X(\tau) \leq x+\epsilon )}{2 \epsilon } \right) 
=  \oplim_{\epsilon^+ \to 0} 
\left( \frac{ O_{[x-\epsilon,x+\epsilon]}(t)}{2 \epsilon } \right) 
\label{localtimexeps}
\end{eqnarray}
As a consequence, the local time $A_{x}(t)$ belongs to $[0,+\infty[$ with no upper bound.
On the other hand, the occupation time $O_{[a,b]}(t) $ 
can be reconstructed from the local time $A_{x}(t) $
 for all the internal positions $x \in [a,b]$
 \begin{eqnarray}
O_{[a,b]}(t)=  \int_{0}^{t} d\tau \int_a^b dx \delta(X(\tau)-x)  = \int_a^b dx A_{x}(t)
\label{occupationA}
\end{eqnarray}


\subsection{ Goals of the present work }

At first sight, the delta function that enters the definition of the local time in Eq. \ref{localtimex}
 might appear as very singular for the purpose of conditioning.
However, as in quantum mechanics where delta impurities are well-known 
to be much simpler than smoother potentials, the delta function in Eq. \ref{localtimex} is actually a huge technical simplification compared to the case of conditioning with respect to an arbitrary general additive observable. Indeed, the exact Dyson equation associated to a single delta impurity allows to analyze the conditioning with respect to the local time for $A_{x=0}(t) =A(t)$ at the origin $x=0$ in terms of the properties of the propagator $G(x,t \vert x_0,t_0)$ of the unconditioned process $X(t)$ alone.
In the present paper, it will be thus interesting to consider that the unconditioned process is a diffusion process $X(t)$ of diffusion coefficient $D=1/2$ with an arbitrary position-dependent drift $\mu(x)$ in order to derive the general properties before the application to various illustrative examples of drifts.
Our goal is to construct various conditioned joint processes $[X^*(t),A^*(t)]$ satisfying certain conditions involving the local time $A^*(T)$ either at the finite time horizon $T$, or in the limit of the infinite time horizon $T \to +\infty$.
For instance, the two basic cases that will be considered for the finite horizon $T$
can be summarized as follows.

(i) The conditioning towards the position $x_T^*$ and the local time $A_T^*$ at the time horizon $T$
involves the conditioned drift
\begin{eqnarray}
  \mu^{[x_T^*,A_T^*]}_T( x,A , t ) = \mu(x) +  \partial_x    \ln P( x_T^*,A_T^*,T \vert  x,A,t)
\nonumber
\end{eqnarray}
where $P( x_T^*,A_T^*,T \vert  x,A,t) $ represents the joint propagator of the unconditioned diffusion.

(ii) The conditioning towards the local time $A_T^*$ at time horizon $T$,
without any condition on the final position $x_T$ involves the conditioned drift
\begin{eqnarray}
  \mu^{[A_T^*]}_T( x,A , t ) = \mu(x) +  \partial_x    \ln \Pi( A_T^*,T \vert  x,A,t)
\nonumber
\end{eqnarray}
in terms of $\Pi( A_T^*,T \vert  x,A,t) = \int d x_T P( x_T,A_T^*,T \vert  x,A,t) $ for the unconditioned diffusion.

When the unconditioned diffusion is the Brownian motion of uniform drift $\mu \geq 0$ or the stochastic process with drift $\mu(x) = - \mu \, {\rm sgn}(x)$ with $\mu > 0$, some examples of 
these conditioned drifts that will be studied are given in the two following tables. 

\begin{table}[!h]
\setcellgapes{4pt}
\begin{tabular}{|>{\centering}p{2.7cm}|c|c|c|}
\hline    
&\multicolumn{2}{c|}{Finite time horizon $T<\infty$}& Infinite time horizon $T=\infty$ \\ 
\cline{2-4}
  & $\begin{aligned} & \mathrm{~~~~~~~Conditioned~drift~}  \mu^{[x_T^*,A_T^*]}_T( x,A , t ) 
   \\ & \mathrm{towards ~the ~position}~x_T^* \mathrm{~and ~the ~local ~time~} A_T^* \end{aligned}  $ & $\begin{aligned} & \mathrm{Conditioned~drift~}  \mu^{[A_T^*]}_T( x,A , t )                \\ & \mathrm{~~~towards ~the ~local ~time~} A_T^* \end{aligned} $ & $\begin{aligned} & \mathrm{Conditioned~drift~}  \mu^{[A_{\infty}^*]}_{\infty}( x,A  )                \\ & \mathrm{~~towards~the ~local ~time~} A_{\infty}^* \end{aligned} $\\
\hline
Region $0<A<A_T^*$ & $  {\rm sgn}(x)
\left[ \frac{1}{\vert x_T^* \vert +\vert x \vert + A_T^*-A } -
 \frac{\vert x_T^* \vert +\vert x \vert + A_T^*-A}{T-t} 
\right] $ 
& 
$ - {\rm sgn} (x) \ \frac{\vert x \vert + A_T^*-A }{ T-t} $
 & $-\mu \, {\rm sgn} (x) $ \\
\hline
Region $A=A_T^*$ & $ \frac{\left( \frac{x_T^*-x}{T-t} \right) e^{- \frac{(x_T^*-x)^2}{2(T-t)} }
  + \left( \frac{x + {\rm sgn}(x) \vert x_T^* \vert}{T-t} \right) e^{ - \frac{(\vert x_T^* \vert +\vert x \vert )^2}{2(T-t)} }       }
{e^{- \frac{(x_T^*-x)^2}{2(T-t)} }- e^{ - \frac{(\vert x_T^* \vert +\vert x \vert )^2}{2(T-t)} }    
}
$
& $\sqrt{\frac{2}{\pi (T-t)}} \frac{e^{-\frac{x^2}{2(T-t)}}}{\erf \left(\frac{\vert x \vert}{\sqrt{2(T-t)}} \right)} {\rm sgn}(x)$  & $ \mu \coth(\mu x) $  \\
\hline
\end{tabular}
\caption{
Some examples of conditioned drifts $\mu_T^*( x,A , t ) $ for the Brownian motion of uniform drift $\mu(x) = \mu > 0$.} 
\label{table1}
\end{table}

\begin{table}[!h]
\setcellgapes{4pt}
\begin{tabular}{|>{\centering}p{2.7cm}|c|c|}
\hline    
&\multicolumn{2}{c|}{Finite time horizon $T<\infty$} \\  
\cline{2-3}
  & $\begin{aligned} & \mathrm{~~~~~~~Conditioned~drift~}  \mu^{[x_T^*,A_T^*]}_T( x,A , t )   
   \\ & \mathrm{towards ~the ~position}~x_T^* \mathrm{~and ~the ~local ~time~} A_T^* \end{aligned}  $ & $\begin{aligned} & \mathrm{Conditioned~drift~}  \mu^{[A_T^*]}_T( x,A , t )                  \\ & \mathrm{~~~towards~the ~local ~time~} A_T^* \end{aligned} $  \\
\hline
Region $0<A<A_T^*$ & $  {\rm sgn}(x)
\left[ \frac{1}{\vert x_T^* \vert +\vert x \vert + A_T^*-A } -
 \frac{\vert x_T^* \vert +\vert x \vert + A_T^*-A}{T-t} 
\right] $ 
& 
$ \scriptstyle{ \mu \, {\rm sgn}(x) - \frac{2}{(T-t)}\frac{(\vert x \vert  + A_T^*-A  ){\rm sgn}(x) }{2 - \mu \sqrt{2 \pi (T-t)} e^{ \frac{(\vert x \vert  + \mu (T-t) + A_T^*-A)^2}{2(T-t)} } \erfc \left( \frac{\vert x \vert +\mu(T-t) + A_T^*-A }{\sqrt{2 (T-t)}}  \right)  }  } $ \\
\hline
Region $A=A_T^*$ & $ \frac{\left( \frac{x_T^*-x}{T-t} \right) e^{- \frac{(x_T^*-x)^2}{2(T-t)} }
  + \left( \frac{x + {\rm sgn}(x) \vert x_T^* \vert}{T-t} \right) e^{ - \frac{(\vert x_T^* \vert +\vert x \vert )^2}{2(T-t)} }       }
{e^{- \frac{(x_T^*-x)^2}{2(T-t)} }- e^{ - \frac{(\vert x_T^* \vert +\vert x \vert )^2}{2(T-t)} }    
}
$
& $\scriptstyle{ 2 \sqrt{\frac{2}{\tau}} 
\frac{
\, {\rm sgn}(x) \left( \frac{1}{\sqrt{\pi }}- \!  e^{\frac{(\vert x \vert + \mu \tau)^2}{2\tau}} \! \mathcal{F} \left( \frac{\vert x \vert + \mu \tau}{\sqrt{2 \tau}}  \right) \right) +  e^{\frac{(x - \mu \tau)^2}{2\tau}} \left(e^{2 \mu x} \! \mathcal{F} \left( \frac{ x + \mu \tau}{\sqrt{2 \tau}}  \right)   -\! \mathcal{F} \left( \frac{ \mu \tau -x}{\sqrt{2 \tau}}  \right)   \right) 
     }
     {
   e^{\frac{(x + \mu \tau)^2}{2 \tau} } \erfc \left( \frac{x + \mu \tau}{\sqrt{2 \tau}}  \right) 
  + e^{\frac{(x - \mu \tau)^2}{2 \tau} } \erfc \left( \frac{\mu \tau - x }{\sqrt{2 \tau}}  \right)
  -2 e^{\frac{(\vert x \vert + \mu \tau)^2}{2 \tau}} \erfc \left( \frac{\vert x \vert + \mu \tau }{\sqrt{2 \tau}} 
   \right)
     } }$    \\
\hline
\end{tabular}
\caption{
Some examples of conditioned drifts $\mu_T^*( x,A , t ) $ for the stochastic process with drift, $\mu(x) = - \mu \, {\rm sgn}(x)$ with $\mu > 0$. Observe that the conditioned drifts are the same as those of the Brownian motion in the case of conditioning towards the position $x_T^*$ and the local time $A_T^*$ at the finite time horizon $T$. We use the notation $\mathcal{F}(x) = x \erfc(x)$ where $\erfc(x)$ is the complementary Error function $\erfc(x) = 1 - \erf(x)$.} 
\label{table2}
\end{table}


\subsection{ Organization of the paper }

The paper is organized as follows.
The properties of the unconditioned diffusion process $X(t)$ with drift $\mu(x)$ 
are recalled in section \ref{sec_x}.
We then analyze the properties of the joint propagator $P(x,A,t \vert x_0,A_0,t_0)$ for the position $x$ and the local time $A$ in section \ref{sec_xa},
as well as the probability $\Pi(A,t \vert x_0,A_0,t_0)=\int dx P(x,A,t \vert x_0,A_0,t_0) $ in section \ref{sec_a}.
The statistical properties of the local time increment $[A(t)-A(t_0)]$
 in the limit of the large time interval $(t-t_0) \to +\infty$ are discussed in section \ref{sec_largetime}
as a function of the recurrence/transience properties of the diffusion process $X(t)$ induced by the drift $\mu(x)$.
In section \ref{sec_Doob}, we construct various conditioned processes $[X^*(t),A^*(t)]$ 
that involve the local time $A^*(T)$ at the finite time horizon $T$
or in the limit of the infinite time horizon $T \to +\infty$.
This general framework is applied to the case of the uniform drift $\mu(x)=\mu$ 
with $\mu=0$ or $\mu>0$ in section \ref{sec_brown},
and to the case $\mu(x)=- \mu \,{\rm sgn}( x)$ of parameter $\mu>0$ in section \ref{sec_sgn},
in order to generate stochastic trajectories of various conditioned processes with respect to the local time.
Monte Carlo simulations illustrate our findings.
Our conclusions are summarized in section \ref{sec_conclusion}.
The three appendices \ref{app_canonical}, \ref{app_canonicalEq} and \ref{app_canonicalnoEq}
are devoted to the canonical conditioned processes $X^*_p(t)$ of parameter $p$,
in order to compare with the microcanonical conditioning described in the main text. 


\section{ Properties of the unconditioned diffusion process $X(t)$ with drift $\mu(x)$}

\label{sec_x}

In this paper, we consider that the unconditioned process $X(t)$ is a diffusion process on 
the whole line $]-\infty,+\infty[$
generated by the Stochastic Differential Equation 
\begin{eqnarray}
dX(t) =  \mu( X(t) ) dt + dB(t)
\label{itox}
\end{eqnarray}
where $B(t)$ is a standard Brownian motion and where the position-dependent drift $\mu(x)$ is the only parameter of the model.
In this section, we recall the recurrence/transience properties that will be useful to analyze 
the statistics of its local time $A(t)$
at the origin in the 
three next sections \ref{sec_xa}, \ref{sec_a} and \ref{sec_largetime}.


\subsection{Propagator $G(x,t \vert x_0,t_0)$ for the diffusion process $X(t)$ }

The propagator $G(x,t \vert x_0,t_0)$ for the diffusion process $X(t)$ generated by Eq. \ref{itox} 
satisfies
the Fokker-Planck dynamics 
\begin{eqnarray}
\partial_t G(  x,t \vert   x_0,t_0) 
= -    \partial_x \left[ \mu(x) G(  x,t \vert   x_0,t_0) \right] +  \frac{1}{2} \partial_x^2  G(  x,t \vert   x_0,t_0) 
\label{forward1d}
\end{eqnarray}
Its Laplace transform with respect to the time interval $(t-t_0)$
\begin{eqnarray}
{\hat G}_s (x \vert x_0) \equiv \int_{t_0}^{+\infty} dt e^{-s (t-t_0) } G(  x,t \vert   x_0,t_0) 
\label{laplace}
\end{eqnarray}
then satisfies
\begin{eqnarray}
- \delta(x-x_0) + s {\hat G}_s (x \vert x_0)
= -       \partial_x \left[ \mu(x) {\hat G}_s (x \vert x_0) \right] +  \frac{1}{2} \partial_x^2  {\hat G}_s (x \vert x_0)
\label{forward1dlaplace}
\end{eqnarray}


\subsection{ Similarity transformation towards an euclidean quantum propagator $\psi(x,t \vert x_0,t_0) $}

As is well-known \cite{risken}, the potential $U(x)$ defined via the following integration of the drift $\mu(y) $
\begin{eqnarray}
U(x) \equiv - 2 \int_0^{x} dy \mu(y) 
\label{potentialU}
\end{eqnarray}
can be used to make the similarity transformation 
\begin{eqnarray}
G(x,t \vert x_0,t_0) = e^{- \frac{  U(x) }{2}} \psi(x,t \vert x_0,t_0)e^{ \frac{  U(x_0) }{2}}
= e^{ \frac{ U(x_0)- U(x) }{2}} \psi(x,t \vert x_0,t_0) = e^{ \int_{x_0}^{x} dy \mu(y)} \psi(x,t \vert x_0,t_0)
\label{defpsi}
\end{eqnarray}
The Fokker-Planck Eq. \ref{forward1d}
for the propagator $G(  x,t \vert   x_0,t_0) $ is then transformed 
 into an Euclidean Schr\"odinger Equation for $\psi(x,t \vert x_0,t_0) $
\begin{eqnarray}
- \partial_t \psi (  x,t \vert   x_0,t_0) = H \psi(  x,t \vert   x_0,t_0) 
\label{schrodinger}
\end{eqnarray}
The corresponding hermitian quantum Hamiltonian 
\begin{eqnarray}
 H =  -  \frac{1}{2} \partial_x^2 +V(x)
\label{hamiltonian}
\end{eqnarray}
involves the quantum potential
\begin{eqnarray}
 V(x) \equiv \frac{ \mu^2(x)}{2} + \frac{\mu'(x)}{2} 
\label{susy}
\end{eqnarray}
This very specific structure of $V(x)$ in terms of the drift $\mu(x)$
allows to factorize the Hamiltonian of Eq. \ref{hamiltonian}
into the supersymmetric form (see the review on supersymmetric quantum mechanics \cite{review_susyquantum} and references therein)
\begin{eqnarray}
H \equiv  \frac{1}{2}  Q^{\dagger} Q
\label{hsusy}
\end{eqnarray}
involving the first-order operator 
\begin{eqnarray}
Q   \equiv   - \partial_x + \mu(x) 
\label{qsusy}
\end{eqnarray}
and its adjoint
\begin{eqnarray}
Q^{\dagger}  \equiv  \partial_x + \mu(x)
\label{qdaggersusy}
\end{eqnarray}
This quantum mapping allows to use all the knowledge on one-dimensional quantum Hamiltonians in general
and on supersymmetric quantum Hamiltonians in particular
to characterize the energy spectrum as follows.


\subsection{ Analysis of the spectrum of the quantum supersymmetric Hamiltonian $H$ }

\subsubsection{ Analysis of the continuous spectrum $]V_{\infty},+\infty[$ of $H$ when it exists }

\label{subsec_continuum}

The minimum of the two limiting values of the quantum potential $V(x)$ of Eq. \ref{susy} as $x \to \pm \infty$
\begin{eqnarray}
 V_{\infty} = \min[ V(x \to +\infty) ; V(x \to - \infty) ]
\label{vinfty}
\end{eqnarray}
determines the lower boundary of the continuous spectrum when it exists.

The discussion is thus as follows :

(i) if $V_{\infty}<+\infty$ is finite, then the continuous spectrum of $H$ is given by $]V_{\infty},+\infty[$.
The physical interpretation is that,
in the infinity region where the asymptotic value of the potential $V(x)$ is $V_{\infty}$,
an eigenstate of energy $E \in ]V_{\infty},+\infty [$
behaves asymptotically like a linear combination of the plane waves $e^{ \pm i k x} $,
where the relation between the wave-number $k$ and the energy $E$ is given by 
the corresponding eigenvalue equation for $H e^{ \pm i k x} = E e^{ \pm i k x}$ in the infinity region
 where the potential is $V_{\infty}$
\begin{eqnarray}
E = \frac{k^2}{2} + V_{\infty}
\label{planewave}
\end{eqnarray}
i.e. the wave-number $k =  \sqrt{2(E-  V_{\infty})}$ is real for any energy $E \in ]V_{\infty},+\infty [$.

The simplest example is the case of the uniform drift $\mu(x)=\mu$,
where the quantum potential of Eq. \ref{susy}
reduces to the constant 
\begin{eqnarray}
 V(x) = \frac{\mu^2}{2} \ \ \ \ \ \ \ {\rm for }  \ \ \ \mu(x)=\mu
\label{vdriftmu}
\end{eqnarray}
so that the continuous spectrum is $] \frac{\mu^2}{2},+\infty[$.

(ii) if $V_{\infty}=+\infty$ is infinite, then there is no continuous spectrum and $H$ has 
only an infinity of bound states.

The simplest example is the case of the Ornstein-Uhlenbeck drift $\mu(x)=- k x$ with $k>0$,
where the quantum potential of Eq. \ref{susy} corresponds to the harmonic oscillator
\begin{eqnarray}
 V(x) = \frac{k^2}{2} x^2 - \frac{k}{2}  \ \ \ \ \ \ \ {\rm for }  \ \ \ \mu(x)=- k x
\label{vOU}
\end{eqnarray}
with its well-known infinite series of discrete levels.


\subsubsection{ Analysis of the normalizable zero-energy ground-state $ \phi^{GS}(  x) $ of $H$ when it exists }

The factorization of Eq. \ref{hsusy} shows that the spectrum of $H$ is positive.
Let us now discuss whether $E=0$ is the ground state energy of $H$.
The wavefunction $ \phi^{E=0}(  x)  $ that is annihilated by the operator $Q$ of Eq. \ref{qsusy}
\begin{eqnarray}
0= Q  \phi^{[E=0]}(  x)  =   - \partial_x \phi^{[E=0]}(  x)+ \mu(x) \phi^{[E=0]}(  x)
\label{qsusyannihilate}
\end{eqnarray}
reads in terms of the potential $U(x)$ of Eq. \ref{potentialU}
\begin{eqnarray}
  \phi^{[E=0]}(  x)  =    \phi^{[E=0]}( 0 ) e^{\int_0^{x} dy \mu(y) } =  \phi^{[E=0]}( 0 )e^{- \frac{ U(x)}{2} }
\label{phiezero}
\end{eqnarray}
This wavefunction can be normalized on $x \in ]-\infty,+\infty[$ if
\begin{eqnarray}
 1= \langle  \phi^{[E=0]} \vert   \phi^{[E=0]} \rangle =  \int_{-\infty}^{+\infty} dx \left[ \phi^{[E=0]}(  x) \right]^2 = \left[ \phi^{[E=0]}(  0) \right]^2  \int_{-\infty}^{+\infty} dx e^{-  U(x) }
\label{phiezeronorma}
\end{eqnarray}

The discussion is thus as follows :

(i) if the integral involving the potential $U(x)$ converges
\begin{eqnarray}
   \int_{-\infty}^{+\infty} dx e^{-  U(x) } <+\infty
\label{qsusyannihilatenorma}
\end{eqnarray}
then $H$ has the following normalizable ground state at zero-energy $E=0$
\begin{eqnarray}
 \phi^{GS}(  x) = \frac{ e^{- \frac{ U(x)}{2} }}{ \sqrt{ \int_{-\infty}^{+\infty} dy e^{-U(y) } } } 
\label{gs}
\end{eqnarray}

(ii) if the integral of Eq. \ref{qsusyannihilatenorma} diverges
\begin{eqnarray}
   \int_{-\infty}^{+\infty} dx e^{-  U(x) } =+\infty
\label{dvnoeq}
\end{eqnarray}
then the zero-energy wavefunction of Eq. \ref{qsusyannihilate} cannot be normalized,
and the Hamiltonian $H$ has no bound state,
but only the continuous spectrum discussed in the previous subsection \ref{subsec_continuum}.


\subsection{ Consequences for the Fokker-Planck propagator $G(x,t \vert x_0,t_0) $ at large time interval $(t-t_0)$ } 

\subsubsection{ When $H$ has the zero-energy ground-state $ \phi^{GS}(  x) $ : 
 $G(x,t \vert x_0,t_0)$ converges towards an equilibrium state $G_{eq}(x)$ }

 When $H$ has the normalizable zero-energy ground-state $ \phi^{GS}(  x) $ of Eq. \ref{gs},
 then the quantum propagator $ \psi (  x,t \vert   x_0,t_0) $ of Eq. \ref{schrodinger} displays the long-time behavior
 \begin{eqnarray}
 \psi (  x,t \vert   x_0,t_0) \opsimeq_{(t-t_0) \to +\infty}  \phi^{GS}(  x) \phi^{GS}(  x_0)
 = \frac{ e^{- \frac{ U(x)}{2} - \frac{ U(x_0)}{2} }}{  \int_{-\infty}^{+\infty} dy e^{-U(y) }  } 
\label{schrodingertlarge}
\end{eqnarray}
So the Fokker-Planck propagator $G(x,t \vert x_0,t_0) $ obtained via the similarity transformation of Eq. \ref{defpsi}
\begin{eqnarray}
G(x,t \vert x_0,t_0) = e^{- \frac{  U(x) }{2}} \psi(x,t \vert x_0,t_0)e^{ \frac{  U(x_0) }{2}}
 \opsimeq_{(t-t_0) \to +\infty}  \frac{ e^{-  U(x) }}{  \int_{-\infty}^{+\infty} dy e^{-U(y) }  } \equiv G_{eq}(x)
\label{boltzmann}
\end{eqnarray}
converges towards the Boltzmann equilibrium $G_{eq}(x) $ 
in the potential $U(x)$.
The equilibrium state $G_{eq}(x) $ is the steady state of the Fokker-Planck dynamics of Eq. \ref{forward1d}
 with no steady current
\begin{eqnarray}
0= \mu(x) G_{eq}(  x)  -  \frac{1}{2} \partial_x  G_{eq}(  x) 
\label{eq}
\end{eqnarray}
For the Laplace transform of Eq. \ref{laplace}, the convergence of Eq. \ref{boltzmann}
means that ${\hat G}_s (x \vert x_0) $ is defined for $s \in ]0,+\infty[$ with the following singularity for $s \to 0^+$
\begin{eqnarray}
{\hat G}_s (x \vert x_0) \opsimeq_{s \to 0^+} \frac{G_{eq}(  x) }{s} 
\label{laplaceseq}
\end{eqnarray}

The simplest example of diffusion converging towards an equilibrium state
is the drift $\mu(x)=-\mu \, {\rm sgn}(x)$ of parameter $\mu>0$
 that will be discussed in section \ref{sec_sgn}.


\subsubsection{ When $H$ has only the continuum $]V_{\infty},+\infty[$ with $V_{\infty}>0$: 
$V_{\infty} $ governs the exponential time decay of $G(x_0,t \vert x_0,t_0)$  }

When $H$ has only the continuous spectrum $]V_{\infty},+\infty[$, where the lower boundary $V_{\infty}$ of Eq. \ref{vinfty} is strictly positive $V_{\infty} >0$,
then the Fokker-Planck propagator $G(x,t \vert x_0,t_0) $ 
and the quantum propagator $\psi(x,t \vert x_0,t_0) $ 
are dominated by the leading exponential time decay involving $V_{\infty} $
\begin{eqnarray}
G(x_0,t \vert x_0,t_0) = e^{ \frac{ U(x_0)- U(x) }{2}} \psi(x_0,t \vert x_0,t_0)
 \oppropto_{(t-t_0) \to +\infty}   e^{- V_{\infty} (t-t_0) }
\label{gtranslarget}
\end{eqnarray}
The physical interpretation is that the diffusion process is transient and flows towards infinity.
For the Laplace transform of Eq. \ref{laplace}, Eq. \ref{gtranslarget} 
means that ${\hat G}_s (x \vert x_0) $ is defined for $s \in ]- V_{\infty},+\infty[$.
In particular, it remains finite for $s=0$ in contrast to the previous case of Eq. \ref{laplaceseq}
\begin{eqnarray}
{\hat G}_{s=0} (x_0 \vert x_0) <+\infty
\label{laplacetrans}
\end{eqnarray}

The simplest example of transient diffusion is the uniform strictly positive drift $\mu(x)=\mu>0$
 that will be discussed in section \ref{sec_brown}.


\subsubsection{ When $H$ has only the continuous spectrum $]V_{\infty}=0 ,+\infty[$
 with the vanishing lower boundary $V_{\infty}=0$ }

When $H$ has only the continuous spectrum $]V_{\infty}=0 ,+\infty[$ with the vanishing lower boundary $V_{\infty}=0$,
then the Fokker-Planck propagator $G(x,t \vert x_0,t_0) $ 
decays in time, but less rapidly than the exponential decay of Eq. \ref{gtranslarget}.

 The simplest example of recurrent diffusion that does not converge towards an equilibrium state
is of course the pure Brownian motion without drift $\mu=0$,
that will be discussed in section \ref{sec_brown}.



\section{ Joint properties of the diffusion process $X(t)$ and its local time $A(t)$ }

\label{sec_xa}

In this section, we focus on the unconditioned joint process $[X(t),A(t)]$ :
the position $X(t)$ and its local time $A(t)$ at the origin satisfy the Ito Stochastic Differential System 
\begin{eqnarray}
dX(t) && =  \mu( X(t) ) dt + dB(t)
\nonumber \\
dA(t) && = \delta ( X(t) ) dt
\label{ito}
\end{eqnarray}
The joint propagator $P(x,A,t \vert x_0,A_0,t_0)$ for the position $x$ and the local time $A$ 
that satisfies the Fokker-Planck dynamics 
\begin{eqnarray}
\partial_t P(x,A,t \vert x_0,A_0,t_0)
= - \delta(x) \partial_A P(x,A,t \vert x_0,A_0,t_0)
 -    \partial_x \left[ \mu(x) P(x,A,t \vert x_0,A_0,t_0) \right] +  \frac{1}{2} \partial_x^2  P(x,A,t \vert x_0,A_0,t_0) 
\label{forwardjoint}
\end{eqnarray}
will be useful to construct conditioned bridges involving both the final position and the final local time,
as will be described in the subsection \ref{subsec_bridgexa}.


\subsection{ Laplace transform ${\tilde P}_{p} (x,t  \vert x_0,t_0) $ with respect to the local time $(A-A_0) \geq 0 $ : Feynman-Kac formula  }

For the Laplace transform ${\tilde P}_{p} (x,t  \vert x_0,t_0) $ of the joint propagator $P(x,A,t \vert x_0,A_0,t_0) $ with respect to the local time increment $(A-A_0) \geq 0 $
\begin{eqnarray}
 {\tilde P}_{p} (x,t  \vert x_0,t_0) \equiv 
\int_{A_0}^{+\infty} dA e^{-p (A-A_0) }P(x,A,t \vert x_0,A_0,t_0)
\label{laplaceA}
\end{eqnarray}
Eq. \ref{forwardjoint}
translates into 
\begin{eqnarray}
\partial_t  {\tilde P}_{p} (x,t  \vert x_0,t_0)
= -  p \delta(x)  {\tilde P}_{p} (0,t  \vert x_0,t_0)
 -    \partial_x \left[ \mu(x)  {\tilde P}_{p} (x,t  \vert x_0,t_0) \right] +  \frac{1}{2} \partial_x^2   {\tilde P}_{p} (x,t  \vert x_0,t_0) 
\label{feynmankac}
\end{eqnarray}
This is a standard example of the Feynman-Kac formula, where the initial Fokker-Planck dynamics 
of Eq. \ref{forward1d} is now supplemented by the additional term in $p \delta(x)$.


\subsection{ Explicit double Laplace transform $ {\hat {\tilde P}}_{s,p} (x \vert x_0)$ of the joint propagator $P(x,A,t \vert x_0,A_0,t_0) $ via the Dyson Eq.}

The further Laplace transform of Eq. \ref{laplaceA}
with respect to the time $(t-t_0)$
\begin{eqnarray}
{\hat {\tilde P}}_{s,p} (x \vert x_0) \equiv \int_{t_0}^{+\infty} dt e^{-s (t-t_0) }  {\tilde P}_{p} (x,t  \vert x_0,t_0)
= \int_{t_0}^{+\infty} dt e^{-s (t-t_0) }\int_{A_0}^{+\infty} dA e^{-p (A-A_0) }P(x,A,t \vert x_0,A_0,t_0)
\label{laplacedouble}
\end{eqnarray}
satisfies
\begin{eqnarray}
- \delta(x-x_0) + s {\hat {\tilde P}}_{s,p}  (x \vert x_0)
= - p \delta(x)  {\hat {\tilde P}}_{s,p}  (0 \vert x_0)
 -       \partial_x \left[ \mu(x) {\hat {\tilde P}}_{s,p}  \right] +  \frac{1}{2} \partial_x^2  {\hat {\tilde P}}_{s,p} 
\label{forward1dlaplacep}
\end{eqnarray}
For $p=0$, Eq. \ref{laplaceA} coincides with the propagator $G(x,t \vert x_0,t_0)$ of the position alone
described in the previous section \ref{sec_x}
\begin{eqnarray}
 {\tilde P}_{p=0} (x,t  \vert x_0,t_0) \equiv 
\int_{A_0}^{+\infty} dA P(x,A,t \vert x_0,A_0,t_0) = G(x,t \vert x_0,t_0)
\label{laplaceApzero}
\end{eqnarray}
As a consequence, Eq. \ref{laplacedouble} becomes
 \begin{eqnarray}
{\hat {\tilde P}}_{s,p=0}  (x \vert x_0) = {\hat G}_s (x \vert x_0)
\label{pzeroG}
\end{eqnarray}
For any $p \ne 0$, the solution ${\hat {\tilde P}}_{s,p}  (x \vert x_0) $ of Eq. \ref{forward1dlaplacep}
satisfies the Dyson equation
\begin{eqnarray}
{\hat {\tilde P}}_{s,p}  (x \vert x_0) = {\hat G}_s (x \vert x_0) -p {\hat G}_s (x \vert 0)  {\hat {\tilde P}}_{s,p}  (0 \vert x_0)
\label{dyson}
\end{eqnarray}
The self-consistency for $x=0$ 
\begin{eqnarray}
{\hat {\tilde P}}_{s,p}  (0 \vert x_0) = {\hat G}_s (0 \vert x_0) -p {\hat G}_s (0 \vert 0)  {\hat {\tilde P}}_{s,p}  (0 \vert x_0)
\label{self}
\end{eqnarray}
yields
\begin{eqnarray}
{\hat {\tilde P}}_{s,p}  (0 \vert x_0) = \frac{{\hat G}_s (0 \vert x_0)}{1 +p {\hat G}_s (0 \vert 0) }
\label{selfzero}
\end{eqnarray}
Plugging this result into Eq. \ref{dyson} yields the final expression of ${\hat {\tilde P}}_{s,p}  (x \vert x_0) $ 
in terms of ${\hat G}_s (. \vert .) $
\begin{eqnarray}
{\hat {\tilde P}}_{s,p}  (x \vert x_0)
=  {\hat G}_s (x \vert x_0) - p   \frac{ {\hat G}_s (x \vert 0) {\hat G}_s (0 \vert x_0)  }{ 1+ p {\hat G}_s (0 \vert 0)}
\label{resum}
\end{eqnarray}


\subsection{ Explicit time Laplace transform 
${\hat P}_{s} (x,A  \vert x_0,A_0) $ of the joint propagator $P(x,A,t \vert x_0,A_0,t_0) $}

The dependence with respect to the parameter $p$ in Eq. \ref{resum}
can be rewritten in terms of a simple pole as
\begin{eqnarray}
{\hat {\tilde P}}_{s,p}  (x \vert x_0)
&&  =  {\hat G}_s (x \vert x_0) - \frac{ {\hat G}_s (x \vert 0){\hat G}_s (0 \vert x_0) }{ {\hat G}_s (0 \vert 0) } 
\left( 1- \frac{ 1   }{ 1+ p {\hat G}_s (0 \vert 0)} \right) 
\nonumber \\
&& = \left[ {\hat G}_s (x \vert x_0) - \frac{ {\hat G}_s (x \vert 0){\hat G}_s (0 \vert x_0) }{ {\hat G}_s (0 \vert 0) } \right]
 + \left[ \frac{ {\hat G}_s (x \vert 0){\hat G}_s (0 \vert x_0) }{ {\hat G}^2_s (0 \vert 0) } \right]
 \frac{ 1   }{  p + \frac{1}{{\hat G}_s (0 \vert 0)} }  
\label{pole}
\end{eqnarray}
So the inverse Laplace transform with respect to $p$ 
yields that the time Laplace transform ${\hat P}_{s} (x,A  \vert x_0,A_0) $
of the joint propagator $P(x,A,t \vert x_0,A_0,t_0) $ reads
\begin{eqnarray}
 {\hat P}_{s} (x,A  \vert x_0,A_0) && \equiv 
 \int_{t_0}^{+\infty} dt e^{-s (t-t_0) }
 P(x,A,t \vert x_0,A_0,t_0)
\nonumber \\
&&  =  \delta(A-A_0)
\left[ {\hat G}_s (x \vert x_0) - \frac{ {\hat G}_s (x \vert 0){\hat G}_s (0 \vert x_0) }{ {\hat G}_s (0 \vert 0) } \right]
 + \theta(A>A_0) \left[ \frac{ {\hat G}_s (x \vert 0){\hat G}_s (0 \vert x_0) }{ {\hat G}^2_s (0 \vert 0) } \right]
 e^{-  \frac{(A-A_0)}{{\hat G}_s (0 \vert 0)} }  
\label{hats}
\end{eqnarray}
The normalization over $A$ corresponding to $p=0$ in Eq. \ref{pzeroG} is given by $ {\hat G}_s (x \vert x_0) $
\begin{eqnarray}
\int_{A_0}^{+\infty} dA {\hat P}_{s} (x,A  \vert x_0,A_0) =  {\hat G}_s (x \vert x_0)
=  \int_{t_0}^{+\infty} dt e^{-s (t-t_0) } G(x,t \vert x_0,t_0)
\label{hatsnorma}
\end{eqnarray}
Let us now explain the physical meaning of the formula of Eq. \ref{hats} in the following subsections.


\subsubsection{ Interpretation of the singular contribution in $\delta(A-A_0)$ of $P (x,A,t  \vert x_0,A_0,t_0) $ in terms of the propagator $G^{abs} (x,t  \vert x_0,t_0) $ }

In Eq. \ref{hats},
the singular contribution involving the delta function $\delta(A-A_0)$ 
\begin{eqnarray}
 {\hat P}^{Singular}_{s} (x,A  \vert x_0,A_0)   =  \delta(A-A_0)
\left[ {\hat G}_s (x \vert x_0) - \frac{ {\hat G}_s (x \vert 0){\hat G}_s (0 \vert x_0) }{ {\hat G}_s (0 \vert 0) } \right]
\label{hatsSingular}
\end{eqnarray}
means that the local time $A$ has kept its initial value $A_0$,
i.e. the diffusion process has not been able to visit the origin $x=0$.
As a consequence, the weight in factor of the delta function $\delta(A-A_0) $
should correspond to the Laplace transform ${\hat G}^{abs}_{s} (x  \vert x_0) $ of the propagator $G^{abs} (x,t \vert x_0,t_0)$ in the presence of an absorbing boundary condition at the origin $x=0$
\begin{eqnarray}
\left[ {\hat G}_s (x \vert x_0) - \frac{ {\hat G}_s (x \vert 0){\hat G}_s (0 \vert x_0) }{ {\hat G}_s (0 \vert 0) } \right]
= {\hat G}^{abs}_{s} (x  \vert x_0) \equiv
 \int_{t_0}^{+\infty} dt e^{-s (t-t_0) } G^{abs}(x,t \vert x_0,t_0) 
\label{laplaceabsorbing}
\end{eqnarray}
This interpretation can be also recovered by considering the limit $p \to +\infty$ in Eq. \ref{pole}
\begin{eqnarray}
{\hat {\tilde P}}_{s,p=+\infty}  (x \vert x_0)
 = \left[ {\hat G}_s (x \vert x_0) - \frac{ {\hat G}_s (x \vert 0){\hat G}_s (0 \vert x_0) }{ {\hat G}_s (0 \vert 0) } \right]
\label{poleinfinity}
\end{eqnarray}
Indeed, in the Feynman-Kac formula of Eq. \ref{feynmankac}, 
the limit of $p \to +\infty$  
amounts to impose the vanishing of the solution ${\tilde P}_{p=+\infty} (x=0,t  \vert x_0,t_0) $ at the origin $x=0$
\begin{eqnarray}
  {\tilde P}_{p=+\infty} (x=0,t  \vert x_0,t_0) =0
\label{pinfty}
\end{eqnarray}
i.e. amounts to impose that the origin $x=0$ is an absorbing boundary condition.

In summary, the singular contribution of Eq. \ref{hatsSingular} can be rewritten as
\begin{eqnarray}
 {\hat P}^{Singular}_{s} (x,A  \vert x_0,A_0)   =  \delta(A-A_0)
 {\hat G}^{abs}_{s} (x  \vert x_0)
\label{hatsSingularAbs}
\end{eqnarray}
and its Laplace inversion with respect to $s$ involves the propagator $G^{abs}(x,t \vert x_0,t_0)  $
in the presence of an absorbing boundary condition at the origin $x=0$
\begin{eqnarray}
 P^{Singular} (x,A,t  \vert x_0,A_0,t_0) 
   =  \delta(A-A_0) G^{abs}(x,t \vert x_0,t_0)
\label{hatsSingularAbsinv}
\end{eqnarray}


\subsubsection{ Corresponding survival probability $S^{abs} (t \vert x_0,t_0) $ and absorption rate $ \gamma^{abs} (t \vert x_0,t_0) $}

The survival probability $S^{abs} (t \vert x_0,t_0) $ at time $t$
in the presence of an absorbing boundary at the origin $x=0$
if one starts at the position $x_0$ at time $t_0$,
can be obtained from the integration of the propagator $G^{abs}(x,t \vert x_0,t_0) $ over the final position $x$
\begin{eqnarray}
S^{abs} (t \vert x_0,t_0) \equiv  \int_{-\infty}^{+\infty} dx G^{abs}(x,t \vert x_0,t_0)
\label{survival}
\end{eqnarray}
 The conservation of probability for the full propagator $G(x,t \vert x_0,t_0) $
 \begin{eqnarray}
\int_{-\infty}^{+\infty} dx G(x,t \vert x_0,t_0) = 1
\label{normaGpropagator}
\end{eqnarray}
 translates for its Laplace transform into
\begin{eqnarray}
\int_{-\infty}^{+\infty} dx  {\hat G}_s (x \vert x_0)
=  \int_{t_0}^{+\infty} dt e^{-s (t-t_0) }
\int_{-\infty}^{+\infty} dx G(x,t \vert x_0,t_0) = \int_{t_0}^{+\infty} dt e^{-s (t-t_0) } = \frac{1}{s}
\label{hatsnormaintegx}
\end{eqnarray}
So the time Laplace transform of Eq. \ref{survival} reads
via the integration of Eq. \ref{laplaceabsorbing} over $x$
\begin{eqnarray}
 {\hat S}^{abs}_s ( x_0) \equiv \int_{t_0}^{+\infty} dt e^{-s (t-t_0) } S^{abs} (t \vert x_0,t_0)
 = \int_{-\infty}^{+\infty} dx {\hat G}^{abs}_{s} (x  \vert x_0)
 =  \frac{1}{s} \left[ 1 - \frac{ {\hat G}_s (0 \vert x_0) }{ {\hat G}_s (0 \vert 0) } \right] 
\label{laplacesurvival}
\end{eqnarray}
It is now useful to introduce the absorption rate $\gamma^{abs}(t \vert x_0,t_0) $ 
at time $t$
when starting at the position $x_0$ at time $t_0$
\begin{eqnarray}
\gamma^{abs}(t \vert x_0,t_0)  \equiv - \partial_t S^{abs} (t \vert x_0,t_0) 
\label{gammaabs}
\end{eqnarray}
Its time Laplace transform is simple, as shown via the following integration by parts
\begin{eqnarray}
{\hat \gamma}^{abs}_s ( x_0) 
&& = - \int_{t_0}^{+\infty} dt e^{-s (t-t_0) } \partial_t S^{abs} (t \vert x_0,t_0) 
= - \left[e^{-s (t-t_0) } S^{abs} (t \vert x_0,t_0)  \right]_{t_0}^{+\infty}
-s  {\hat S}^{abs}_s ( x_0) = 1 - s {\hat S}^{abs}_s ( x_0)
\nonumber \\
&&  = \frac{ {\hat G}_s (0 \vert x_0) }{ {\hat G}_s (0 \vert 0) }
\label{laplacegammaabs}
\end{eqnarray}


\subsubsection{ Interpretation of the regular contribution in $\theta(A>A_0)$ 
for $P (x,A ,t \vert x_0,A_0,t_0) $}

For the special case where the initial and the final positions are at the origin $x=0=x_0$, Eq. \ref{hats} reduces to
\begin{eqnarray}
 {\hat P}_{s} (x=0,A  \vert x_0=0,A_0) 
   =  \theta(A>A_0) e^{-  \frac{(A-A_0)}{{\hat G}_s (0 \vert 0)} }  
\label{hats00}
\end{eqnarray}
Its normalization over $A$  
\begin{eqnarray}
\int dA  {\hat P}_{s} (x=0,A  \vert x_0=0,A_0) = {\hat G}_s (0 \vert 0) \equiv \int_{t_0}^{+\infty} dt e^{-s (t-t_0) } G(  0,t \vert   0,t_0) 
\label{laplace00}
\end{eqnarray}
involves the propagator $G(  0,t \vert   0,t_0) $ from the origin to the origin.

The regular contribution of Eq. \ref{hats}
\begin{eqnarray}
 {\hat P}^{Regular}_{s} (x,A  \vert x_0,A_0)  = \theta(A>A_0) \left[ \frac{ {\hat G}_s (x \vert 0){\hat G}_s (0 \vert x_0) }{ {\hat G}^2_s (0 \vert 0) } \right]
 e^{-  \frac{(A-A_0)}{{\hat G}_s (0 \vert 0)} }  
\label{hatsRegular}
\end{eqnarray}
can be thus rewritten as the product of the three following functions using Eqs \ref{laplacegammaabs}
and \ref{hats00}
\begin{eqnarray}
 {\hat P}^{Regular}_{s} (x,A  \vert x_0,A_0)  =
   {\hat \gamma}^{abs}_s ( x)
    {\hat P}_{s} (0,A  \vert 0,A_0) 
   {\hat \gamma}^{abs}_s ( x_0)
\label{hatsRegular3terms}
\end{eqnarray}
Its Laplace inversion involves the time-convolution of the three functions
\begin{eqnarray}
 P^{Regular} (x,A,t  \vert x_0,A_0,t_0) 
   =  \int_{t_0}^t dt_1 \int_{t_1}^t dt_2 
      \gamma^{abs} ( t \vert x, t_2)
    P (0,A , t_2 \vert 0,A_0, t_1) 
   \gamma^{abs} (t_1 \vert  x_0, t_0)
\label{hatsRegular3termsinv}
\end{eqnarray}
with the following physical meaning.

(i) The time $t_1$ is the first passage time at the origin if one starts at the initial position $x_0$ at time $t_0$,
whose statistics is governed by the absorption rate $\gamma^{abs} (t_1 \vert  x_0, t_0) $.

(ii) The time $t_2$ is the last passage time at the origin before reaching the final point $x$ at time $t$,
where the statistics of the time interval $(t-t_2)$ is governed by 
absorption rate $\gamma^{abs} (t \vert  x, t_2) $ of the alternative 
problem when one starts at position $x$ at time $t_2$.

(iii) Between the first-passage-time $t_1$ and the last-passage-time $t_2$ at the origin,
the statistics of the local time increment $(A-A_0)$ is governed by the probability $P (0,A , t_2 \vert 0,A_0, t_1) $.


\section{ Probability distribution $\Pi(A,t \vert x_0,A_0,t_0)$ of the local time $A$ at time $t$}

\label{sec_a}

In this section, we focus on the
the probability $\Pi(A,t \vert x_0,A_0,t_0)  $ to see the local time $A$ at time $t$ 
if one starts at position $x_0$ with the local time $A_0$ at time $t_0$.
It can be obtained from the integration  over the final position $x$ of 
the joint propagator $P(x,A,t \vert x_0,A_0,t_0) $ studied in the previous section
and it can be thus decomposed
 into a singular contribution in $\delta(A-A_0)$ and a regular contribution in $\theta(A>A_0)$
\begin{eqnarray}
 \Pi(A,t \vert x_0,A_0,t_0) &&  \equiv \int_{-\infty}^{+\infty} dx  P(x,A,t \vert x_0,A_0,t_0)
 \nonumber \\
 && = \Pi^{Singular } (A,t \vert x_0,A_0,t_0) + \Pi^{Regular } (A,t \vert x_0,A_0,t_0)
\label{propagAalone}
\end{eqnarray}
This probability of Eq. \ref{propagAalone}
will be useful to construct conditioned bridges involving the local time,
as will be described in the subsection \ref{subsec_bridgea}.


\subsection{  Explicit Laplace transform ${\hat \Pi}_{s} (A  \vert x_0,A_0) $ 
of the probability $\Pi (A,t  \vert x_0,A_0,t_0) $}

The Laplace transform of Eq. \ref{propagAalone} with respect to the time interval $(t-t_0)$
\begin{eqnarray}
 {\hat \Pi}_{s} (A  \vert x_0,A_0) \equiv \int_{t_0}^{+\infty} dt e^{-s (t-t_0) } \Pi(A,t \vert x_0,A_0,t_0)
 =  \int_{-\infty}^{+\infty} dx  {\hat P}_s(x,A \vert x_0,A_0) 
\label{laplaceAalone}
\end{eqnarray}
can be obtained via the integration over $x$ of ${\hat P}_s(x,A \vert x_0,A_0)  $ given by Eq. \ref{hats}
using Eq. \ref{hatsnormaintegx}
\begin{eqnarray}
{\hat \Pi}_{s} (A  \vert x_0,A_0) 
 && =  \Pi^{Singular }_s (A \vert x_0,t_0) + \Pi^{Regular }_s (A \vert x_0,A_0)
 \nonumber \\
 \Pi^{Singular }_s (A \vert x_0,t_0)  && =  \delta(A-A_0) \frac{1}{s} 
\left[ 1 - \frac{{\hat G}_s (0 \vert x_0) }{ {\hat G}_s (0 \vert 0) } \right]
 \nonumber \\
\Pi^{Regular }_s (A \vert x_0,A_0) && =  \theta(A>A_0) \left[ \frac{ {\hat G}_s (0 \vert x_0) }{ s {\hat G}^2_s (0 \vert 0) } \right]
 e^{-  \frac{(A-A_0)}{{\hat G}_s (0 \vert 0)} }  
\label{hatsintegrate}
\end{eqnarray}


\subsubsection{ Interpretation of the singular contribution in $\delta(A-A_0) $ }

The singular contribution $ \Pi^{Singular }_s (A \vert x_0,t_0) $ of Eq. \ref{hatsintegrate} involves 
the time Laplace transform $ {\hat S}^{abs}_s ( x_0) $ of Eq. \ref{laplacesurvival}
\begin{eqnarray}
{\hat \Pi}^{Singular}_{s} (A  \vert x_0,A_0) 
  =  \delta(A-A_0) \frac{1}{s} 
\left[ 1 - \frac{{\hat G}_s (0 \vert x_0) }{ {\hat G}_s (0 \vert 0) } \right] 
=  \delta(A-A_0)
{\hat S}^{abs}_s ( x_0)  
\label{hatsintegratesingular}
\end{eqnarray}
and its Laplace inversion involves the survival probability $S^{abs} ( t \vert x_0, t_0) $ of Eq. \ref{survival}
\begin{eqnarray}
 \Pi^{Singular} (A ,t \vert x_0,A_0,t_0) 
=  \delta(A-A_0)
S^{abs} ( t \vert x_0, t_0)  
\label{hatsintegratesingularinv}
\end{eqnarray}


\subsubsection{ Interpretation of the regular contribution in $\theta(A>A_0)$ }

For the special case where the initial position vanishes $x_0=0$,
Eq. \ref{hatsintegrate} reduces to
\begin{eqnarray}
 {\hat \Pi}_{s} (A  \vert x_0=0,A_0)   =  {\hat \Pi}^{Regular}_{s} (A  \vert x_0=0,A_0)
= \frac{\theta(A>A_0)}{ s {\hat G}_s (0 \vert 0) } e^{-  \frac{(A-A_0)}{{\hat G}_s (0 \vert 0)} }  
\label{hatsintegrate00}
\end{eqnarray}
As a consequence, the regular contribution ${\hat \Pi}^{Regular}_{s} (A  \vert x_0,A_0) $ of Eq. \ref{hatsintegrate}
 can be rewritten as the product of two functions using Eq. \ref{laplacegammaabs}
 and \ref{hatsintegrate00}
\begin{eqnarray}
 {\hat \Pi}^{Regular}_{s} (A  \vert x_0,A_0)   = \theta(A>A_0)  \frac{ {\hat G}_s (0 \vert x_0) }{ s {\hat G}^2_s (0 \vert 0) } 
 e^{-  \frac{(A-A_0)}{{\hat G}_s (0 \vert 0)} }  
 =
   {\hat \Pi}_{s} (A  \vert x_0=0,A_0)   {\hat \gamma}^{abs}_s ( x_0) 
\label{hatsintegrateregular}
\end{eqnarray}
Its Laplace inversion with respect to $s$ can be written as the time-convolution of two functions
\begin{eqnarray}
  \Pi^{Regular}(A ,t \vert x_0,A_0,t_0)   = 
    \int_{t_0}^t d t_1   \Pi (A, t  \vert 0,A_0,t_1)  \gamma^{abs} ( t_1 \vert x_0,t_0) 
\label{hatsintegrateregularinv}
\end{eqnarray}
with the following physical meaning :

(i) as in Eq. \ref{hatsRegular3termsinv},
the time $t_1$ is the first passage time at the origin if one starts at the initial position $x_0$ at time $t_0$,
whose statistics is governed by the absorption rate $\gamma^{abs} (t_1 \vert  x_0, t_0) $.

(ii) for the remaining time interval $(t-t_1)$, 
$\Pi (A, t  \vert 0,A_0,t_1)$ represents the probability to see the local time increment $(A-A_0)$  when starting at the origin $x_0=0$.



\subsection{  Moments $m^{[k]}(t \vert x_0,t_0)$ of order $k=1,2,.. $ of the local time increment $[A(t)-A(t_0)]$  }

\subsubsection{ Computation of the moments $m^{[k]}(t \vert x_0,t_0) $ from the probability distribution $ \Pi(A,t \vert x_0,A_0,t_0) $ }

The moments of order $k =1,2,... $ of the local time increment $[A(t)-A(t_0)]$  
only involve the regular part  $\Pi^{Regular } (A,t \vert x_0,A_0,t_0)$ of the probability distribution
$ \Pi(A,t \vert x_0,A_0,t_0) $ of Eq. \ref{propagAalone}
\begin{eqnarray}
m^{[k]}(t \vert x_0,t_0) && \equiv \int_{A_0}^{+\infty} dA \left[ A-A_0 \right]^k\Pi(A,t \vert x_0,A_0,t_0) 
\nonumber \\
&& = \int_{A_0}^{+\infty} dA \left[ A-A_0 \right]^k\Pi^{Regular}(A,t \vert x_0,A_0,t_0) 
\label{mkdef}
\end{eqnarray}
Their Laplace transforms with respect to the time interval $(t-t_0)$ can be obtained from ${\hat \Pi}^{Regular}_{s} (A  \vert x_0,A_0) $ of Eq. \ref{hatsintegrateregular}
\begin{eqnarray}
m^{[k]}_s( x_0) && \equiv \int_{t_0}^{+\infty} dt e^{-s (t-t_0)} m^{[k]}(t \vert x_0,t_0) 
=\int_{A_0}^{+\infty} dA \left[ A-A_0 \right]^k  {\hat \Pi}^{Regular}_{s} (A  \vert x_0,A_0) 
\nonumber \\
&&   = \frac{ {\hat G}_s (0 \vert x_0) }{ s {\hat G}^2_s (0 \vert 0) }  \int_{A_0}^{+\infty} dA \left[ A-A_0 \right]^k 
 e^{-  \frac{(A-A_0)}{{\hat G}_s (0 \vert 0)} }  
 =  \frac{ k! \left[ {\hat G}_s (0 \vert 0) \right]^{k-1} {\hat G}_s (0 \vert x_0) }{ s  } 
\label{mklaplace}
\end{eqnarray}

\subsubsection{ Alternative computation of the moments $m^{[k]}(t \vert x_0,t_0) $ from the definition of Eq. \ref{localtimex} }

Alternatively, one can use the definition of Eq. \ref{localtimex}
\begin{eqnarray}
A(t) -A(t_0) = \int_{t_0}^{t} d\tau \delta(X(\tau)) 
\label{localtime0}
\end{eqnarray}
to compute the moment of order $k$ for trajectories starting at $X(t_0)=x_0$ 
 in terms of the propagator $G(x, t \vert y, t')$ 
\begin{eqnarray}
&& m^{[k]}(t \vert x_0,t_0)  
=\int_{t_0}^{t} dt_k \int_{t_0}^{t} dt_{k-1}...   \int_{t_0}^{t} dt_2  \int_{t_0}^{t} dt_1
\langle  \delta(X(t_k))\delta(X(t_{k-1})) ...
\delta(X(t_2)) \delta(X(t_1))  \delta(X(t_0)-x_0) \rangle
\nonumber \\
&& = k! \int_{t_0}^{t} dt_k \int_{t_0}^{t_k} dt_{k-1}...   \int_{t_0}^{t_3} dt_2  \int_{t_0}^{t_2} dt_1
\langle  \delta(X(t_k))\delta(X(t_{k-1})) ...
\delta(X(t_2)) \delta(X(t_1))  \delta(X(t_0)-x_0) \rangle
\nonumber \\
&& = k! \int_{t_0}^{t} dt_k \int_{t_0}^{t_k} dt_{k-1}...   \int_{t_0}^{t_3} dt_2  \int_{t_0}^{t_2} dt_1
G(0, t_k \vert 0,t_{k-1}) 
G(0, t_{k-1} \vert 0,t_{k-2}) 
...
G(0, t_2 \vert 0,t_1)
G(0, t_1 \vert x_0,t_0)
\label{mkdirect}
\end{eqnarray}
so that its Laplace transform
\begin{eqnarray}
&& m^{[k]}_s( x_0)  = \int_{t_0}^{+\infty} dt e^{-s (t-t_0)}  m^{[k]}(t \vert x_0,t_0)  
\nonumber \\
&& = k! \int_{t_0}^{+\infty} dt e^{-s (t-t_k)-s(t_k-t_{k-1}) -... -s(t_2-t_1)-s(t_1-t_0)} 
\nonumber \\
&& \int_{t_0}^{t} dt_k \int_{t_0}^{t_k} dt_{k-1}...   \int_{t_0}^{t_3} dt_2  \int_{t_0}^{t_2} dt_1
G(0, t_k \vert 0,t_{k-1}) G(0, t_{k-1} \vert 0,t_{k-2}) 
...
G(0, t_2 \vert x_0,t_1)
G(0, t_1 \vert x_0,t_0)
\nonumber \\
&& = k!  \int_{t_0}^{+\infty} dt_1e^{-s(t_1-t_0) } G(0, t_1 \vert x_0,t_0) 
... \int_{t_0}^{+\infty} dt_k e^{-s(t_k-t_{k-1})} 
G(0, t_k \vert 0,t_{k-1}) \int_{t_k}^{+\infty} dt e^{-s (t-t_k)}
\nonumber \\
&& =
\frac{ k! \left[ {\hat G}_s (0 \vert 0) \right]^{k-1} {\hat G}_s (0 \vert x_0) }{ s  } 
\label{mkdirectlaplace}
\end{eqnarray}
coincides with Eq. \ref{mklaplace} as it should.

\subsubsection{ Example of the two first moments for $k=1$ and $k=2$ }

The first moment $k=1$ of Eq. \ref{mkdirect} reduces to the single time integral 
\begin{eqnarray}
 m^{[k=1]}(t \vert x_0,t_0)  
 =  \int_{t_0}^{t}  dt_1
G(0, t_1 \vert x_0,t_0)
\label{m1}
\end{eqnarray}
Its growth is thus directly governed by the propagator $G(0, t \vert x_0,t_0) $
\begin{eqnarray}
\partial_t m^{[k=1]}(t \vert x_0,t_0)   = G(0, t \vert x_0,t_0) >0
\label{m1deri}
\end{eqnarray}
The Laplace transform of Eq. \ref{mklaplace} reads
\begin{eqnarray}
 m^{[k=1]}_s( x_0)  = \frac{  {\hat G}_s (0 \vert x_0) }{ s  } 
\label{m1laplace}
\end{eqnarray}

The second moment $k=2$ of Eq. \ref{mkdirect} reads  
\begin{eqnarray}
 m^{[k=2]}(t \vert x_0,t_0)  
 =  2   \int_{t_0}^{t} dt_2  \int_{t_0}^{t_2} dt_1 G(0, t_2 \vert 0,t_1) G(0, t_1 \vert x_0,t_0)
\label{m2}
\end{eqnarray}
with its Laplace transform of Eq. \ref{mklaplace}
\begin{eqnarray}
 m^{[k=2]}_s( x_0)  =
\frac{ 2  {\hat G}_s (0 \vert 0) {\hat G}_s (0 \vert x_0) }{ s  } 
\label{m2laplace}
\end{eqnarray}


\section{ Statistics of the local time increment $(A-A_0)$ for large time interval $(t-t_0)$}

\label{sec_largetime}

In this section, we describe how
the recurrence/transience properties of the diffusion process $X(t)$ induced by the drift $\mu(x)$ 
produce very different behaviors for the scaling of the 
 local time increment $(A-A_0)$ with respect to the large time interval $(t-t_0)$.


\subsection{When $X(t)$ is transient  : the local time increment $(A-A_0)$ remains finite for $(t-t_0) \to +\infty$  }  

\label{subsec_transient}

Among transient diffusions, the simplest example is the uniform strictly positive drift $\mu(x)=\mu>0$
 that will be discussed in section \ref{sec_brown}.
 
When the diffusion process $X(t)$ is transient with the exponential time decay of Eq. \ref{gtranslarget} for the propagator $G(x,t \vert x_0,t)$, then the first moment $m^{[k=1]}(t \vert x_0,t_0) $ of the 
local time increment $[A(t)-A(t_0)]$ of Eq. \ref{m1} 
converges towards the finite value $m^{[k=1]}(\infty \vert x_0) $ for $(t-t_0) \to +\infty$
\begin{eqnarray}
 m^{[k=1]}(t \vert x_0,t_0)  \opsimeq_{(t-t_0) \to + \infty} m^{[k=1]}(\infty \vert x_0) = \int_{0}^{+\infty}  dt_1
G(0, t_1 \vert x_0,0)< +\infty
\label{m1finite}
\end{eqnarray}

More generally, the local time increment $(A-A_0)$ will remain 
a finite random variable for $(t-t_0) \to +\infty$
with the following notation for the limit of the distribution $\Pi( A,t \vert  x_0,A_0,t_0) $ of Eq. \ref{propagAalone}
\begin{eqnarray}
\Pi( A, \infty \vert  x_0,A_0) && = \lim_{(t-t_0) \to +\infty} \Pi( A,t \vert  x_0,A_0,t_0)
\nonumber \\
&&  = \Pi^{Singular } (A,\infty \vert x_0,A_0) + \Pi^{Regular } (A,\infty \vert x_0,A_0)
\label{piAlimit}
\end{eqnarray}
Let us now discuss its singular and regular contributions.

(i) The singular contribution $\Pi^{Singular } (A,\infty \vert x_0,A_0) $ involves the infinite-time limit of Eq. \ref{hatsintegratesingularinv}
\begin{eqnarray}
 \Pi^{Singular} (A ,\infty \vert x_0,A_0,t_0) 
=  \delta(A-A_0)
S^{abs} ( \infty \vert x_0)  
\label{hatsintegratesingularinvinfty}
\end{eqnarray}
that involves the probability to survive forever 
\begin{eqnarray}
S^{abs} ( \infty \vert x_0)= \lim_{(t-t_0) \to +\infty} S^{abs} ( t \vert x_0, t_0)
\label{survivallimit}
\end{eqnarray}
This probability to escape towards infinity without visiting the origin $x=0$
can be obtained from the Laplace transform $ {\hat S}^{abs}_s ( x_0) $ of Eq. \ref{laplacesurvival}
by considering the limit $s \to 0$ of
\begin{eqnarray}
S^{abs} ( \infty \vert x_0)  = \lim_{s \to 0} \left[ s  {\hat S}^{abs}_s ( x_0)\right]
=  \lim_{s \to 0} \left[ 1 - \frac{ {\hat G}_s (0 \vert x_0) }{ {\hat G}_s (0 \vert 0) } \right]
=1 - \frac{ {\hat G}_0 (0 \vert x_0) }{ {\hat G}_0 (0 \vert 0) }
\label{survivallimitfromlaplace}
\end{eqnarray}

(ii) The Regular contribution $\Pi^{Regular } (A,\infty \vert x_0,A_0) $
can be obtained from the Laplace transform ${\hat \Pi}^{Regular}_{s} (A  \vert x_0,A_0) $ of Eq. \ref{hatsintegrateregular}
by considering the limit $s \to 0$ of
\begin{eqnarray}
\Pi^{Regular}( A, \infty \vert  x_0,A_0) 
&& = \lim_{s \to 0} \left[ s {\hat \Pi}^{Regular}_{s} (A  \vert x_0,A_0)\right]
= \lim_{s \to 0} \left[   \theta(A>A_0)  \frac{ {\hat G}_s (0 \vert x_0) }{  {\hat G}^2_s (0 \vert 0) } 
 e^{-  \frac{(A-A_0)}{{\hat G}_s (0 \vert 0)} }   \right]
 \nonumber \\
 && = \theta(A>A_0)  \frac{ {\hat G}_0 (0 \vert x_0) }{  {\hat G}^2_0 (0 \vert 0) } 
 e^{-  \frac{(A-A_0)}{{\hat G}_0 (0 \vert 0)} }  
 \nonumber \\
 && = \theta(A>A_0) \left[ 1- S^{abs} ( \infty \vert x_0) \right]  \frac{ 1 }{  {\hat G}_0 (0 \vert 0) } 
 e^{-  \frac{(A-A_0)}{{\hat G}_0 (0 \vert 0)} }  
\label{piAlimitregularfromlaplace}
\end{eqnarray}
Its physical meaning can be understood as follows : with the complementary probability 
$\left[ 1- S^{abs} ( \infty \vert x_0) \right] $ with respect to Eq. \ref{survivallimitfromlaplace},
the diffusion process $X(t)$ visits the origin before escaping towards infinity,
and then the local time $(A-A_0) $ is an exponential random variable with the finite scale 
\begin{eqnarray}
 {\hat G}_{s=0} (0 \vert 0) \equiv \int_{0}^{+\infty} dt  G(  0,t \vert   0,0) 
\label{laplace000}
\end{eqnarray}


\subsection{ When $X(t)$ converges towards an equilibrium : the increment $(A-A_0)$ grows extensively in $(t-t_0)$  }  

\label{subsec_equilibrium}

Among diffusions converging towards equilibrium, one simple example is
is the drift $\mu(x)=-\mu \, {\rm sgn}(x)$ of parameter $\mu>0$
 that will be discussed in section \ref{sec_sgn}.

When the diffusion process $X(t)$ converges towards the 
Boltzmann equilibrium state of Eq. \ref{boltzmann}
\begin{eqnarray}
 G_{eq}(  x) = \frac{ e^{-U(x) }}{ \int_{-\infty}^{+\infty} dy e^{-U(y) } } = \left[  \phi^{GS}_0(  x)\right]^2
\label{boltzmanngs}
\end{eqnarray}
then the first moment $m^{[k=1]}(t \vert x_0,t_0) $ of the local time increment $[A(t)-A(t_0)]$ 
discussed in Eqs \ref{m1}, \ref{m1deri}
is extensive with respect to the time interval $(t-t_0)$
\begin{eqnarray}
 m^{[k=1]}(t \vert x_0,t_0)  \opsimeq_{(t-t_0) \to + \infty} (t-t_0) G_{eq}(x=0)
\label{m1extensive}
\end{eqnarray}
The corresponding intensive local time
\begin{eqnarray}
a \equiv \frac{A-A_0}{t-t_0}
\label{additiveIntensive}
\end{eqnarray}
then converges in the thermodynamic limit $(t-t_0) = +\infty$
 towards its equilibrium value 
\begin{eqnarray}
 a_{eq} = G_{eq}(  0) = \frac{ e^{-U(0) }}{ \int_{-\infty}^{+\infty} dy e^{-U(y) } } = \left[  \phi^{GS}_0(  0)\right]^2
\label{g1}
\end{eqnarray} 
For large but finite $(t-t_0)$, it is thus interesting to analyze its large deviations properties.

\subsubsection { Large deviations properties of the intensive local time $a = \frac{A-A_0}{t-t_0} $ }

\label{subsec_largedeva}

The probability 
$\Pi (A=A_0+(t-t_0) a ,t \vert x_0,A_0,t_0)$ to see the intensive local time
$a$ different from its 
equilibrium value $a_{eq}$
will display the large deviation form with respect to $(t-t_0)$
 \begin{eqnarray}
 \Pi (A=A_0+(t-t_0) a ,t \vert x_0,A_0,t_0) \oppropto_{(t-t_0) \to +\infty} e^{- (t-t_0) I ( a )} 
\label{level1def}
\end{eqnarray} 
The positive rate function $I(a) \geq 0 $ is defined for $a \in [0,+\infty[$ 
and vanishes only for the equilibrium value $a_{eq}$ of Eq. \ref{g1}
where it is minimum
 \begin{eqnarray}
 I ( a_{eq} ) =0 =  I' ( a_{eq} )
\label{iaeqvanish}
\end{eqnarray}
The Central Limit theorem governing the small Gaussian fluctuations around $a_{eq}$
can be recovered via the Taylor expansion at second order of the rate function $I(a)$ around $a_{eq}$
 \begin{eqnarray}
 I ( a ) = \frac{(a-a_{eq})^2}{2} I''( a_{eq} )+ o \left( (a-a_{eq})^2 \right)
\label{taylor}
\end{eqnarray}

The link with the singular and regular contributions of $ \Pi (A ,t \vert x_0,A_0,t_0) $ can be understood as follows.

(i) The survival probability $S^{abs} ( t \vert x_0, t_0)   $ representing the weight of the singular contribution 
$ \Pi^{Singular} (A ,t \vert x_0,A_0,t_0) $ of Eq. \ref{hatsintegratesingularinv} 
corresponds to the value $a=0$
and will thus display the following exponential decay with respect to $(t-t_0)$
that involves the boundary value $I(a=0)$
\begin{eqnarray}
S^{abs} ( t \vert x_0, t_0)   \oppropto_{(t-t_0) \to +\infty} e^{- (t-t_0) I ( a=0 )}
\label{sabsI0}
\end{eqnarray}

(ii) For the regular contribution, the compatibility at leading order in the exponentials
between the large deviation form of Eq. \ref{level1def}
and the Laplace transform $ {\hat \Pi}^{Regular}_{s} (A  \vert x_0,A_0)  $ of Eq. \ref{hatsintegrateregular}
yields \cite{occupationsinai} 
\begin{eqnarray}
 e^{-  \frac{ (A-A_0) }{{\hat G}_s (0 \vert 0)} }  
&& \opsimeq_{(A-A_0) \to +\infty } \int_{t_0}^{+\infty} dt e^{-s (t-t_0)} e^{- (t-t_0) I \left( \frac{A-A_0}{t-t_0}\right) }
\label{regularlargedevcompatibility}
\end{eqnarray}
It is thus convenient to make a change of variable in the integral from the time $t$ towards the intensive local time $a=\frac{A-A_0}{t-t_0}$
\begin{eqnarray}
 e^{-  \frac{ (A-A_0) }{{\hat G}_s (0 \vert 0)} }  
 \opsimeq_{(A-A_0) \to +\infty } \int_{0}^{+\infty} \frac{ da}{a^2} (A-A_0) e^{- (A-A_0) \frac{ s+ I (a)}{a} }
\label{regularlargedevcompatibilitya}
\end{eqnarray}
For large increment $(A-A_0) \to +\infty$, the evaluation of this integral via the saddle-point method 
 allows to obtain the following link between ${\hat G}_s (0 \vert 0) $ and the rate function $I(a)$ \cite{occupationsinai} 
\begin{eqnarray}
  \frac{ 1 }{{\hat G}_s (0 \vert 0)}   && =  \frac{ s+ I (a)}{a}
  \nonumber \\
  0 && = \partial_a \left[  \frac{ s+ I (a)}{a} \right] = \frac{ s+ I' (a)}{a} - \frac{ s+ I (a)}{a^2}  
\label{regularlargedevcompatibilityasaddle}
\end{eqnarray}
This quasi-Legendre transform can be written in reciprocal form in order to compute 
the rate function $I(a)$ from the knowledge of ${\hat G}_s (0 \vert 0) $, as discussed in detail in the next subsection.

\subsubsection { Evaluation of the leading order of $\Pi^{Regular} (A=A_0+(t-t_0) a,t  \vert x_0,A_0,t_0)  $ with the prefactors }

For the Doob conditioned processes that will be discussed in section \ref{sec_Doob},
one needs to compute the dependence with respect to the initial position $x_0$,
so that one needs to
include the prefactors in the reciprocal calculation
concerning the Laplace inverse of ${\hat \Pi}^{Regular}_{s} (A  \vert x_0,A_0) $ of Eq. \ref{hatsintegrateregular}
\begin{eqnarray}
\Pi^{Regular} (A,t  \vert x_0,A_0,t_0) 
&& = \int_{c-i\infty}^{c+i\infty} \frac{ds}{2 i \pi}  e^{s (t-t_0)} {\hat \Pi}^{Regular}_{s} (A  \vert x_0,A_0)
\nonumber \\
&&  = \int_{c-i\infty}^{c+i\infty} \frac{ds}{2 i \pi}  e^{s (t-t_0)}  \frac{ {\hat G}_s (0 \vert x_0) }{ s {\hat G}^2_s (0 \vert 0) } 
 e^{-  \frac{(A-A_0)}{{\hat G}_s (0 \vert 0)} }  
\label{laplaceinverse}
\end{eqnarray}
Here the goal is to evaluate this regular contribution for $A=A_0+(t-t_0)a$
\begin{eqnarray}
\Pi^{Regular} (A=A_0+(t-t_0) a,t  \vert x_0,A_0,t_0) 
&&  = \int_{c-i\infty}^{c+i\infty} \frac{ds}{2 i \pi}   \frac{ {\hat G}_s (0 \vert x_0) }{ s {\hat G}^2_s (0 \vert 0) } 
 e^{- (t-t_0) \left[   \frac{a}{{\hat G}_s (0 \vert 0)} - s \right] }  
\label{laplaceinversea}
\end{eqnarray}
For large $(t-t_0)$, the saddle-point evaluation of this integral will be governed by the solution $s_a$ of
the following equation in $s$ as a function of the parameter $a$
\begin{eqnarray}
  0  =   \partial_s \left[  \frac{ a }{{\hat G}_s (0 \vert 0)}  -s   \right] = a   \partial_s \left[  \frac{ 1 }{{\hat G}_s (0 \vert 0)}     \right] -1 \ \ \ \ \ {\rm with \ \ solution} \ \ s=s_a
\label{saddlesa}
\end{eqnarray}
In the integral of Eq. \ref{laplaceinversea}, one then needs to make the change of variable
around this saddle-point value $s_a$
\begin{eqnarray}
 s = s_a + i \omega
\label{complex}
\end{eqnarray}
The Taylor expansion at second order in $\omega$ of the function in the exponential
\begin{eqnarray}
 \left[ \frac{ a }{{\hat G}_s (0 \vert 0)}  -s   \right]_{s=s_a + i \omega}  
 = I(a) + 0 + \frac{\omega^2}{2} K(a) + o\left( \omega^2 \right)
\label{saddlesaexp}
\end{eqnarray}
involves the two functions
\begin{eqnarray}
I(a) && =   \left( \frac{ a }{{\hat G}_s (0 \vert 0)}  -s   \right)\bigg\vert_{s=s_a}
\nonumber \\
K(a) && = -  \left( \partial^2_s \left[  \frac{ a }{{\hat G}_s (0 \vert 0)}  -s   \right] \right) \bigg\vert_{s=s_a}
\label{iafromsaddle}
\end{eqnarray}
In particular, the rate function $I(a)$ can be computed from the knowledge of ${\hat G}_s (0 \vert 0) $
via Eq. \ref{iafromsaddle} using the saddle-point value $s_a$ determined by Eq. \ref{saddlesa}.
As it should for consistency, Eqs \ref{iafromsaddle} and \ref{saddlesa} correspond to the reciprocal quasi-Legendre transform of Eq. \ref{regularlargedevcompatibilityasaddle}.
Simple examples will be given in Eqs \ref{Iamu0} \ref{laplacefreemuzero} \ref{saddlesamuzero}
\ref{saddlesamuzerosol}, as well as
in Eqs \ref{laplacesgn00} \ref{Iasgn}
\ref{saddlesasgn}
\ref{saddlesasgnsol}.

Putting everything together, one obtains the final result for the leading order of Eq. \ref{laplaceinversea} based on the 
remaining Gaussian integral over $\omega$
\begin{eqnarray}
\Pi^{Regular} (A=A_0+(t-t_0) a,t  \vert x_0,A_0,t_0) 
&&  \opsimeq_{(t-t_0) \to +\infty}  
   \frac{ {\hat G}_{s_a} (0 \vert x_0) }{ s_a {\hat G}^2_{s_a} (0 \vert 0) } 
   e^{- (t-t_0) I(a) } \int_{\infty}^{\infty} \frac{d\omega}{2  \pi}   
 e^{- (t-t_0)  \frac{ K(a)}{2} \omega^2   }  
\nonumber \\
&&   \opsimeq_{(t-t_0) \to +\infty}     \frac{ {\hat G}_{s_a} (0 \vert x_0) }{ s_a {\hat G}^2_{s_a} (0 \vert 0) } 
\frac{ e^{- (t-t_0) I(a) }  }{ \sqrt{ 2 \pi (t-t_0) K(a) } }
\label{laplaceinverseasaddle}
\end{eqnarray}
Note that the dependence with respect to the initial position $x_0$ is only in the function 
${\hat G}_{s_a} (0 \vert x_0) $ evaluated for the saddle-point value $s_a$ determined by Eq. \ref{saddlesa}.


\subsubsection { Evaluation of the leading order of $P^{Regular} (x,A=A_0+(t-t_0) a,t  \vert x_0,A_0,t_0)  $ with the prefactors }

Similarly, let us now consider the Laplace inverse of $P^{Regular}_{s} (x,A  \vert x_0,A_0) $ of Eq. \ref{hatsRegular}
\begin{eqnarray}
P^{Regular} (x,A,t  \vert x_0,A_0,t_0) 
&& = \int_{c-i\infty}^{c+i\infty} \frac{ds}{2 i \pi}  e^{s (t-t_0)}  {\hat P}^{Regular}_{s} (x,A  \vert x_0,A_0)
\nonumber \\
&&  = \int_{c-i\infty}^{c+i\infty} \frac{ds}{2 i \pi}  e^{s (t-t_0)} 
 \left[ \frac{ {\hat G}_s (x \vert 0){\hat G}_s (0 \vert x_0) }{ {\hat G}^2_s (0 \vert 0) } \right]
 e^{-  \frac{(A-A_0)}{{\hat G}_s (0 \vert 0)} }  
\label{laplaceinverseP}
\end{eqnarray}
Again we are interested into the value $A=A_0+(t-t_0)a$
\begin{eqnarray}
P^{Regular} (x, A=A_0+(t-t_0) a,t  \vert x_0,A_0,t_0) 
&&  = \int_{c-i\infty}^{c+i\infty} \frac{ds}{2 i \pi}  \left[ \frac{ {\hat G}_s (x \vert 0){\hat G}_s (0 \vert x_0) }{ {\hat G}^2_s (0 \vert 0) } \right]
 e^{- (t-t_0) \left[   \frac{a}{{\hat G}_s (0 \vert 0)} - s \right] }  
\label{laplaceinverseaP}
\end{eqnarray}
So we can use the same saddle-point method described in the previous subsection to obtain
the final result analog to Eq. \ref{laplaceinverseasaddle}
\begin{eqnarray}
P^{Regular} (x, A=A_0+(t-t_0) a,t  \vert x_0,A_0,t_0) 
   \opsimeq_{(t-t_0) \to +\infty}     \frac{{\hat G}_{s_a} (x \vert 0) {\hat G}_{s_a} (0 \vert x_0) }{  {\hat G}^2_{s_a} (0 \vert 0) } 
\frac{ e^{- (t-t_0) I(a) }  }{ \sqrt{ 2 \pi (t-t_0) K(a) } }
\label{laplaceinverseasaddleP}
\end{eqnarray}
Note that the dependence with respect to the initial position $x_0$ and to the final position $x$
are only in the functions ${\hat G}_{s_a} (0 \vert x_0) $ and $ {\hat G}_{s_a} (x \vert 0)$
evaluated for the saddle-point value $s_a$ determined by Eq. \ref{saddlesa}.


\subsection{When $X(t)$ is recurrent but does not converge towards an equilibrium state  } 

\label{subsec_recnoneq}

Among recurrent diffusions that do not converge towards an equilibrium state,
the simplest case is the pure Brownian motion without drift $\mu=0$,
that will be discussed in section \ref{sec_brown}.

When the diffusion process $X(t)$ is recurrent but does not converge towards an equilibrium state,
then the first moment $m^{[k=1]}(t \vert x_0,t_0) $ of the local time increment $[A(t)-A(t_0)]$ of Eq. \ref{m1} will
diverge for $(t-t_0) \to +\infty$ in contrast to Eq. \ref{m1finite}
\begin{eqnarray}
 m^{[k=1]}(t \vert x_0,t_0)  \opsimeq_{(t-t_0) \to + \infty}  +\infty
\label{m1dv}
\end{eqnarray}
However this divergence will be weaker than the extensive behavior of Eq. \ref{m1extensive}
 \begin{eqnarray}
\frac{ m^{[k=1]}(t \vert x_0,t_0) }{(t-t_0)}  \opsimeq_{(t-t_0) \to + \infty}  0
\label{m1lessextensive}
\end{eqnarray}
Nevertheless, the saddle-point evaluations of Eq. \ref{laplaceinversea}
can still be performed to obtain as in Eq. \ref{laplaceinverseasaddle} the leading behavior
 \begin{eqnarray}
 \Pi^{Regular} (A=A_0+(t-t_0) a ,t \vert x_0,A_0,t_0) \oppropto_{(t-t_0) \to +\infty} 
    \frac{ {\hat G}_{s_a} (0 \vert x_0) }{ s_a {\hat G}^2_{s_a} (0 \vert 0) } 
\frac{ e^{- (t-t_0) I(a) }  }{ \sqrt{ 2 \pi (t-t_0) K(a) } }
\label{laplaceinverseasaddlecriti}
\end{eqnarray} 
The only important difference is that the rate function $I(a)$ defined for $a \in [0,+\infty[$ 
will now vanish at the boundary $a=0$ where it is minimum
 \begin{eqnarray}
 I ( a=0 ) =0 =  I' ( a=0^+ )
\label{ia0vanish}
\end{eqnarray}
instead of the finite value $a_{eq}>0$ of Eq. \ref{g1} discussed in the previous subsection.


\section{ Construction of conditioned processes involving the local time}

\label{sec_Doob}

In this section, the goal is to construct various conditioned joint processes $[X^*(t),A^*(t)] $
satisfying certain conditions involving the local time.

\subsection{ Conditioned Bridge towards the position $x_T^*$ and the local time $A_T^*$ at the time horizon $T$ }

\label{subsec_bridgexa}

When the initial position is $x_0$ with $A_0=0$ at time $t=0$,
the conditioning towards the final position $x_T^*$ and the final local time $A_T^*$ at the time horizon $T$
leads to 
the following conditioned probability $P^{[x_T^*,A_T^*]}_T( x,A,t) $ for the position $x$ and the local time $A$ 
at some interior time $t \in ]0,T[ $ in terms of the unconditioned joint propagator $P(x_2,A_2,t_2 \vert x_1,A_1,t_1) $ 
described in section \ref{sec_xa}
\begin{eqnarray}
P^{[x_T^*,A_T^*]}_T( x,A,t) = \frac{P( x_T^*,A_T^*,T \vert  x,A,t) P(x,A,t \vert x_0,A_0=0,0)}
{P( x_T^*,A_T^*,T \vert  x_0,A_0=0,0)} 
\label{pstarBridgexA}
\end{eqnarray}
As described in detail in \cite{refdeBruyne2021,c_microcanonical},
the corresponding conditioned process $[X^*(t),A^*(t)]$ then satisfies the Ito system analog to Eq. \ref{ito}
\begin{eqnarray}
dX^*(t) && =  \mu^{[x_T^*,A_T^*]}_T( X^*(t),A^*(t),t ) dt + dB(t)
\nonumber \\
dA^*(t) && = \delta ( X^*(t) ) dt
\label{Doob}
\end{eqnarray}
where the conditioned drift $\mu^{[x_T^*,A_T^*]}_T( x,A , t ) $ involves the unconditioned drift $\mu(x)$
and the logarithmic derivative of the unconditioned propagator $P( x_T^*,A_T^*,T \vert  x,A,t) $  with respect to $x$
\begin{eqnarray}
  \mu^{[x_T^*,A_T^*]}_T( x,A , t ) = \mu(x) +  \partial_x    \ln P( x_T^*,A_T^*,T \vert  x,A,t)
\label{mustarbridge}
\end{eqnarray}


The decomposition of the unconditioned joint propagator $P( x_T^*,A_T^*,T \vert  x,A,t) $
into its singular contribution of Eq. \ref{hatsSingularAbsinv} corresponding to $A_T^*= A$
and its regular contribution $P^{Regular}( x_T^*,A_T^*,T \vert  x,A,t) $ of Eq. \ref{hatsRegular3termsinv} 
corresponding to $A_T^*>A$
\begin{eqnarray}
 P( x_T^*,A_T^*,T \vert  x,A,t) = \delta(A_T^*-A) G^{abs}(x_T^*,T \vert x,t) + \theta(A_T^*>A ) P^{Regular}( x_T^*,A_T^*,T \vert  x,A,t)
\label{prosingreg}
\end{eqnarray}
yields that the conditioned dynamics can be decomposed into the two following regions.

(i) In the region $A_0=0 \leq A<A_T^*$ where the local time $A$ has not yet reached its conditioned final value $A_T^*$, the conditioned drift of Eq. \ref{mustarbridge} involves the regular contribution of the propagator
\begin{eqnarray}
  \mu^{[x_T^*,A_T^*]}_T( x,A<A_T^* , t )   = \mu(x) +  \partial_x    \ln P^{Regular}( x_T^*,A_T^*,T \vert  x,A,t)
\label{mustarbridgebelow}
\end{eqnarray}

(ii) In the region $A=A_T^*$ where the local time $A$ has already reached its conditioned final value $A_T^*$,
and where the position $x$ cannot visit the origin $x=0$ anymore,
the conditioned drift of Eq. \ref{mustarbridge} involves the singular contribution of the propagator
\begin{eqnarray}
  \mu^{[x_T^*,A_T^*]}_T( x , A=A_T^*, t )   = 
 \mu(x) +  \partial_x    \ln G^{abs}(x_T^*,T \vert x,t) 
\label{mustarbridgeabove}
\end{eqnarray}
As a consequence, it only depends on the propagator $G^{abs}(x_T^*,T \vert x,t)  $ in the presence of an absorbing
boundary at the origin so that one recovers the standard problem of a diffusion conditioned to avoid the origin.
Note that for the special case where the final conditioned position is at the origin $x_T^* = 0$,
this region (ii) does not exist, and the local time $A$ should reach its final conditioned value $A_T^*$
only at the final time $T$.

Examples of conditioned bridges towards the position $x_T^*$ and the local time $A_T^*$ at the time horizon $T$
will be given in subsections \ref{subsec_bridgexabrown} and \ref{subsec_bridgexasgn}.


\subsubsection*{ Generalization : conditioning towards some joint distribution $P^*_T(x_T, A_T) $ of the position $x_T$ and of the local time $A_T$ }

If instead of the bridge described above, one wishes
to impose some joint distribution $P^*_T( x_T,A_T) $ of the final position $x_T$ and of the
final local time $A_T$ at the time horizon $T$, 
the conditioned probability $P^*( x,A,t) $ for the position and the local time $A$ 
at some interior time $t \in ]0,T[ $ reads
\begin{eqnarray}
P^*_T( x,A,t) =Q_T( x,A,t)  P(x,A,t \vert x_0,A_0=0,0)
\label{pstarQP}
\end{eqnarray}
where the function $Q_T( x,A,t) $ reads in terms of the final distribution $P^*_T( x_T,A_T) $ that one wishes to impose
\begin{eqnarray}
Q_T( x,A,t) \equiv \int_{-\infty}^{+\infty} dx_T  \int_{0}^{+\infty} dA_T P^*(x_T,A_T,T) 
\frac{P( x_T,A_T,T \vert  x,A,t) }
{P( x_T,A_T,T \vert  x_0,A_0=0,0)}
\label{qfull}
\end{eqnarray}
As a consequence, the conditioned drift $\mu^*( x,A , t ) $ now involves
 the logarithmic derivative of the function $ Q(  x,A,t)$ of Eq. \ref{qfull} with respect to $x$
\begin{eqnarray}
  \mu^*_T( x,A , t ) = \mu(x) +  \partial_x    \ln Q_T(  x,A,t)
\label{mustar}
\end{eqnarray}


\subsection{ Conditioned Bridge towards the local time $A_T^*$ at the time horizon $T$ }

\label{subsec_bridgea}

If the conditioning is towards the local time $A_T^*$ at time horizon $T$,
without any condition on the final position $x_T$,
the conditioned probability for the position $x$ and the local time $A$ at some interior time $t \in ]0,T[ $
involves the unconditioned 
probability $\Pi( A_2,t_2 \vert  x_1,A_1,t_1) $ described in section \ref{sec_a}
\begin{eqnarray}
P^{[A_T^*]}_T( x,A,t) =\frac{\Pi( A_T,T \vert  x,A,t) P(x,A,t \vert x_0,A_0=0,0)}
{\Pi( A_T,T \vert  x_0,A_0=0,0)}  
\label{pstarbridgea}
\end{eqnarray}
The corresponding conditioned drift 
then involves the logarithmic derivative of the unconditioned 
probability $\Pi( A_T^*,T \vert  x,A,t)$ with respect to $x$
\begin{eqnarray}
  \mu^{[A_T^*]}_T( x,A , t ) = \mu(x) +  \partial_x    \ln \Pi( A_T^*,T \vert  x,A,t)
\label{mustarbridgepi}
\end{eqnarray}


Again, the decomposition of $\Pi( A_T^*,T \vert  x,A,t) $
into its singular contribution of Eq. \ref{hatsintegratesingularinv} corresponding to $A_T^*= A$
and its regular contribution $\Pi^{Regular}( A_T^*,T \vert  x,A,t) $ of Eq. \ref{hatsintegrateregular} 
corresponding to $A_T^*>A$
\begin{eqnarray}
\Pi( A_T^*,T \vert  x,A,t)  = \delta(A_T^*-A) S^{abs}(T \vert x,t) + \theta(A_T^*>A ) \Pi^{Regular}( A_T^*,T \vert  x,A,t)
\label{pisingreg}
\end{eqnarray}
yields that the conditioned dynamics can be decomposed into the two following regions.

(i) In the region $A_0=0 \leq A<A_T^*$ where the local time $A$ has not yet reached its conditioned final value $A_T^*$, the conditioned drift of Eq. \ref{mustarbridgepi} involves the regular contribution
\begin{eqnarray}
  \mu^{[A_T^*]}_T( x,A<A_T^* , t )   = \mu(x) +  \partial_x    \ln \Pi^{Regular}( A_T^*,T \vert  x,A,t)
\label{mustarbridgepibelow}
\end{eqnarray}

(ii) In the region $A=A_T^*$ where the local time $A$ has already reached its conditioned final value $A_T^*$,
and where the position $x$ cannot visit the origin $x=0$ anymore,
the conditioned drift of Eq. \ref{mustarbridgepi} involves the singular contribution 
\begin{eqnarray}
  \mu^{[A_T^*]}_T( x , A=A_T^*, t )   =  \mu(x) +  \partial_x    \ln \Pi^{Singular}( A_T^*,T \vert  x,A,t)
=
 \mu(x) +  \partial_x    \ln S^{abs}(T \vert x,t) 
\label{mustarbridgepiabove}
\end{eqnarray}
It depends only on the survival probability $S^{abs}(x_T^*,T \vert x,t)  $ in the presence of an absorbing
boundary at the origin, so that one recovers the standard problem of a diffusion conditioned to survive up to time $T$.

Example of the conditioned bridge towards the local time $A_T^*$ at the time horizon $T$
will be given in subsections \ref{subsec_bridgeabrown} and \ref{subsec_bridgeasgn}.

It is now interesting to consider two possibilities in the limit of the infinite horizon $T \to +\infty$,
as described in the two next subsections.


\subsubsection{ Conditioning towards the finite asymptotic local time $A_{\infty} <+\infty$ for the infinite horizon $T \to +\infty$  }

\label{subsec_doobinfinity}

If one wishes to impose the finite asymptotic local time $A_{\infty} <+\infty$ for the infinite horizon $T \to +\infty$,
one needs to analyze the limit of the infinite horizon $T \to +\infty$
for the conditioned drift of Eq. \ref{mustarbridgepibelow} 
\begin{eqnarray}
  \mu^{[A_{\infty}^*]}_{\infty}( x,A<A_{\infty}^* , t )   =
   \mu(x) + \lim_{T \to +\infty}  \left[ \partial_x    \ln \Pi^{Regular}( A_{\infty}^*,T \vert  x,A,t) \right]
\label{mustarbridgepibelowinfty}
\end{eqnarray}
and for the conditioned drift of Eq. \ref{mustarbridgepiabove} 
\begin{eqnarray}
  \mu^{[A_{\infty}^*]}_{\infty}( x , A=A_{\infty}^*, t )  
   =   \mu(x) + \lim_{T \to +\infty}  \left[  \partial_x    \ln \Pi^{Singular}( A_{\infty}^*,T \vert  x,A,t) \right]
= \mu(x) + \lim_{T \to +\infty}  \left[  \partial_x    \ln S^{abs}(T \vert x,t) \right]
\label{mustarbridgepiaboveinfty}
\end{eqnarray}

An example of the conditioning towards the finite asymptotic local time $A_{\infty} <+\infty$ for the infinite horizon $T \to +\infty$
will be given in subsection \ref{subsec_brownainfty}.


\subsubsection{ Conditioning towards the intensive local time $a^* = \frac{A_T^*}{T}$ 
for large time horizon $T \to + \infty$  }

\label{subsec_doobintensive}

If one wishes to impose instead the fixed intensive local time $a^*$ for large time horizon $T \to + \infty $, 
one needs to plug the value
\begin{eqnarray}
 A_T^* = T a^*
 \label{atstarintensive}
\end{eqnarray}
into the conditioned drift of Eq. \ref{mustarbridgepibelow}
\begin{eqnarray}
\mu^{[T a^*]}_T( x,A<T a^* , t )   = \mu(x) +  \partial_x    \ln \Pi^{Regular}( T a^*,T \vert  x,A,t)
\label{mustarbridgepibelowintensive}
\end{eqnarray}
The asymptotic form of Eq. \ref{laplaceinverseasaddle}
for the propagator $\Pi^{Regular}( T a^*,T \vert  x,A,t) $ in the region $t \ll T$
\begin{eqnarray}
\Pi^{Regular} (T a^*,T \vert  x,A,t) 
&&   \opsimeq_{(T-t) \to +\infty}     \frac{ {\hat G}_{s_{a_t}} (0 \vert x) }{ s_{a_t} {\hat G}^2_{s_{a_t}} (0 \vert 0) } 
\frac{ e^{- (T-t) I(a_t) }  }{ \sqrt{ 2 \pi (T-t) K(a_t) } }
\label{laplaceinverseasaddledoob}
\end{eqnarray}
involves the corresponding intensive local time $a_t$ on the time interval $(T-t)$ 
\begin{eqnarray}
a_t \equiv \frac{ T a^* - A}{ T-t} \opsimeq_{T \to +\infty } a^*
\label{aeffective}
\end{eqnarray}
that reduces to $a^*$ at leading order when $T \to +\infty$.
So at leading order for the large time horizon $T \to +\infty$, 
the conditioned drift of Eq. \ref{mustarbridgepibelowintensive}
reduces 
\begin{eqnarray}
\mu^{[T a^*]}_T( x,A<T a^* , t )    \opsimeq_{T \to +\infty} \mu(x) +  \partial_x  
  \ln {\hat G}_{s_{a^*}} (0 \vert x) \equiv \mu^{[a^*]}_{\infty}( x )
\label{mustarbridgepibelowintensiveasympto}
\end{eqnarray}
to the time-independent drift $\mu^{[a^*]}_{\infty}( x )$ where $s_{a^*} $ 
should be computed as the solution of the saddle-point Equation \ref{saddlesa}
\begin{eqnarray}
  0  =    a^*   \partial_s \left[  \frac{ 1 }{{\hat G}_s (0 \vert 0)}     \right] -1
\label{saddlesastar}
\end{eqnarray}

Using the similarity transformation of Eq. \ref{defpsi} for time-Laplace transforms
\begin{eqnarray}
{\hat G}_s (0 \vert x)= e^{ - \int_{0}^{x} dy \mu(y)} {\hat \psi}_s(0 \vert x)
\label{defpsilaplace}
\end{eqnarray}
one obtains that the conditioned drift of Eq. \ref{mustarbridgepibelowintensiveasympto}
only involves the logarithmic derivation with respect to $x$ of 
the time Laplace transform ${\hat \psi}_{s_{a^*}} (0 \vert x)  $ of the quantum propagator 
\begin{eqnarray}
 \mu^{[a^*]}_{\infty}( x ) \equiv \mu(x) +  \partial_x    \ln {\hat G}_{s_{a^*}} (0 \vert x) 
 = \partial_x    \ln {\hat \psi}_{s_{a^*}} (0 \vert x) 
\label{mustarbridgepibelowintensiveasymptopsi}
\end{eqnarray}
while Eq. \ref{saddlesastar}
for $s_a^*$ becomes
\begin{eqnarray}
  0  =    a^*   \partial_s \left[  \frac{ 1 }{{\hat \psi}_s (0 \vert 0)}     \right] -1
\label{saddlesastarpsi}
\end{eqnarray}

The conditioned potential $ U^{[a^*]}_{\infty}( x ) $
 associated to the conditioned drift $ \mu^{[a^*]}_{\infty}( x ) $ of
Eq. \ref{mustarbridgepibelowintensiveasymptopsi} via Eq. \ref{potentialU}
\begin{eqnarray}
U^{[a^*]}( x ) \equiv - 2 \int_0^{x} dy  \mu^{[a^*]}_{\infty}( x ) 
= \ln \left( \frac{  {\hat \psi}_{s_{a^*}} (0 \vert 0)} {{\hat \psi}_{s_{a^*}} (0 \vert x)} \right)^2
\label{potentialUastar}
\end{eqnarray}
corresponds to the conditioned Boltzmann equilibrium of Eq. \ref{boltzmann}
\begin{eqnarray}
G^{[a^*]}_{eq}(x) = \frac{ e^{-  U^{[a^*]}(x) }}{  \int_{-\infty}^{+\infty} dy e^{-U^{[a^*]}(y) }  } 
=\frac{  \left[ {\hat \psi}_{s_{a^*}} (0 \vert x)\right]^2 } {
\int_{-\infty}^{+\infty} dy \left[{\hat \psi}_{s_{a^*}} (0 \vert y)\right]^2 }
\label{boltzmannastar}
\end{eqnarray}

Example of the conditioning towards the intensive local time $a^* = \frac{A_T^*}{T}$ 
for large time horizon $T \to + \infty$
will be given in subsections \ref{subsec_brownintensive} and \ref{subsec_sgnintensive}.


\subsubsection{ Generalization : conditioning towards the distribution $\Pi^*_T( A_T) $ of the local time $A_T$ at the time horizon $T$  }

If instead of the bridge corresponding to the single value $A_T^*$,
one wishes to impose some distribution $\Pi^*_T( A_T) $ of the local time $A_T$ at the time horizon $T$,
the conditioned probability for the position $x$ and the local time $A$ at some interior time $t \in ]0,T[ $
is given by
\begin{eqnarray}
P_T^*( x,A,t) =Q_T( x,A,t)  P(x,A,t \vert x_0,A_0=0,0)
\label{pstarQPpi}
\end{eqnarray}
where the function $Q_T( x,A,t) $ involves the final distribution $\Pi^*_T( A_T) $ that one wishes to impose
\begin{eqnarray}
Q_T( x,A,t) \equiv   \int_{0}^{+\infty} dA_T \Pi^*(A_T,T) 
\frac{\Pi( A_T,T \vert  x,A,t) }
{\Pi( A_T,T \vert  x_0,A_0=0,0)}
\label{qpi}
\end{eqnarray}
Its logarithmic derivative with respect to $x$ allows to compute
the corresponding conditioned drift 
\begin{eqnarray}
  \mu_T^*( x,A , t ) = \mu(x) +  \partial_x    \ln Q_T(  x,A,t)
\label{mustarpi}
\end{eqnarray}


\section{Application to the uniform drift $\mu \geq 0 $  }

\label{sec_brown}

In this section, the unconditioned process is the Brownian motion with uniform drift $\mu(x)=\mu \geq 0$, so the Ito system of Eq. \ref{ito} reads
\begin{eqnarray}
dX(t) && =  \mu dt + dB(t)
\nonumber \\
dA(t) && = \delta ( X(t) ) dt
\label{itofree}
\end{eqnarray}
Note that in the transient cases $\mu>0$, the local time increment $(A-A_0)$ of this unconditioned process
remains finite for $(t-t_0) \to +\infty$ as described in subsection \ref{subsec_transient}, while $\mu=0$ corresponds to
the case of recurrent diffusion that does not converge towards an equilibrium state discussed in subsection \ref{subsec_recnoneq}.


\subsection{ Properties of the unconditioned diffusion process $X(t)$ alone }


\subsubsection{ Propagator $G(x,t \vert x_0,t_0) $ for the position alone}

The propagator $G(x,t \vert x_0,t_0) $ discussed in section \ref{sec_x} is Gaussian
\begin{eqnarray}
G(x,t \vert x_0,t_0) = \frac{1}{\sqrt{2 \pi (t-t_0)}}  e^{- \frac{[x-x_0-\mu(t-t_0) ]^2}{2(t-t_0)}} 
=\frac{1}{\sqrt{2 \pi (t-t_0)}}  
e^{- \frac{(x-x_0)^2}{2(t-t_0)} 
+\mu (x-x_0)
- \frac{\mu^2 }{2}(t-t_0)
}
\label{free}
\end{eqnarray}
and its time Laplace transform reads
\begin{eqnarray}
{\tilde G}_s (x \vert x_0) && \equiv \int_{t_0}^{+\infty} dt e^{-s (t-t_0) } G(  x,t \vert   x_0,t_0) 
= \frac{ e^{\mu (x-x_0)} } { \sqrt{2 \pi }} 
\int_{0}^{+\infty} d\tau \tau^{-\frac{1}{2}} e^{-\left(s + \frac{\mu^2}{2} \right) \tau }
e^{- \frac{(x-x_0)^2}{2\tau} }
 \nonumber \\ &&
 = \frac{ e^{\mu (x-x_0) - \sqrt{\mu^2+2 s } \vert x-x_0 \vert } } { \sqrt{\mu^2+2 s }} 
\label{laplacefree}
\end{eqnarray}


\subsubsection{ Properties in the presence of an absorbing boundary at the origin $x=0$ }

The Laplace transform ${\hat G}^{abs}_s (x \vert x_0) $ of Eq. \ref{laplaceabsorbing}
reads using Eq. \ref{laplacefree}
\begin{eqnarray}
{\hat G}^{abs}_s (x \vert x_0) 
&& = {\hat G}_s (x \vert x_0) -
 \frac{ {\hat G}_s (x \vert 0){\hat G}_s (0 \vert x_0) }{ {\hat G}_s (0 \vert 0) }  
=  \frac{ e^{\mu (x-x_0) - \sqrt{\mu^2+2 s } \vert x-x_0 \vert } } { \sqrt{\mu^2+2 s }} 
-  \frac{ e^{\mu (x-x_0) - \sqrt{\mu^2+2 s } (\vert x \vert  +\vert x_0 \vert )}  } { \sqrt{\mu^2+2 s }}  
\nonumber \\
&&  =  \frac{ e^{\mu (x-x_0)  }} { \sqrt{\mu^2+2 s }} 
\left[ e^{- \sqrt{\mu^2+2 s } \vert x-x_0 \vert }
- e^{ - \sqrt{\mu^2+2 s } (\vert x \vert  +\vert x_0 \vert )}
 \right]
 \label{laplacecomplementaryfree}
\end{eqnarray}
Its Laplace inversion with respect to $s$ yields the propagator $G^{abs} (x,t \vert x_0,t_0)  $ in the presence of an absorbing boundary at the origin $x=0$
\begin{eqnarray}
G^{abs}(x,t \vert x_0,t_0) 
  =\frac{e^{ \mu (x-x_0)
- \frac{\mu^2 }{2}(t-t_0) } }{\sqrt{2 \pi (t-t_0)}}  
\left[ e^{- \frac{(x-x_0)^2}{2(t-t_0)} }
-e^{- \frac{(\vert x \vert  +\vert x_0 \vert)^2}{2(t-t_0)} 
}
\right] 
\label{absorbingfree}
\end{eqnarray}
in agreement with the method of images.

The Laplace transform $ {\hat \gamma}^{abs}_s ( x_0) $ of Eq. \ref{laplacegammaabs}
reads using Eq. \ref{laplacefree}
\begin{eqnarray}
{\hat \gamma}^{abs}_s ( x_0) 
  = \frac{ {\hat G}_s (0 \vert x_0) }{ {\hat G}_s (0 \vert 0) } = e^{- \mu x_0 - \sqrt{\mu^2+2 s } \vert x_0 \vert }
\label{laplacegammaabsfree}
\end{eqnarray}
The rewriting of the Laplace transform of Eq. \ref{laplacefree} in terms of the parameter $\alpha>0$
\begin{eqnarray}
\frac{ e^{ -  \alpha \sqrt{\mu^2+2 s }  } } { \sqrt{\mu^2+2 s }}  = 
\int_{0}^{+\infty}  \frac{d \tau } { \sqrt{2 \pi \tau }}  e^{-\left(s + \frac{\mu^2}{2} \right) \tau }
e^{- \frac{\alpha^2}{2\tau} } 
\label{laplaceinter}
\end{eqnarray}
allows to obtain via the derivation with respect to $\alpha$
\begin{eqnarray}
e^{ -  \alpha \sqrt{\mu^2+2 s } }&& = -\partial_{\alpha} \left( \frac{ e^{ -  \alpha \sqrt{\mu^2+2 s }  } } { \sqrt{\mu^2+2 s }}  \right)
= - 
\int_{0}^{+\infty}  \frac{d \tau } { \sqrt{2 \pi \tau }}
 e^{-\left(s + \frac{\mu^2}{2} \right) \tau }
\partial_{\alpha} e^{- \frac{\alpha^2}{2\tau} } 
\nonumber \\
&& = 
\int_{0}^{+\infty} d\tau  e^{-\left(s + \frac{\mu^2}{2} \right) \tau }
\left( \frac{\alpha}{ \sqrt{2 \pi }\tau^{\frac{3}{2}}} \right)  e^{- \frac{\alpha^2}{2\tau} } 
\label{laplacederi}
\end{eqnarray}
The Laplace inversion of Eq. \ref{laplacegammaabsfree}
thus yields the absorption rate $\gamma^{abs}(t \vert x_0,t_0) $ 
at time $t$ if one starts at position $x_0$ at time $t_0$
\begin{eqnarray}
\gamma^{abs}(t \vert x_0,t_0)  = 
\frac{ \vert x_0 \vert  } { \sqrt{2 \pi } (t-t_0)^{\frac{3}{2}}} 
  e^{- \mu x_0- \frac{\mu^2}{2} (t-t_0) - \frac{ x_0^2}{2(t-t_0)} } 
\label{gammaabsfree}
\end{eqnarray}

Finally the survival probability $S^{abs} (t \vert x_0,t_0) $ of Eq. \ref{survival}
can be obtained from the 
integral over the final position $x$ of the propagator $G^{abs}(x,t \vert x_0,t_0) $ of Eq. \ref{absorbingfree}
\begin{eqnarray}
S^{abs} (t \vert x_0,t_0) \equiv  \int_{-\infty}^{+\infty} dx G^{abs}(x,t \vert x_0,t_0)
=  \int_{-\infty}^{+\infty} dx
\frac{e^{ \mu (x-x_0)
- \frac{\mu^2 }{2}(t-t_0) } }{\sqrt{2 \pi (t-t_0)}}  
\left[ e^{- \frac{(x-x_0)^2}{2(t-t_0)} }
-e^{- \frac{(\vert x \vert  +\vert x_0 \vert)^2}{2(t-t_0)} 
}
\right] 
\label{survivalfree}
\end{eqnarray}
Its Laplace transform of Eq. \ref{laplacesurvival} reads using Eq. \ref{laplacegammaabsfree}
\begin{eqnarray}
 {\hat S}^{abs}_s ( x_0) 
 =  \frac{1}{s} \left[ 1 - \frac{ {\hat G}_s (0 \vert x_0) }{ {\hat G}_s (0 \vert 0) } \right] 
 =  \frac{1}{s} \left[ 1 - 
 e^{- \mu x_0 - \sqrt{\mu^2+2 s } \vert x_0 \vert }
  \right] 
\label{laplacesurvivalfree}
\end{eqnarray}

For $\mu>0$, the forever-survival probability $S^{abs[\mu>0]}(\infty \vert x_0)$ at infinite time $(t-t_0) \to +\infty$ 
when starting at the position $x_0$
can be recovered via the limit $s \to 0$ of
\begin{eqnarray}
S^{abs[\mu>0]}(\infty \vert x_0) = \lim_{s \to 0} \left[ s  {\hat S}^{abs}_s ( x_0)   \right] 
 =  1 -  e^{- \mu (x_0 +  \vert  x_0 \vert )  }
 = 
 \left\lbrace
  \begin{array}{lll}
    0  
    &~~\mathrm{if~~} x_0 \le 0
    \\
    1- e^{-2 \mu x_0 }  
    &~~\mathrm{if~~} x_0 > 0  
  \end{array}
\right.
\label{survivalfreeinfty}
\end{eqnarray}
It remains finite for $x_0>0$, since the particle can escape towards $(+ \infty)$ without touching the origin $x=0$. The finite-time survival probability 
$ S^{abs} (t \vert x_0,t_0) $ given by Eq. \ref{survivalfree} can be expressed 
in terms of the Error function $\erf(x)$ and
and in terms of the complementary Error function $\erfc(x) = 1 - \erf(x)$
as
\begin{eqnarray}
 S^{abs}(t \vert x_0,t_0) 
 = 
 \left\lbrace
  \begin{array}{lll}
      \frac{1}{2} \left[ 1+ \erf \left(\frac{x_0 +\mu  (t-t_0)}{\sqrt{2(t-t_0)}}\right) - e^{-2 \mu x_0} \erfc \left(\frac{x_0 -\mu  (t-t_0)}{\sqrt{2(t-t_0)}}\right) \right]
    &~~\mathrm{if~~} x_0 > 0
    \\
      \frac{1}{2} \left[ e^{-2 \mu x_0} \left( -2 + \erfc \left(\frac{ x_0 -\mu  (t-t_0)}{\sqrt{2(t-t_0)}}\right) \right)+ \erfc \left(\frac{x_0 +\mu  (t-t_0)}{\sqrt{2(t-t_0)}}\right)  \right]
    &~~\mathrm{if~~} x_0 < 0  
  \end{array}
\right.
\end{eqnarray}
In the region $x_0 \leq 0$ where the forever-survival $S^{abs[\mu>0]}(\infty \vert x_0) $ vanishes, using the asymptotic behavior of the complementary Error function
\begin{eqnarray}
 \erfc (x)  
  \simeq
 \left\lbrace
  \begin{array}{lll}
      e^{-x^2} \left( \frac{1}{\sqrt{\pi} x} -  \frac{1}{2 \sqrt{\pi} x^3} \right)
    &~~\mathrm{when~~} x \to \infty
    \\
      2 +  e^{-x^2} \left( \frac{1}{\sqrt{\pi} x} -  \frac{1}{2 \sqrt{\pi} x^3} \right)
    &~~\mathrm{when~~} x \to -\infty 
  \end{array}
\right.
 \label{erfcasymptotic}
\end{eqnarray}
the asymptotic behavior of $S^{abs}(t \vert x_0,t_0)$ for large time $(t-t_0)$ and fixed $x_0$ reads
\begin{eqnarray}
 S^{abs[\mu>0]}(t \vert x_0 < 0,t_0) 
 \underset{(t-t_0) \to +\infty}{\simeq}
 \sqrt\frac{2}{\pi} \frac{\vert x_0\vert e^{\mu \vert x_0\vert -\frac{\mu^2}{2} (t-t_0)}}{\mu^2 (t-t_0)^{\frac{3}{2}} }
\end{eqnarray}

For $\mu=0$, the forever-survival probability $S^{abs[\mu=0]}(\infty \vert x_0) $ of Eq. \ref{survivalfreeinfty} vanishes for any $x_0$. The leading singularity of Eq. \ref{laplacesurvivalfree} for $s \to 0^+$
\begin{eqnarray}
 {\hat S}^{abs[\mu=0]}_s ( x_0)   =  \frac{1}{s} \left[ 1 -  e^{ - \sqrt{2 s } \vert x_0 \vert }  \right] 
 \opsimeq_{s \to 0^+} \vert x_0 \vert \sqrt{ \frac{2 }{s}  }  
\label{laplacesurvivalfreemuzero}
\end{eqnarray}
allows to recover the dominant asymptotic behavior for large time
\begin{eqnarray}
S^{abs[\mu=0]} ( t \vert x_0,t_0)    
 \opsimeq_{(t-t_0) \to + \infty} \vert x_0 \vert \sqrt{ \frac{2 }{\pi (t-t_0)}  } 
\label{laplacesurvivalfreemuzerolarget}
\end{eqnarray}


\subsection{ Joint propagator $  P(  x,A,t \vert   x_0,A_0,t_0) $ for the unconditioned joint process $[X(t),A(t)]$  }

The singular contribution of Eq. \ref{hatsSingularAbsinv}
involves the propagator $G^{abs}(x,t \vert x_0,t_0)  $ of Eq. \ref{absorbingfree}
\begin{eqnarray}
 P^{Singular} (x,A,t  \vert x_0,A_0,t_0) 
 &&  =  \delta(A-A_0) G^{abs}(x,t \vert x_0,t_0)
 \nonumber \\
 && =    \delta(A-A_0)
\frac{e^{ \mu (x-x_0)- \frac{\mu^2 }{2}(t-t_0) } }{\sqrt{2 \pi (t-t_0)}}  
\left[ e^{- \frac{(x-x_0)^2}{2(t-t_0)} }
-e^{- \frac{(\vert x \vert  +\vert x_0 \vert)^2}{2(t-t_0)} 
}
\right]
\label{hatsSingularAbsinvfree}
\end{eqnarray}
The Laplace transform $ {\hat P}^{Regular}_{s} (x,A  \vert x_0,A_0)  $ of Eq. \ref{hatsRegular}
reads using Eq. \ref{laplacefree}
\begin{eqnarray}
 {\hat P}^{Regular}_{s} (x,A  \vert x_0,A_0)  && = 
 \theta(A>A_0) \left[ \frac{ {\hat G}_s (x \vert 0){\hat G}_s (0 \vert x_0) }{ {\hat G}^2_s (0 \vert 0) } \right]
 e^{-  \frac{(A-A_0)}{{\hat G}_s (0 \vert 0)} }  
 \nonumber \\
 && = \theta(A>A_0)  e^{\mu (x-x_0)  - \sqrt{\mu^2+2 s } (\vert x \vert  +\vert x_0\vert +  A-A_0 ) }  
\label{hatsRegularfree}
\end{eqnarray}
Its Laplace inversion using Eq. \ref{laplacederi} yields
\begin{eqnarray}
  P^{Regular}(  x,A,t \vert   x_0,A_0,t_0)
= \theta(A>A_0) e^{\mu (x-x_0)- \frac{\mu^2}{2} (t-t_0) }
\left( \frac{\vert x \vert  +\vert x_0\vert +  A-A_0 }{ \sqrt{2 \pi }
(t-t_0)^{\frac{3}{2}}} \right)  e^{- \frac{(\vert x \vert  +\vert x_0\vert +  A-A_0 )^2}{2(t-t_0)} } 
\label{hatsRegularfreeinv}
\end{eqnarray}

In summary, the joint propagator $ P(  x,A,t \vert   x_0,A_0,t_0) $
involving the two contributions of Eq. \ref{hatsSingularAbsinvfree}
and Eq. \ref{hatsRegularfreeinv}
 reads
\begin{eqnarray}
  P(  x,A,t \vert   x_0,A_0,t_0) && =  P^{Singular} (x,A,t  \vert x_0,A_0,t_0) 
  +P^{Regular}(  x,A,t \vert   x_0,A_0,t_0)
  \nonumber \\
&&  =  \delta(A-A_0)
\frac{e^{ \mu (x-x_0)- \frac{\mu^2 }{2}(t-t_0) } }{\sqrt{2 \pi (t-t_0)}}  
\left[ e^{- \frac{(x-x_0)^2}{2(t-t_0)} }
-e^{- \frac{(\vert x \vert  +\vert x_0 \vert)^2}{2(t-t_0)} 
}
\right]
\nonumber \\
&& + \theta(A>A_0) e^{\mu (x-x_0)- \frac{\mu^2}{2} (t-t_0) }
\left( \frac{\vert x \vert  +\vert x_0\vert +  A-A_0 }{ \sqrt{2 \pi }
(t-t_0)^{\frac{3}{2}}} \right)  e^{- \frac{(\vert x \vert  +\vert x_0\vert +  A-A_0 )^2}{2(t-t_0)} } 
\label{jointpropagatorfree}
\end{eqnarray}
For $\mu = 0$ and $x_0 = 0$ (i.e. a standard Brownian motion), the regular part of the propagator $P(  x,A,t \vert   x_0,A_0,t_0)$ reduces to
\begin{eqnarray}
	P^{Regular}(  x,A,t \vert   0,0,0) = \frac{\vert x \vert + A }{\sqrt{2 \pi t^3}} 
	e^{-\frac{(\vert x \vert + A)^2}{2 t } }
\end{eqnarray}
a result that can be found in the mathematical literature \cite{Karatzas}.


\subsection{  Probability $\Pi (A,t  \vert x_0,A_0,t_0) $ to see the local time $A$ at time $t$ }

The probability $\Pi(A,t \vert x_0,A_0,t_0)  $ of Eq. \ref{propagAalone}
can be obtained via the integration of the joint propagator $ P(  x,A,t \vert   x_0,A_0,t_0) $ of Eq. \ref{jointpropagatorfree}
over the final position $x$
\begin{eqnarray}
 \Pi(A,t \vert x_0,A_0,t_0) \equiv \int_{-\infty}^{+\infty} dx  P(A,t \vert x_0,A_0,t_0)
\label{propagAalonefree}
\end{eqnarray}
Its singular contribution of Eq. \ref{hatsintegratesingularinv} involves 
the survival probability $S^{abs} ( t \vert x_0, t_0)   $ of Eq. \ref{survivalfree}
\begin{eqnarray}
 \Pi^{Singular} (A ,t \vert x_0,A_0,t_0) 
=  \delta(A-A_0)
S^{abs} ( t \vert x_0, t_0)  
\label{hatsintegratesingularinvfree}
\end{eqnarray}
The Laplace transform of Eq. \ref{hatsintegrateregular} reads using Eq. \ref{laplacefree}
\begin{eqnarray}
 {\hat \Pi}^{Regular}_{s} (A  \vert x_0,A_0)  
 && = \theta(A>A_0)  \frac{ {\hat G}_s (0 \vert x_0) }{ s {\hat G}^2_s (0 \vert 0) } 
 e^{-  \frac{(A-A_0)}{{\hat G}_s (0 \vert 0)} }  
  = \theta(A>A_0)  \frac{ \sqrt{\mu^2+2 s }}{s}  e^{- \mu x_0 - \sqrt{\mu^2+2 s } ( \vert x_0 \vert + A-A_0 ) }  
\label{hatsintegrateregularfree}
\end{eqnarray}


\subsubsection{  Case $\mu=0$ }

For the case $\mu=0$, Eq. \ref{hatsintegrateregularfree} reduces to
\begin{eqnarray}
 {\hat \Pi}^{Regular[\mu=0]}_{s} (A  \vert x_0,A_0)    
 = \theta(A>A_0)   \sqrt{\frac{2}{s}}  e^{ - \sqrt{2 s } ( \vert x_0 \vert + A-A_0 ) }  
\label{hatsintegrateregularfreemuzeros}
\end{eqnarray}
 so that Eq. \ref{laplaceinter} for $\mu=0$
can be used to obtain the Laplace inversion of Eq. \ref{hatsintegrateregularfreemuzeros}
\begin{eqnarray}
 \Pi^{Regular[\mu=0]} (A ,t \vert x_0,A_0,t_0)    
 = \theta(A>A_0)  
 \sqrt{ \frac{2 } {  \pi (t-t_0) }  }
e^{- \frac{( \vert x_0 \vert + A-A_0 )^2}{2 (t-t_0)} }  
\label{hatsintegrateregularfreemuzerosinv}
\end{eqnarray}

The singular contribution of Eq. \ref{hatsintegratesingularinvfree}
 involves the survival probability of Eq. \ref{survivalfree}
\begin{eqnarray}
 \Pi^{Singular[\mu=0]} (A ,t \vert x_0,A_0,t_0) 
&& =  \delta(A-A_0)S^{abs[\mu=0]} ( t \vert x_0, t_0)  
\nonumber \\
&&  =  \delta(A-A_0) \int_{-\infty}^{+\infty} dx
\frac{ e^{- \frac{x^2+x_0^2}{2(t-t_0)} } }{\sqrt{2 \pi (t-t_0)}}  
\left[ e^{ \frac{xx_0}{(t-t_0)} }
-e^{- \frac{\vert x  x_0 \vert}{(t-t_0)} 
}
\right] 
\nonumber \\
&&  =  \delta(A-A_0) e^{- \frac{x_0^2}{2(t-t_0)} }
\int_{-\infty}^{+\infty} dz
\frac{ e^{- \frac{z^2}{2} } }{\sqrt{2 \pi }}  
\left[ e^{ \frac{z x_0}{ \sqrt{t-t_0}} }
-e^{- \frac{\vert z  x_0 \vert}{\sqrt{t-t_0}} 
}
\right] 
\nonumber \\
&&  =  \delta(A-A_0) \erf \left(\frac{\vert x_0 \vert}{\sqrt{2(t-t_0)}} \right)
 \label{hatsintegratesingularinvfreemuzerotlarge}
\end{eqnarray} 

Putting together the two contributions, one obtains that 
 $\Pi^{[\mu=0]} (A ,t \vert x_0,A_0,t_0) $ reads 
 \begin{eqnarray}
  \Pi^{[\mu=0]} (A ,t \vert x_0,A_0,t_0) 
 && = \delta(A-A_0)S^{abs[\mu=0]} ( t \vert x_0, t_0)  +  \Pi^{Regular[\mu=0]} (A ,t \vert x_0,A_0,t_0)
 \nonumber \\
 && 
 =  \delta(A-A_0)\erf \left(\frac{\vert x_0 \vert}{\sqrt{2(t-t_0)}} \right)
 + \theta(A>A_0)  
 \sqrt{ \frac{2 } {  \pi (t-t_0) }  }
e^{- \frac{( \vert x_0 \vert + A-A_0 )^2}{2 (t-t_0)} }  
\label{piamuzerototal}
\end{eqnarray} 
Since $\erf(x) \underset{x \to 0}{\simeq} 2 x/\sqrt{\pi}$, we get the asymptotic behavior for large $(t-t_0)$
\begin{eqnarray}
 \Pi^{[\mu=0]} (A ,t \vert x_0,A_0,t_0) 
 \opsimeq_{(t-t_0) \to + \infty} 
 \sqrt{ \frac{2 }{\pi (t-t_0)}  } 
 \left[   \delta(A-A_0) \vert x_0 \vert 
 +  \theta(A>A_0)  
e^{- \frac{( \vert x_0 \vert + A-A_0 )^2}{2 (t-t_0)} }  \right]
\label{piamuzerolarget}
\end{eqnarray} 

In the case where $x_0 = 0$, $A_0 = 0$ and $t_0 = 0$, the propagator of Eq. \ref{piamuzerototal}
reduces to the half-Gaussian distribution
\begin{eqnarray}
  \Pi^{[\mu=0]} (A ,t \vert 0,0,0) =  \theta(A>0)  \sqrt{ \frac{2} {\pi t}} e^{- \frac{A^2}{2 t} }  
\end{eqnarray}
a result that can be found in \cite{occupationsinai,bressloff}.

The first moment $m^{[k=1]}(t \vert x_0,t_0=0) $ of the local time increment 
can be computed via Eq. \ref{mkdef} or via Eq. \ref{m1}
\begin{eqnarray}
m^{[k=1]}(t \vert x_0,t_0=0)  && = \int_0^{\infty} dA \, A \, \Pi^{[\mu=0]} (A ,t \vert x_0,A_0=0,t_0=0) 
\nonumber \\
&& = \int_{0}^{t}  dt_1 G(0, t_1 \vert x_0,t_0=0)
 = \int_{0}^{t}  dt_1   \frac{e^{-\frac{x_0^2}{2 t_1}}}{\sqrt{2 \pi t_1}}
\nonumber \\
                                       && = \sqrt{\frac{2 t}{\pi}} e^{-\frac{x_0^2}{2 t}}
                       -\vert x_0 \vert  \erfc \left(\frac{\vert x_0 \vert}{\sqrt{2 t}} \right)
\label{m1brownmuzero}
\end{eqnarray}
and displays the power-law asymptotic growth independent of $x_0$
\begin{eqnarray}
m^{[k=1]}(t \vert x_0,t_0=0) \underset{t \to +\infty}{\simeq} \sqrt{\frac{2 t}{ \pi}}
\label{m1brownmuzerolargetime}
\end{eqnarray}
that is intermediate as it should between the finite case of Eq. \ref{m1finite}
and the extensive case of Eq. \ref{m1extensive},
since the Brownian motion without drift $\mu=0$ is recurrent but does not converge towards an equilibrium distribution.

The second moment $m^{[k=2]}(t \vert x_0,t_0=0) $ of the local time increment 
can be computed via Eq. \ref{mkdef} or Eq. \ref{m2}
\begin{eqnarray}
m^{[k=2]}(t \vert x_0,t_0=0)  &&  = \int_0^{\infty} dA \, A^2 \, \Pi^{[\mu=0]} (A ,t \vert x_0,A_0=0,t_0=0) 
=  2   \int_{0}^{t} dt_2  \int_{0}^{t_2} dt_1 G(0, t_2 \vert 0,t_1) G(0, t_1 \vert x_0,t_0=0)
\nonumber \\
&& =  \frac{1}{\pi}    \int_{0}^{t} dt_2  \int_{0}^{t_2} dt_1  
 \frac{e^{-\frac{x_0^2}{2 t_1}}}{\sqrt{(t_2-t_1) t_1}}
=  \frac{1}{\pi}    \int_{0}^{t} dt_1  \frac{e^{-\frac{x_0^2}{2 t_1}}}{\sqrt{ t_1}} \int_{t_1}^{t} \frac{ dt_2  }{\sqrt{t_2-t_1 }} 
 =  \frac{2}{\pi}    \int_{0}^{t} dt_1 e^{-\frac{x_0^2}{2 t_1}} \frac{\sqrt{t-t_1 }}{\sqrt{ t_1}}    
\nonumber \\
                    && =\left(t  + x_0^2 \right) \erfc \left(\frac{\vert x_0 \vert}{\sqrt{2 t}} \right)  - \vert x_0 \vert \sqrt{\frac{2 t}{\pi}} e^{-\frac{x_0^2}{2 t}}                       
\label{m2brownmuzero}
\end{eqnarray}
with the following asymptotic growth independent of $x_0$
\begin{eqnarray}
m^{[k=2]}(t \vert x_0,t_0=0) \underset{t \to +\infty}{\simeq} t
\label{m2brownmuzerolargetime}
\end{eqnarray}


\subsubsection{  Case $\mu>0$ }
For the case $\mu>0$, integrating the singular and regular part of the joint propagator $P(  x,A,t \vert   x_0,A_0,t_0) $ of Eq. \ref{jointpropagatorfree} with respect to the final position $x$, gives respectively

\begin{eqnarray}
\label{pisingularinversefree}
 \Pi^{Singular} (A ,t \vert x_0,A_0,t_0) 
&& =  \delta(A-A_0)
	\int_{-\infty}^{+\infty} dx  \frac{e^{ \mu (x-x_0)- \frac{\mu^2 }{2}(t-t_0) } }{\sqrt{2 \pi (t-t_0)}}  
\left[ e^{- \frac{(x-x_0)^2}{2(t-t_0)} }
-e^{- \frac{(\vert x \vert  +\vert x_0 \vert)^2}{2(t-t_0)} 
}
\right]  \\ 
&& =  \delta(A-A_0) \left[1- \frac{1}{2} e^{-\mu x_0} \left(e^{-\mu \vert x_0 \vert} \erfc \left( \frac{\vert x_0 \vert-\mu(t-t_0)}{\sqrt{2 (t-t_0)}}  \right) + e^{\mu \vert x_0 \vert} \erfc \left( \frac{\vert x_0 \vert+\mu(t-t_0)}{\sqrt{2 (t-t_0)}}  \right) \right) \right] \nonumber
\end{eqnarray}
and 
\begin{eqnarray}
\label{piregularinversefree}
 && \Pi^{Regular} (A ,t \vert x_0,A_0, t_0)  
  =  \theta(A>A_0) \int_{-\infty}^{+\infty} dx \,  e^{\mu (x-x_0)- \frac{\mu^2}{2} (t-t_0) }
\left( \frac{\vert x \vert  +\vert x_0\vert +  A-A_0 }{ \sqrt{2 \pi }
(t-t_0)^{\frac{3}{2}}} \right)  e^{- \frac{(\vert x \vert  +\vert x_0\vert +  A-A_0 )^2}{2(t-t_0)} }\nonumber \\  
 && ~~~~~~~~~~~~~~~~~~~~~~~~~~~~~~~~ =   \theta(A>A_0) \left[ \sqrt{\frac{2}{\pi(t-t_0)}} e^{-\mu x_0} e^{-\frac{\mu^2}{2} (t-t_0)} e^{-\frac{(\vert x_0\vert +  A-A_0 )^2}{2(t-t_0)}} \right. \\ 
 && 
 \left. + \frac{1}{2} \mu e^{-\mu x_0 }\left(e^{-\mu (\vert x_0\vert +  A-A_0)} \erfc \left( \frac{\vert x_0\vert-\mu(t-t_0) +A-A_0 }{\sqrt{2 (t-t_0)}}  \right)  -e^{\mu (\vert x_0\vert +  A-A_0)} \erfc \left( \frac{\vert x_0\vert+\mu(t-t_0) +A-A_0 }{\sqrt{2 (t-t_0)}}  \right)   \right) \right] \nonumber 
\end{eqnarray}

The limit of the infinite time interval $(t-t_0) \to +\infty$ yields
\begin{eqnarray}
\Pi (A ,\infty \vert x_0,A_0) && = \Pi^{Singular} (A ,\infty \vert x_0,A_0) + \Pi^{Regular}(A,\infty \vert x_0,A_0)
\nonumber \\
&& = \delta(A-A_0) \left[ 1 -  e^{- \mu (x_0 +  \vert  x_0 \vert )  }  \right]
+  \theta(A>A_0)   \mu e^{ - \mu ( x_0+\vert x_0 \vert + A-A_0 ) }   
\nonumber \\
&& 
= \left\lbrace
  \begin{array}{lll}
     \theta(A>A_0)   \mu e^{ - \mu (A-A_0)  }                     &~~\mathrm{if~~} x_0 < 0
    \\
     \delta(A-A_0) \left[ 1 -  e^{- 2 \mu x_0   } \right]
+  e^{- 2 \mu x_0} \theta(A>A_0)   \mu e^{ - \mu (A-A_0)  }     &~~\mathrm{if~~} x_0 > 0 
  \end{array}
\right.
\label{pitotfreeinfty}
\end{eqnarray}

In the region $x_0<0$ where the limit of the singular contribution vanishes $\Pi^{Singular} (A ,\infty \vert x_0<0,A_0) =0$,
one can use the asymptotic behaviors of the $\erfc$ function given by Eq. \ref{erfcasymptotic}
to obtain that the leading contribution to Eq. \ref{pisingularinversefree} reads  for large time $(t-t_0)$ 
 \begin{eqnarray}
\Pi^{Singular}(A,t \vert x_0<,A_0,t_0) 
\opsimeq_{(t-t_0) \to +\infty} 
 \delta(A-A_0)   \vert  x_0 \vert \sqrt{\frac{2}{\pi}} \frac{e^{-\frac{(\vert  x_0 \vert+\mu (t-t_0))^2}{2(t-t_0)}}}{\mu^2 (t-t_0)^{3/2}} 
\label{hatsintegratesingularfreeinfty}
\end{eqnarray}

\begin{figure}[h]
\centering
\includegraphics[width=4.2in,height=3.2in]{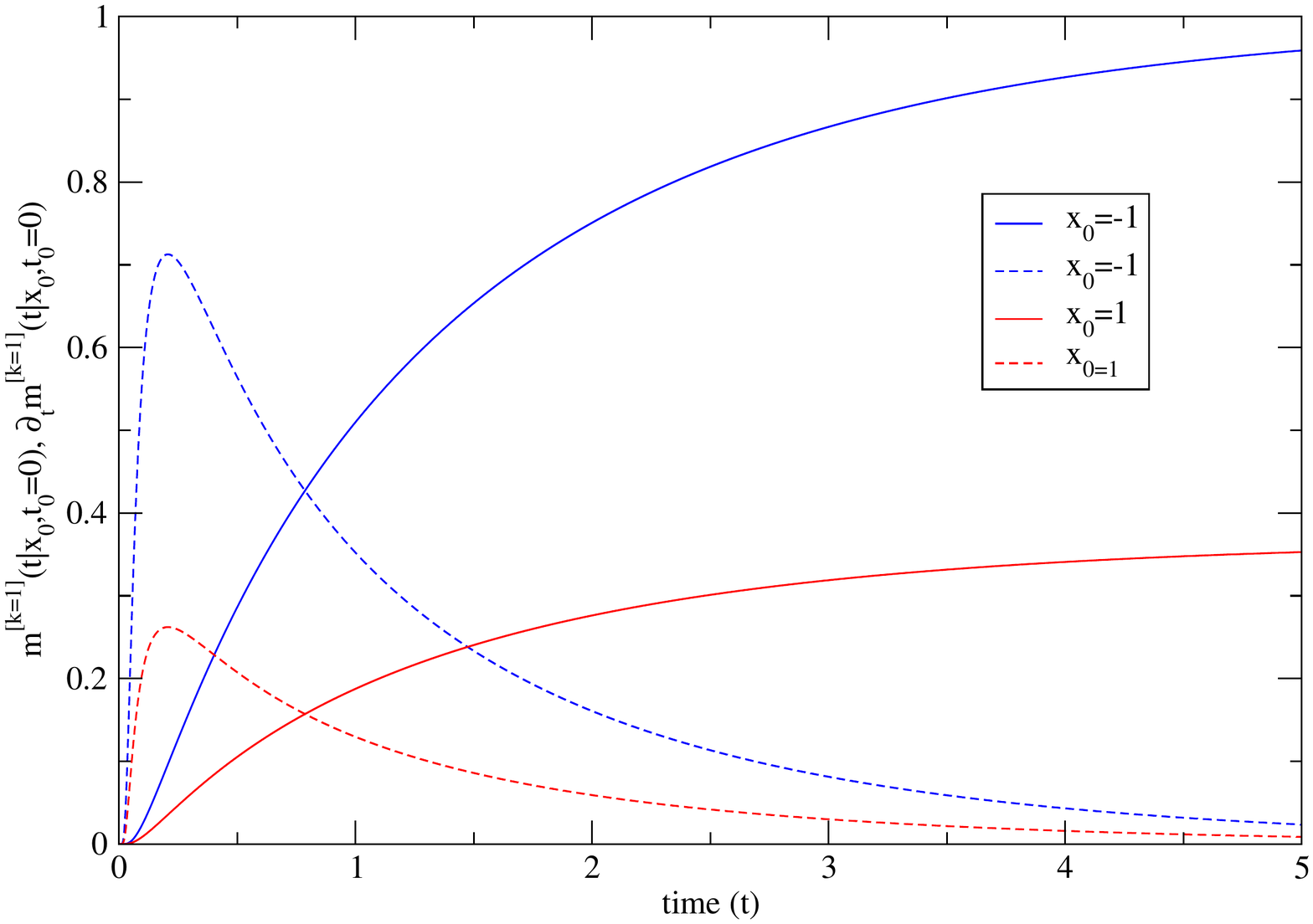}
\setlength{\abovecaptionskip}{15pt}  
\caption{Plain curve: mean local time, dash curve: derivative of the mean local time. Blue lines correspond to $x_0 = -1$, Red lines correspond to $x_0 = 1$. The constant drift $\mu$ is equal to $1$.}
\label{fig1}
\end{figure}

The first moment $m^{[k=1]}(t \vert x_0,t_0=0) $ of the local time increment of Eqs \ref{mkdef} and \ref{m1}
\begin{eqnarray}
m^{[k=1]}(t \vert x_0,t_0=0)  && = \int_0^{\infty} dA \, A \, \Pi^{[\mu]} (A ,t \vert x_0,A_0=0,t_0=0) 
= \int_{0}^{t}  dt_1 G(0, t_1 \vert x_0,t_0=0)
\nonumber \\
&& = \int_{0}^{t} dt_1   \frac{e^{- \frac{(x_0-\mu t_1 )^2}{2t_1}} }{\sqrt{2 \pi t_1}}  
\nonumber \\
                                       && =  \frac{e^{-\mu(\vert x_0 \vert + x_0)} \erfc \left(\frac{\vert x_0 \vert - \mu t}{\sqrt{2 t}} \right) - e^{\mu(\vert x_0 \vert - x_0)} \erfc \left(\frac{\vert x_0 \vert + \mu t}{\sqrt{2 t}} \right) }{2 \mu}
\label{m1brownmu}
\end{eqnarray}
converges to a finite asymptotic value which depends strongly on the sign of the initial position $x_0$ of the process
\begin{eqnarray}
	\lim_{t \to \infty}  m^{[k=1]}(t \vert x_0,t_0=0)  =  \frac{e^{-\mu(x_0-\vert x_0 \vert)}}{\mu}  = 
 \left\lbrace
  \begin{array}{lll}
     \frac{1}{\mu}                  &~~\mathrm{if~~} x_0 < 0
    \\
     \frac{e^{-2\mu x_0}}{\mu}      &~~\mathrm{if~~} x_0 > 0 
  \end{array}
\right.
\end{eqnarray}
In the region $x_0 > 0$ where the process has a finite probability to escape towards ($+\infty$) without visiting the origin, the mean local time is of course smaller than in the region $x_0 < 0$ where the process is certain to cross the origin. In the latter case, as expected, its average asymptotic value is the same as if the process started from 0.\\

The two first time-derivatives of the first moments
\begin{eqnarray}
 \partial_t m^{[k=1]}(t \vert x_0,t_0=0) && = \frac{e^{-\frac{(x_0+ \mu t)^2}{2 t}}}{\sqrt{2 \pi t}} 
\nonumber \\
  \partial_t^2  m^{[k=1]}(t \vert x_0,t_0=0) && = \frac{e^{-\frac{(x_0+ \mu t)^2}{2 t}}}{2 \sqrt{2 \pi t^5}} \left( x_0^2 - t(1 + t \mu^2) \right)
\label{secondderivativemeanA}
\end{eqnarray}
shows that the mean local time increases the most at a time $T = \frac{\sqrt{1 + 4 x_0^2 \mu^2}-1}{2 \mu^2}$ which is independent of the sign of $x_0$. The behavior of the mean local time and its derivative is shown in Fig.\ref{fig1}.\\

The second moment $m^{[k=2]}(t \vert x_0,t_0=0) $ of the local time increment of Eqs \ref{mkdef} and \ref{m2}
\begin{eqnarray}
&& m^{[k=2]}(t \vert x_0,t_0=0)    = \int_0^{\infty} dA \, A^2 \, \Pi^{[\mu=0]} (A ,t \vert x_0,A_0=0,t_0=0) 
=  2   \int_{0}^{t} dt_2  \int_{0}^{t_2} dt_1 G(0, t_2 \vert 0,t_1) G(0, t_1 \vert x_0,t_0=0)
\nonumber \\
&& = 
 \left\lbrace
  \begin{array}{lll}
       \frac{1}{\mu^2}  \erfc \left(\frac{x_0 +\mu t}{\sqrt{2 t}}\right) + 
      \frac{1}{\mu^2}  e^{-\frac{\mu}{2}(6 x_0 + \mu t)} \left[-2 e^{2 \mu x_0} \erfc \left(\frac{x_0 }{\sqrt{2 t}}\right) + e^{\mu x_0 + \frac{\mu^2}{2} t} \erfc \left(\frac{x_0 -\mu t}{\sqrt{2 t}}\right) \right]
    &~~\mathrm{if~~} x_0 > 0
    \\ \\
        \frac{1}{\mu^2} \erfc \left(\frac{-x_0 -\mu t}{\sqrt{2 t}}\right) + 
      \frac{1}{\mu^2}  e^{-\frac{\mu}{2}(6 x_0 + \mu t)} \left[-2 e^{2 \mu x_0} \erfc \left(\frac{- x_0 }{\sqrt{2 t}}\right) + e^{\mu x_0 + \frac{\mu^2}{2} t} \erfc \left(\frac{x_0 -\mu t}{\sqrt{2 t}}\right) \right]   
    &~~\mathrm{if~~} x_0 < 0  
  \end{array}
\right.                      
\label{secondmomentlocaltime}
\end{eqnarray}


\subsection{ Large deviations of the intensive local time $a$ for the Brownian motion without drift $\mu=0$ }

\subsubsection{ Rate function $I(a) $ for the intensive local time $a= \frac{A-A_0}{t-t_0} \in [0,+\infty[$ }

The probability distribution $\Pi^{[\mu=0]} (A ,t \vert x_0,A_0,t_0)  $ of Eq. \ref{piamuzerototal} 
allows to evaluate the probability to see $A=A_0+a(t-t_0)$
\begin{eqnarray}
 \Pi^{[\mu=0]} (A =A_0+a(t-t_0),t \vert x_0,A_0,t_0) 
&& =  \delta(a(t-t_0))e^{- \frac{x_0^2}{2(t-t_0)} }
\int_{-\infty}^{+\infty} dz
\frac{ e^{- \frac{z^2}{2} } }{\sqrt{2 \pi }}  
\left[ e^{ \frac{z x_0}{ \sqrt{t-t_0}} }
-e^{- \frac{\vert z  x_0 \vert}{\sqrt{t-t_0}} } \right]
\nonumber \\
&& + \theta(a>0)  
 \sqrt{ \frac{2 } {  \pi (t-t_0) }  }
e^{- \frac{ x_0^2 }{2 (t-t_0)} - \vert x_0 \vert a- \frac{a^2}{2} (t-t_0)}  
\label{piamuzerototalintensive}
\end{eqnarray} 
So the large deviations of the intensive local time $a$
are governed by the simple rate function \cite{occupationsinai}
 \begin{eqnarray}
 I ( a ) = \frac{a^2}{2} \ \ {\rm for } \ \ a \in [0,+\infty[
\label{Iamu0}
\end{eqnarray} 
that vanishes and is minimum at its boundary value $a=0$ in agreement with Eq. \ref{ia0vanish}.

If one includes the prefactors, the leading order of the regular contribution of Eq. \ref{piamuzerototalintensive}
reads
\begin{eqnarray}
 \Pi^{Regular[\mu=0]} (A =A_0+a(t-t_0),t \vert x_0,A_0,t_0) 
&& \opsimeq_{(t-t_0) \to + \infty}    
 \sqrt{ \frac{2 } {  \pi (t-t_0) }  }
e^{ - \vert x_0 \vert a-  (t-t_0) I(a) }  
\label{piamuzerototalintensiveReg}
\end{eqnarray} 
The agreement with the general formula of Eq. \ref{laplaceinverseasaddlecriti}
can be checked using Eq. \ref{laplacefree} for $\mu=0$
\begin{eqnarray}
{\tilde G}^{[\mu=0]}_s (x \vert x_0) 
 = \frac{ e^{ - \sqrt{2 s } \vert x-x_0 \vert } } { \sqrt{2 s }} 
\label{laplacefreemuzero}
\end{eqnarray}
and Eq. \ref{saddlesa}
\begin{eqnarray}
  0  =    a   \partial_s \left[   \sqrt{2 s }    \right] -1 = \frac{a}{ \sqrt{2 s }} -1
\label{saddlesamuzero}
\end{eqnarray}
that leads to the saddle-point
\begin{eqnarray}
  s_a  =   \frac{a^2}{2}
\label{saddlesamuzerosol}
\end{eqnarray}


\subsubsection{ Rate function $I(a,v)$ for the intensive local time $a= \frac{A-A_0}{t-t_0} $ and the intensive displacement $v =\frac{x-x_0}{t-t_0} $  }

The joint propagator of Eq. \ref{jointpropagatorfree} for the case $\mu=0$
\begin{eqnarray}
  P^{[\mu=0]}(  x,A,t \vert   x_0,A_0,t_0) &&   =  \delta(A-A_0)
\frac{ e^{- \frac{x^2+x_0^2}{2(t-t_0)}}  }{\sqrt{2 \pi (t-t_0)}}  
\left[ e^{ \frac{xx_0}{2(t-t_0)} }
-e^{- \frac{\vert x  x_0 \vert}{(t-t_0)} }
\right]
\nonumber \\
&& + \theta(A>A_0) 
\left( \frac{\vert x \vert  +\vert x_0\vert +  A-A_0 }{ \sqrt{2 \pi }
(t-t_0)^{\frac{3}{2}}} \right)
e^{- \frac{x^2  +x_0^2 +  (A-A_0)^2}{2(t-t_0)} } 
  e^{- \frac{\vert x  x_0 \vert +  (\vert x \vert  +\vert x_0\vert )  (A-A_0) }{(t-t_0)} } 
\label{jointpropagatorfreemuzero}
\end{eqnarray}
allows to evaluate the joint probability to see $x=x_0+v (t-t_0)$ and $A=A_0+a(t-t_0)$
\begin{eqnarray}
&&  P^{[\mu=0]}(  x=x_0+v (t-t_0),A=A_0+a(t-t_0),t \vert   x_0,A_0,t_0)    =  \delta(a(t-t_0))
\frac{ e^{- \frac{\left[ x_0+v (t-t_0)\right]^2+x_0^2}{2(t-t_0)}}  }{\sqrt{2 \pi (t-t_0)}}  
\left[ e^{ \frac{\left[ x_0+v (t-t_0)\right]x_0}{2(t-t_0)} }
-e^{- \frac{\vert \left[ x_0+v (t-t_0)\right]  x_0 \vert}{(t-t_0)} }
\right]
\nonumber \\
&& + \theta(a>0) 
\left( \frac{\vert  x_0+v (t-t_0)\vert  +\vert x_0\vert +  a(t-t_0) }{ \sqrt{2 \pi }
(t-t_0)^{\frac{3}{2}}} \right)
e^{- \frac{\left[ x_0+v (t-t_0)\right]^2  +x_0^2 +  a^2(t-t_0)^2}{2(t-t_0)} } 
  e^{- \frac{\vert \left[ x_0+v (t-t_0)\right]  x_0 \vert +  (\vert  x_0+v (t-t_0) \vert  +\vert x_0\vert )  a(t-t_0) }{(t-t_0)} } 
  \nonumber \\
  &&=  \delta(a(t-t_0))
\frac{ e^{ - \frac{v^2}{2} (t-t_0) - x_0 v - \frac{ x_0^2}{(t-t_0)}}  }{\sqrt{2 \pi (t-t_0)}}  
\left[ e^{ \frac{ vx_0}{2} + \frac{ x_0^2}{2(t-t_0)} }
-e^{- \frac{\vert x_0^2 +v x_0 (t-t_0)  \vert}{(t-t_0)} }
\right]
\nonumber \\
&& + \theta(a>0) 
\left( \frac{\vert  x_0+v (t-t_0)\vert  +\vert x_0\vert +  a(t-t_0) }{ \sqrt{2 \pi }
(t-t_0)^{\frac{3}{2}}} \right)
e^{ - \frac{v^2+a^2}{2} (t-t_0) - x_0 v - \frac{ x_0^2 }{(t-t_0)} 
- \frac{\vert  x_0^2+x_0 v (t-t_0)  \vert  }{(t-t_0)} 
- (\vert  x_0+v (t-t_0) \vert  +\vert x_0\vert )  a }
\label{jointpropagatorfreemuzerointensive}
\end{eqnarray}
So the large deviations for the joint probability of
the intensive local time $a= \frac{A-A_0}{t-t_0} \in [0,+\infty[$
and of the intensive displacement $v =\frac{x-x_0}{t-t_0} \in ]-\infty,+\infty[$
 \begin{eqnarray}
P^{[\mu=0]}(  x=x_0+v (t-t_0),A=A_0+a(t-t_0),t \vert   x_0,A_0,t_0) 
   \oppropto_{(t-t_0) \to +\infty} e^{- (t-t_0) I(a,v)}
\label{Pjointvamuzero}
\end{eqnarray} 
are governed by the rate function 
 \begin{eqnarray}
 I(a,v) = \frac{v^2+a^2}{2} + \vert v \vert a = \frac{ ( \vert v \vert +a)^2}{2}
 \ \ \ \ {\rm for } \ \ a \in [0,+\infty[ \ \ \ {\rm and } \ \ \ v  \in ]-\infty,+\infty[
\label{Ivamu0}
\end{eqnarray} 
For any $a  \in [0,+\infty[$, the joint rate function
$ I(a,v)$ is minimum for $v=0$ where one recovers $I(a)$ of Eq. \ref{Iamu0}.

As a final remark, let us stress that the joint rate function $I(a,v)$ only occurs
for diffusion processes that are recurrent but do not converge towards an equilibrium
for the following reasons.

(i)  For transient processes, the local time $(A_T-A_0)$ remains a finite random variable for $T \to +\infty$,
while the large deviations properties of the intensive displacement $v =\frac{x-x_0}{t-t_0} $
are governed by some rate function $I(v)$.

(ii) For processes converging towards an equilibrium, 
the total displacement $(x-x_0)$ remains a finite random variable for $T \to +\infty$,
while the large deviations properties of the intensive local time $a= \frac{A-A_0}{t-t_0} $
are governed by some rate function $I(a)$ as described in subsection \ref{subsec_largedeva}.


\subsection{  Conditioning towards the position $x_T^*$ and the local time $A_T^*$ at the finite time horizon $T$}

\label{subsec_bridgexabrown}

Let us now apply the framework described in the subsection \ref{subsec_bridgexa}.
Using the explicit joint propagator of Eq. \ref{jointpropagatorfree}
\begin{eqnarray}
 && \ln  P(  x_T,A_T,T \vert   x,A,t) 
 = \mu (x_T-x)- \frac{\mu^2 }{2}(T-t) - \ln \left(\sqrt{2 \pi (T-t)}  \right)
\nonumber \\
&& + \ln \left[ 
 \delta(A_T-A) \left( e^{- \frac{(x_T-x)^2}{2(T-t)} }- e^{ - \frac{(\vert x_T \vert +\vert x \vert )^2}{2(T-t)} }    \right)  
 +  \theta(A_T>A) 
 \frac{(\vert x_T \vert +\vert x \vert +A_T-A)}{(T-t)} e^{ - \frac{(\vert x_T \vert +\vert x \vert +A_T-A)^2}{2(T-t)} }   
 \right]
\label{propagatorend}
\end{eqnarray}
one obtains that
the conditioned drift of Eq. \ref{mustarbridge}
\begin{eqnarray}
&&  \mu^{[x_T^*,A_T^*]}_T( x,A , t )  = \mu +  \partial_x    \ln P( x_T^*,A_T^*,T \vert  x,A,t)
  \nonumber \\
  && = \partial_x \ln \left[ 
\delta(A_T^*-A) \left( e^{- \frac{(x_T^*-x)^2}{2(T-t)} }- e^{ - \frac{(\vert x_T^* \vert +\vert x \vert )^2}{2(T-t)} }    \right)  
 +    \theta(A_T^*>A) 
 \frac{(\vert x_T^* \vert +\vert x \vert +A_T^*-A)}{(T-t)} e^{ - \frac{(\vert x_T^* \vert +\vert x \vert + A_T^*-A)^2}{2(T-t)} }   
 \right]
\label{mustarbridgebrown}
\end{eqnarray}
does not depend on the initial unconditioned drift $\mu$ anymore 
and
can be decomposed into the two following regions  for $x_T^* \ne 0$.

(i) In the region $A_0=0 \leq A<A_T^*$ where the local time $A$ has not yet reached its conditioned final value $A_T^*$, Eq. \ref{mustarbridgebrown} reduces to
\begin{eqnarray}
  \mu^{[x_T^*,A_T^*]}_T( x,A<A_T^* , t )   = 
  {\rm sgn}(x)
\left[ \frac{1}{\vert x_T^* \vert +\vert x \vert + A_T^*-A } -
 \frac{\vert x_T^* \vert +\vert x \vert + A_T^*-A}{T-t} 
\right] 
\label{mustarbridgebrownbelow}
\end{eqnarray}

(ii) In the region $A=A_T^*$ where the local time $A$ has already reached its conditioned final value $A_T^*$,
and where the position $x$ cannot visit the origin $x=0$ anymore,
the drift of Eq. \ref{mustarbridgebrown} reduces to
\begin{eqnarray}
  \mu^{[x_T^*,A_T^*]}_T( x , A=A_T^*, t )   = 
  \frac{\left( \frac{x_T^*-x}{T-t} \right) e^{- \frac{(x_T^*-x)^2}{2(T-t)} }
  + \left( \frac{x + {\rm sgn}(x) \vert x_T^* \vert}{T-t} \right) e^{ - \frac{(\vert x_T^* \vert +\vert x \vert )^2}{2(T-t)} }       }
{e^{- \frac{(x_T^*-x)^2}{2(T-t)} }- e^{ - \frac{(\vert x_T^* \vert +\vert x \vert )^2}{2(T-t)} }    
}
\label{mustarbridgebrownabove}
\end{eqnarray}

The fact that the initial unconditioned drift $\mu$ does not appear in the conditioned drift 
of Eqs \ref{mustarbridgebrownbelow}
and \ref{mustarbridgebrownabove} is actually an immediate consequence of a more general result stating that the constraints can be imposed one after the other \cite{refMazzoloJstat,us_DoobFirstEncounter}. By first imposing the final position of the process, one obtains a Brownian bridge which does not depend on the original drift. Then, imposing additional constraints on this process (whatever they are: local time, area under the curve, etc...) will not change this result. In addition, in the particular case where $A_T^*=0$ (in other word the process cannot cross the origin) then only the singular part of the propagator contributes to the conditioned drift. Assume moreover that $x_T^*$ and $x$ are positive, then Eq. \ref{mustarbridgebrownabove} reduces to
\begin{eqnarray}
  \mu^{[x_T^*,A_T^*]}_T( x , A=A_T^*=0, t )   = 
  \frac{  \left( \frac{x_T^*-x}{T-t} \right) e^{- \frac{(x_T^*-x)^2}{2(T-t)} }
  + \left( \frac{x_T^* +x}{T-t} \right) e^{ - \frac{( x_T^*  + x  )^2}{2(T-t)} }       }
{e^{- \frac{(x_T^*-x)^2}{2(T-t)} }- e^{ - \frac{( x_T^* + x )^2}{2(T-t)} }  }
\label{mustarbridgebrownaboveAstarzero}
\end{eqnarray}
which is the drift of a Brownian bridge conditioned to stay positive, as it should be. This equation can be found in \cite{refMajumdarOrland,refbookRogers}.

\begin{figure}[h]
\centering
\includegraphics[width=5.2in,height=4.5in]{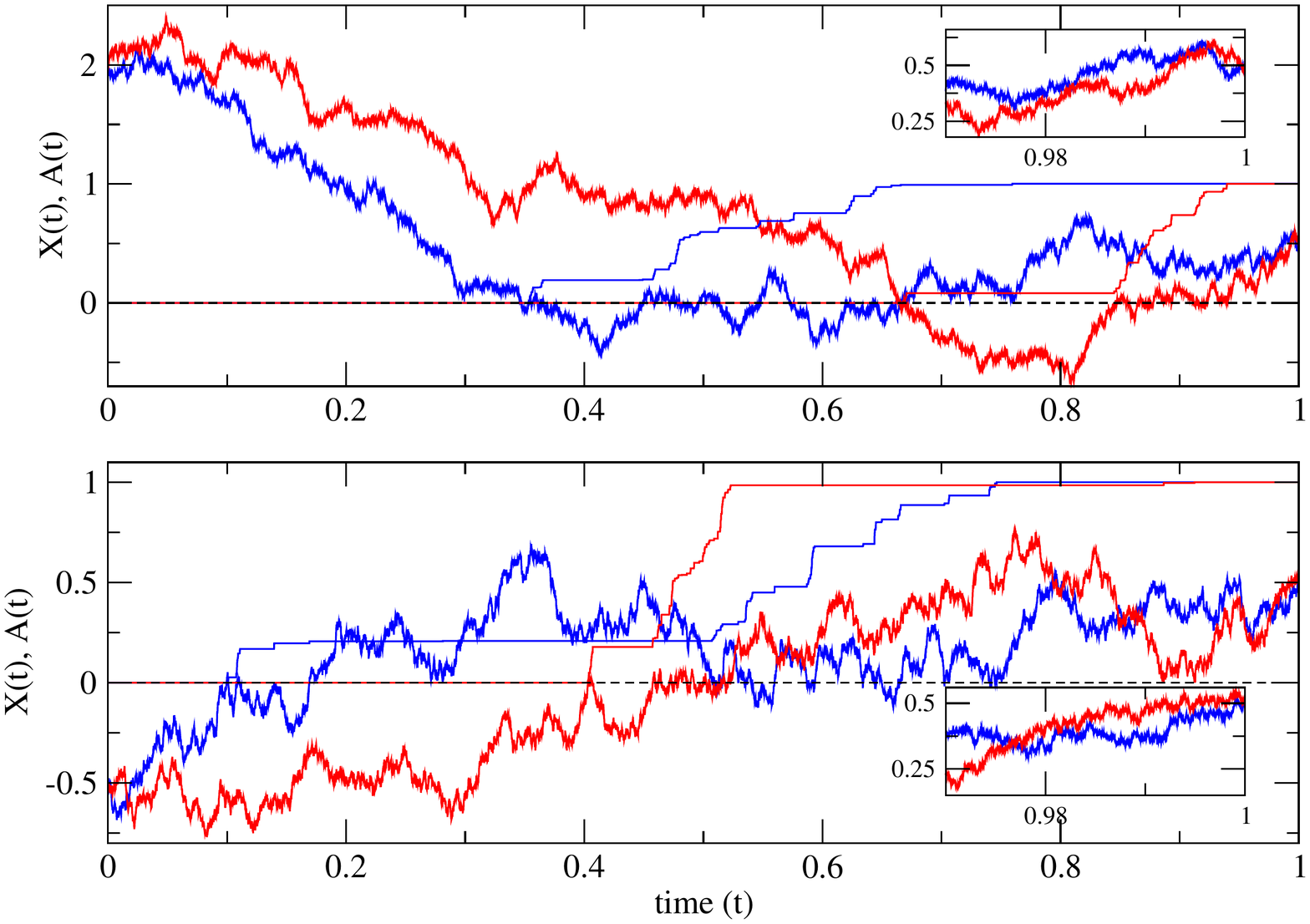}
\setlength{\abovecaptionskip}{15pt}  
\caption{Examples of realizations of the Brownian bridge conditioned to end at the final position $x_T^* = 0.5$ and to have the final local time $A_T^*=1$ at the finite time horizon $T=1$ (see the conditioned drift of Eqs. \ref{mustarbridgebrownbelow} and \ref{mustarbridgebrownabove}). For each trajectory, the associated local time $A(t)$ is shown as a function of the time $t \in [0,T]$. Top figure: the process begins at the position $x_0=2$. Bottom figure: the process begins at position $x_0=-0.5$. The encapsulated graphs show the convergence of both processes to the desired final value $x_T^* = 0.5$. The time step used in the discretization is $dt = 10^{-5}$.}
\label{fig2}
\end{figure}


\subsection{ Case $\mu=0$ : Conditioning towards the local time $A_T^*$ at the finite time horizon $T$  }

\label{subsec_bridgeabrown}

Let us now apply the framework described in the subsection \ref{subsec_bridgea}.
Using the explicit probability of Eq. \ref{piamuzerototal} 
\begin{eqnarray}
 \Pi^{[\mu=0]} (A_T^* ,T \vert x,A,t) 
 =  \delta(A_T^*-A) \erf \left(\frac{\vert x \vert}{\sqrt{2(T-t)}} \right)
  + \theta(A_T^*>A)  
 \sqrt{ \frac{2 } {  \pi (T-t) }  }
e^{- \frac{( \vert x \vert + A_T^*-A )^2}{2 (T-t)} }  
\label{piamuzerototalstar}
\end{eqnarray} 
one obtains that the conditioned drift of Eq. \ref{mustarbridgepi}
\begin{eqnarray}
  \mu_T^{[A_T^*]}( x,A , t ) =   \partial_x    \ln \Pi^{[\mu=0]} ( A_T^*,T \vert  x,A,t)
\label{mustarbridgepimu0}
\end{eqnarray}
can be decomposed into the two following regions.

(i) In the region $A_0=0 \leq A<A_T^*$ where the local time $A$ has not yet reached its conditioned final value $A_T^*$, the conditioned drift of Eq. \ref{mustarbridgepimu0} reduces to
\begin{eqnarray}
  \mu_T^{[A_T^*]}( x,A<A_T^* , t )  && =  \partial_x    \ln \left[  \sqrt{ \frac{2 } {  \pi (T-t) }  }
e^{- \frac{( \vert x \vert + A_T^*-A )^2}{2 (T-t)} }\right]
\nonumber \\
&&  =  - {\rm sgn} (x) \ \frac{\vert x \vert + A_T^*-A }{ T-t} 
\label{mustarbridgepibelowmu0}
\end{eqnarray}

(ii) In the region $A=A_T^*$ where the local time $A$ has already reached its conditioned final value $A_T^*$,
and where the position $x$ cannot visit the origin $x=0$ anymore,
the conditioned drift of Eq. \ref{mustarbridgepimu0} reads
\begin{eqnarray}
  \mu_T^{[A_T^*]}( x , A=A_T^*, t )  
  &&  = 
   \partial_x    \ln \left[ \erf \left(\frac{\vert x \vert}{\sqrt{2(T-t)}} \right)
\right]
\nonumber \\
&& = \sqrt{\frac{2}{\pi (T-t)}} \frac{e^{-\frac{x^2}{2(T-t)}}}{\erf \left(\frac{\vert x \vert}{\sqrt{2(T-t)}} \right)} {\rm sgn}(x)
\label{mustarbridgepiabovemu0}
\end{eqnarray}
The asymptotic behavior near the origin $x \to 0$ is given by
\begin{eqnarray}
  \mu_T^{[A_T^*]}[ x  , A=A_T^*, t ] \opsimeq_{x \to 0} \frac{1}{x} - \frac {x}{3(T-t)}
\label{mustarbridgepiabovemu0xto0}
\end{eqnarray}
Due to the $1/x$ term, the origin $x = 0$ is an entrance boundary that the process cannot cross, therefore in the second region the local time can no longer increase, as wished.

\begin{figure}[h]
\centering
\includegraphics[width=5.2in,height=4.5in]{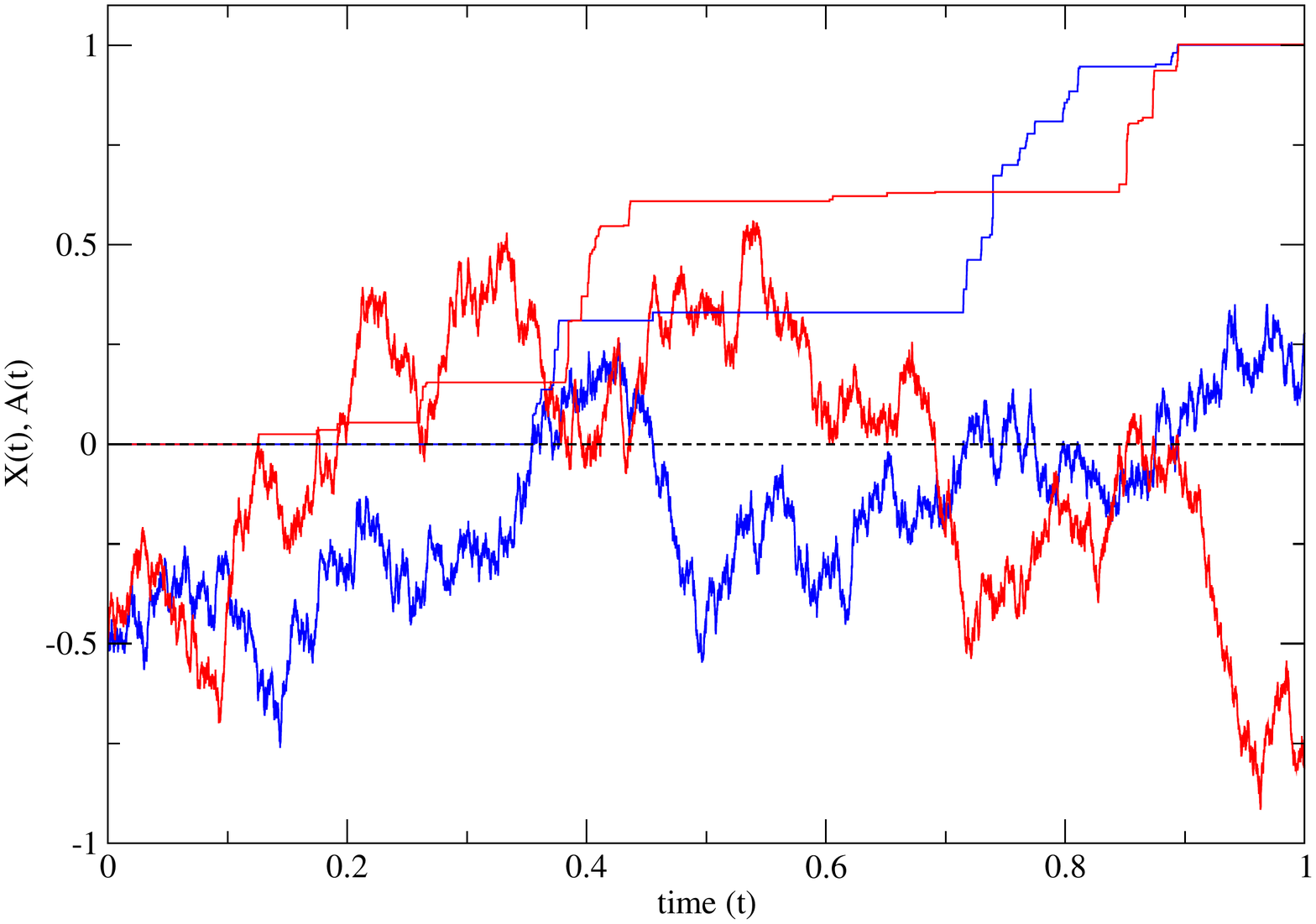}
\setlength{\abovecaptionskip}{15pt}  
\caption{Examples of realizations of the Brownian process conditioned 
to have the final local time $A_T^*=1$ at the finite time horizon $T=1$ (see the conditioned drift of Eqs. \ref{mustarbridgepibelowmu0} and \ref{mustarbridgepiabovemu0}). For each trajectory, the associated local time $A(t)$ is shown as a function of the time $t \in [0,T]$. The process can end at any final position $x_T$, while the initial position is $x_0=-0.5$ here. The time step used in the discretization is $dt = 10^{-5}$.}
\label{fig3}
\end{figure}


\subsection{ Case $\mu=0$ : Conditioning towards the intensive local time $a^*=\frac{A_T^*}{T}$ in the limit $T\to +\infty$ }

\label{subsec_brownintensive}

In order to impose the intensive local time $a^*=\frac{A_T^*}{T}$ in the limit $T\to +\infty$,
one can plug $A_T^* = T a^* $
into the conditioned drift of Eq. \ref{mustarbridgepibelowmu0}
to obtain at leading order for $T \to +\infty$ while $t$ remains finite
\begin{eqnarray}
  \mu_T^{[T a^*]}( x,A<T a^* , t )    =  - {\rm sgn} (x) \ \frac{\vert x \vert + (T a^*-A) }{ T-t} 
  \opsimeq_{T \to +\infty } - {\rm sgn} (x) a^* \equiv  \mu_{\infty}^{[a^*]}( x )
\label{mustarbridgepibelowmu0intensive}
\end{eqnarray}

The agreement with the general formula of Eq. \ref{mustarbridgepibelowintensiveasympto} for the drift $\mu_{\infty}^{[a^*]}(x ) $
can be checked using Eq. \ref{laplacefree}
\begin{eqnarray}
{\tilde G}^{[\mu=0]}_s (0 \vert x) 
 = \frac{ e^{ - \sqrt{2 s } \vert x \vert } } { \sqrt{2 s }}  
 \label{laplacefreeintense}
\end{eqnarray}
   and the saddle-point value $s_{a^*}=\frac{[a^*]^2}{2} $ of Eq. \ref{saddlesamuzerosol} to obtain
\begin{eqnarray}
\mu_{\infty}^{[a^*]}( x ) =    \partial_x    \ln {\hat G}_{s_{a^*}} (0 \vert x)  
= \partial_x    \left( - \sqrt{2 s_{a^*} } \vert x \vert - \ln (\sqrt{2 s_{a^*} }) \right) 
= - \sqrt{2 s_{a^*} }  {\rm sgn} (x)
= -  {\rm sgn} (x) a^*
\label{mustarbridgepibelowintensiveasymptobrown}
\end{eqnarray}
As explained in the Appendices,
this result can be also recovered via the appropriate canonical conditioning leading to Eq. \ref{mustarpnegtrada}.


\subsection{ Case $\mu>0$ : Conditioning towards the finite local time $A_{\infty}^*<+\infty$ at the infinite time horizon $T= + \infty$ }

\begin{figure}[h]
\centering
\includegraphics[width=5.2in,height=4.in]{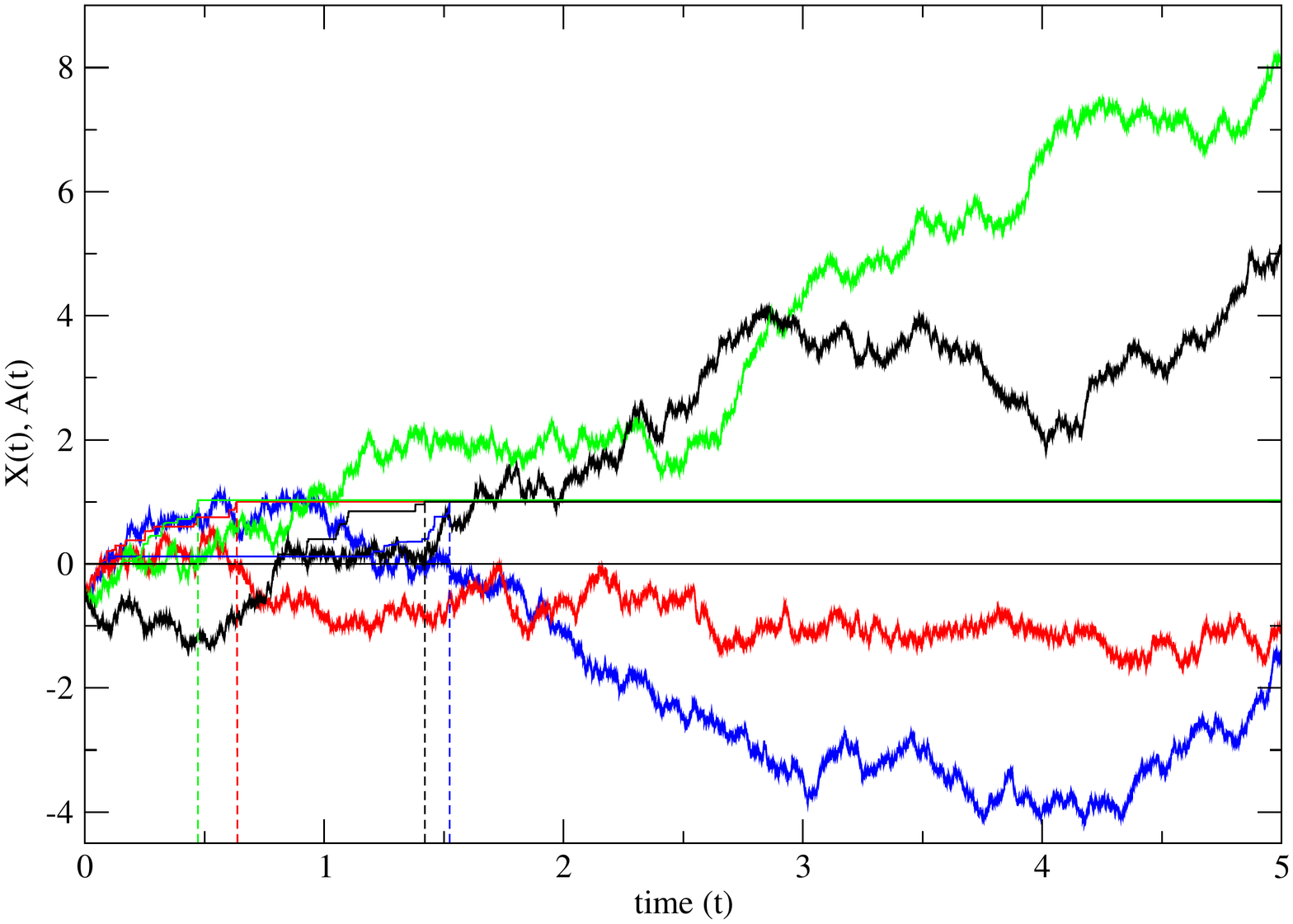}
\setlength{\abovecaptionskip}{15pt}  
\caption{Examples of realization of the Brownian process with drift $\mu>0$ where the local time is conditioned 
towards the finite asymptotic value $A_{\infty}^* = 1 $ at the infinite time horizon $T= + \infty$ (see the conditioned drift of Eqs. \ref{mustarpiinftybridgebelowfree} and \ref{mustarpiinftybridgeabovefreeregionpos}). For each trajectory, the associated local time $A(t)$ is shown as a function of the time $t $. The dashed vertical lines indicate the time when the local time reaches the finite asymptotic value $A_{\infty}^* = 1 $. The time step used in the discretization is $dt = 10^{-5}$.}
\label{fig4}
\end{figure}

\label{subsec_brownainfty}

The framework described in the subsection \ref{subsec_doobinfinity} can be applied as follows.

(i) In the region $A_0=0 \leq A<A_{\infty}^*$ where the local time $A$ has not yet reached its conditioned 
asymptotic value $A_T^*$, 
the conditioned drift $ \mu_{A_{\infty}^*}^*( x,A< A_{\infty}^*  )$ of Eq. \ref{mustarbridgepibelowinfty}
can be obtained from $\Pi^{Regular} (A ,\infty \vert x_0,A_0)$ given in Eq. \ref{pitotfreeinfty} 
\begin{eqnarray}
  \mu_{A_{\infty}^*}^*( x,A< A_{\infty}^*  ) 
  && = \mu +  \partial_x    \ln \Pi^{Regular}( A_{\infty}^*,\infty \vert  x,A) 
 = \mu +  \partial_x    \ln 
  \left[   \mu e^{ - \mu ( x+\vert x \vert + A_{\infty}^*-A ) } \right] 
    \nonumber \\
  && = -\mu \, {\rm sgn} (x)
\label{mustarpiinftybridgebelowfree}
\end{eqnarray}
So in the region $x<0$, the conditioned drift coincide with the initial unconditioned drift $\mu$,
while in the region $x>0$, the conditioned drift is opposite to the initial unconditioned drift $\mu$
in order to visit again the origin and to increase the local time.

(ii) In the region $A=A_{\infty}^*$ where the local time $A$ has already reached its conditioned 
asymptotic value $A_T^*$,
the conditioned drift of Eq. \ref{mustarbridgepiaboveinfty}
can be obtained from $\Pi^{Singular}( A_{\infty}^*,\infty \vert  x,A)$ given in Eq. \ref{pitotfreeinfty} 
for $x>0$
\begin{eqnarray}
 \mu_{A_{\infty}^*}^*( x>0,A_{\infty}^*  ) 
 && =   \mu +   \partial_x    \ln \Pi^{Singular}( A_{\infty}^*,\infty \vert  x,A)  = \mu +  \partial_x    \ln    \left[ 1 -  e^{- \mu 2x  } \right]
 \nonumber \\
 &&
  =   \mu \left(   \frac{e^{ 2 \mu x  } + 1    } {  e^{ 2 \mu x   } -1 } \right) 
 =    \mu \coth(\mu x) 
\label{mustarpiinftybridgeabovefreeregionpos}
\end{eqnarray}
For $x<0$, the conditioned drift of Eq. \ref{mustarbridgepiaboveinfty}
should be computed using the leading asymptotic form of $\Pi^{Singular}( A_{\infty}^*,T \vert  x<0,A,t)  $ 
for large time interval $(T-t)$ given in Eq. \ref{hatsintegratesingularfreeinfty}
\begin{eqnarray}
 \mu_{A_{\infty}^*}^*( x<0,A=A_{\infty}^*  )  
 && =   \mu + \lim_{T \to +\infty} \left(  \partial_x    \ln \Pi^{Singular}( A_{\infty}^*,T \vert  x<0,A,t) \right)
  = \mu +  \lim_{T \to +\infty} \left( \partial_x    \ln    \left[   (-x) \sqrt{\frac{2}{\pi}} \frac{e^{-\frac{(-x+\mu T)^2}{2 T}}}{\mu^2 T^{3/2}}\right] \right) 
\nonumber \\
 &&  = \lim_{T \to +\infty} \left(     \frac{1}{x} - \frac{x}{T} \right) 
=    \frac{1}{x} 
\label{mustarpiinftybridgeabovefreeregionneg}
\end{eqnarray}
In both cases, whether $x>0$ or $x<0$, when $x \to 0$, the conditioned drift behaves as $\mu_{A_{\infty}^*}^*( x \to 0,A=A_{\infty}^*  ) \simeq 1/x$. The origin $x = 0$ is thus an entrance boundary that the process cannot cross, 
therefore in the second region (ii), the local time cannot increase any further, as expected.


\section{ Application to the drift $\mu(x)=- \mu \, {\rm sgn}( x)$ of parameter $\mu>0$   }

\label{sec_sgn}

In this section, we consider the drift directed towards the origin $x=0$ of amplitude $\mu>0$ 
\begin{eqnarray}
 \mu(x)=- \mu \, {\rm sgn}( x)
\label{musgn}
\end{eqnarray}
This process is sometimes called Brownian motion with alternating drift (or bang-bang process \cite{BorodinHandbook,touchette_bang_bang}) and was originally introduced by de Gennes to study dry friction \cite{deGennes}.

\noindent The associated potential $U(x)$ of Eq. \ref{potentialU}
\begin{eqnarray}
U(x) =  2 \mu \int_0^{x} dy \, {\rm sgn}( y) = 2 \mu \vert x \vert
\label{potentialUsgn}
\end{eqnarray}
corresponds to the normalizable equilibrium Boltzmann distribution of Eq. \ref{boltzmann}
\begin{eqnarray}
 G_{eq}(  x) = \frac{ e^{-U(x) }}{ \int_{-\infty}^{+\infty} dy e^{-U(y) } } =  \mu  e^{ - 2 \mu  \vert x \vert }
\label{boltzmannsgn}
\end{eqnarray}
So the unconditioned dynamics converges towards this equilibrium distribution for $t \to +\infty$,
and the local time increment $(A-A_0)$ grows extensively in $(t-t_0)$  as discussed in subsection \ref{subsec_equilibrium}.


\subsection{ Properties of the unconditioned diffusion process $X(t)$ alone }

\subsubsection{ Propagator $G(x,t \vert x_0,t_0) $ for the position alone}

Via the similarity transformation of Eq. \ref{defpsi} based on the potential $U(x)$ of Eq. \ref{potentialUsgn}
\begin{eqnarray}
G(x,t \vert x_0,t_0) = e^{ \mu (\vert x_0 \vert - \vert x \vert) } \psi(x,t \vert x_0,t_0) 
\label{psisgn}
\end{eqnarray}
the Fokker-Planck Eq. \ref{forward1d}
for the propagator $G(  x,t \vert   x_0,t_0) $ 
becomes the Schr\"odinger Equation of Eq. \ref{schrodinger}
for $\psi(x,t \vert x_0,t_0) $,
where the quantum Hamiltonian of Eq. \ref{hamiltonian}
involves the potential of Eq. \ref{susy} that reads for the drift of Eq. \ref{musgn}
\begin{eqnarray}
 V(x) = \frac{ \mu^2}{2}  - \mu \delta(x)
\label{susysgn}
\end{eqnarray}
As a consequence, the Hamiltonian of Eq. \ref{hamiltonian} can be decomposed 
\begin{eqnarray}
H && = H_0+H_1
\label{Hsgn01}
\end{eqnarray}
into
the two contributions
\begin{eqnarray}
H_0 && \equiv -  \frac{1}{2} \partial_x^2 + \frac{ \mu^2}{2}
\nonumber \\
H_1 && \equiv   - \mu \delta(x)
\label{Hsgn}
\end{eqnarray}

When the contribution $H_1$ is absent, the Schr\"odinger propagator $\psi^{[0]}(x,t \vert x_0,t_0) $
associated to the Hamiltonian $H_0$ whose potential reduces to the constant $\frac{ \mu^2}{2} $ 
is given by
\begin{eqnarray}
\psi^{[0]}(x,t \vert x_0,t_0) = \frac{1}{\sqrt{2 \pi (t-t_0)}}  
e^{- \frac{(x-x_0)^2}{2(t-t_0)} - \frac{\mu^2 }{2}(t-t_0)}
\label{psizero}
\end{eqnarray}
while its time Laplace transform reads
\begin{eqnarray}
{\tilde \psi}^{[0]}_s (x \vert x_0) && \equiv \int_{t_0}^{+\infty} dt e^{-s (t-t_0) } \psi^{[0]}(  x,t \vert   x_0,t_0) 
= \frac{ 1 } { \sqrt{2 \pi }} 
\int_{0}^{+\infty} d\tau \tau^{-\frac{1}{2}} e^{-\left(s + \frac{\mu^2}{2} \right) \tau }
e^{- \frac{(x-x_0)^2}{2\tau} }
 = \frac{ e^{ - \sqrt{\mu^2+2 s } \vert x-x_0 \vert } } { \sqrt{\mu^2+2 s }} 
\label{laplacepsizero}
\end{eqnarray}

When the contribution $H_1$ is present, the Laplace transform ${\tilde \psi}_s (x \vert x_0)$
of the Schr\"odinger propagator $\psi(x,t \vert x_0,t_0) $ for the full Hamiltonian $H$
can be computed from ${\tilde \psi}^{[0]}_s (x \vert x_0) $ of Eq. \ref{laplacepsizero}
via the Dyson formula analog to Eq. \ref{resum}
to obtain
\begin{eqnarray}
 {\tilde \psi}_s (x \vert x_0)
&&  =  {\tilde \psi}^{[0]}_s (x \vert x_0) 
+ {\tilde \psi}^{[0]}_s (x \vert 0)  
\frac{ \mu   }{ 1-\mu {\tilde \psi}^{[0]}_s (0 \vert 0)}
{\tilde \psi}^{[0]}_s (0 \vert x_0)
\nonumber \\
&& =\frac{ e^{ - \sqrt{\mu^2+2 s } \vert x-x_0 \vert } } { \sqrt{\mu^2+2 s }}  
+ \frac{ e^{ - \sqrt{\mu^2+2 s } \vert x \vert } } { \sqrt{\mu^2+2 s }}  
\left( \frac{ \mu   }{ 1- \frac{ \mu } { \sqrt{\mu^2+2 s }} } \right)
\frac{ e^{ - \sqrt{\mu^2+2 s } \vert x_0 \vert } } { \sqrt{\mu^2+2 s }} 
\nonumber \\
&& =\frac{ 1 } { \sqrt{\mu^2+2 s }}  
\left[ e^{ - \sqrt{\mu^2+2 s } \vert x-x_0 \vert }
+  \frac{ \mu   }{ \sqrt{\mu^2+2 s }-  \mu } 
 e^{ - \sqrt{\mu^2+2 s } (\vert x \vert +  \vert x_0 \vert) } 
\right]
\label{dysonsgn}
\end{eqnarray}

So the Laplace transform ${\tilde G}_s (x \vert x_0) $ of the Fokker-Planck propagator 
$G(  x,t \vert   x_0,t_0) $ of Eq. \ref{psisgn}
reads
\begin{eqnarray}
{\tilde G}_s (x \vert x_0) && \equiv \int_{t_0}^{+\infty} dt e^{-s (t-t_0) } G(  x,t \vert   x_0,t_0) 
= \int_{t_0}^{+\infty} dt e^{-s (t-t_0) } e^{ \mu (\vert x_0 \vert - \vert x \vert) } \psi(x,t \vert x_0,t_0) 
= e^{ \mu (\vert x_0 \vert - \vert x \vert) }  {\tilde \psi}_s (x \vert x_0)
\nonumber \\
&& =
\frac{ e^{ \mu (\vert x_0 \vert - \vert x \vert) } } { \sqrt{\mu^2+2 s }}  
\left[ e^{ - \sqrt{\mu^2+2 s } \vert x-x_0 \vert }
+  \frac{ \mu   }{ \sqrt{\mu^2+2 s }-  \mu }  
 e^{ - \sqrt{\mu^2+2 s } (\vert x \vert +  \vert x_0 \vert) } 
\right]
\label{laplacesgn}
\end{eqnarray}

The limit $s \to 0$ of
\begin{eqnarray}
\lim_{s \to 0} \left[ s  {\tilde G}_s (x \vert x_0)  \right]
&& =  \lim_{s \to 0} \left( s  
\frac{ e^{ \mu (\vert x_0 \vert - \vert x \vert) } } { \sqrt{\mu^2+2 s }}  
\left[ e^{ - \sqrt{\mu^2+2 s } \vert x-x_0 \vert }
+  \frac{ \mu  (\sqrt{\mu^2+2 s }+  \mu )  }{ 2s }  
 e^{ - \sqrt{\mu^2+2 s } (\vert x \vert +  \vert x_0 \vert) } 
\right]
  \right)
  \nonumber \\
  && =  \mu  e^{ - 2 \mu  \vert x \vert } = G_{eq}(x)
\label{sto0eq}
\end{eqnarray}
allows to recover the equilibrium distribution $ G_{eq}(x)$ of Eq. \ref{boltzmannsgn} as it should.


\subsubsection{ Properties in the presence of an absorbing boundary at the origin $x=0$ }

In the presence of an absorbing boundary at the origin $x=0$, 
the present model $\mu(x) = -\mu \, {\rm sgn}(x)$ is of course very similar 
to the previous section concerning the model $\mu(x)=\mu$ : the two models coincide for $x_0<0$,
and the region $x_0>0$ could be obtained by symmetry for the present model.
However, one can also use the general formula as follows.

The evaluation of Eq. \ref{laplacesgn}
for the special case $x=0$
\begin{eqnarray}
{\tilde G}_s (0 \vert x_0)  =
\frac{ e^{ \mu \vert x_0 \vert  } } { \sqrt{\mu^2+2 s }}  
\left[ e^{ - \sqrt{\mu^2+2 s } \vert x_0 \vert }
+  \frac{ \mu   }{ \sqrt{\mu^2+2 s }-  \mu }  
 e^{ - \sqrt{\mu^2+2 s }   \vert x_0 \vert } 
\right]
 = \frac{ e^{ (\mu -\sqrt{\mu^2+2 s } )\vert x_0 \vert   } } { \sqrt{\mu^2+2 s }-  \mu}  
\label{laplacesgnxeq0}
\end{eqnarray}
for the special case  $x_0=0$
\begin{eqnarray}
{\tilde G}_s (x \vert 0) && =
\frac{ e^{ - \mu  \vert x \vert } } { \sqrt{\mu^2+2 s }}  
\left[ e^{ - \sqrt{\mu^2+2 s } \vert x \vert }
+  \frac{ \mu   }{ \sqrt{\mu^2+2 s }-  \mu } 
 e^{ - \sqrt{\mu^2+2 s }   \vert x \vert } 
\right]  = \frac{ e^{ - (\mu +\sqrt{\mu^2+2 s } )\vert x \vert   } } { \sqrt{\mu^2+2 s }-  \mu}  
\label{laplacesgnx0eq0}
\end{eqnarray}
and for the special case  $x=0=x_0$
\begin{eqnarray}
{\tilde G}_s (0 \vert 0)  
=  \frac{ 1   }{ \sqrt{\mu^2+2 s }-  \mu }  
\label{laplacesgn00}
\end{eqnarray}
allows to compute the Laplace transform ${\hat G}^{abs}_{s} (x  \vert x_0)  $
via Eq. \ref{laplaceabsorbing}
\begin{eqnarray}
{\hat G}^{abs}_s (x \vert x_0) && \equiv {\hat G}_s (x \vert x_0) 
- \frac{ {\hat G}_s (x \vert 0){\hat G}_s (0 \vert x_0) }{ {\hat G}_s (0 \vert 0) }  
 =  \frac{ e^{ \mu (\vert x_0 \vert - \vert x \vert) } } { \sqrt{\mu^2+2 s }}  
\left[ e^{ - \sqrt{\mu^2+2 s } \vert x-x_0 \vert }
- e^{ - \sqrt{\mu^2+2 s } (\vert x \vert +  \vert x_0 \vert) } 
\right]
\label{laplaceabssgn}
\end{eqnarray}
 The Laplace inversion using Eq. \ref{laplaceinter}
 yields the propagator $G^{abs} (x,t \vert x_0,t_0)  $ in the presence of an absorbing boundary at the origin $x=0$
\begin{eqnarray}
  G^{abs} (  x,A,t \vert   x_0,A_0,t_0)
&&  = 
 e^{ \mu (\vert x_0 \vert - \vert x \vert) }
 \frac{e^{- \frac{\mu^2}{2}  (t-t_0) }  } { \sqrt{2 \pi (t-t_0)}} 
\left[ e^{- \frac{ ( x-x_0 )^2}{2(t-t_0)} } 
- e^{- \frac{(\vert x \vert +  \vert x_0 \vert)^2}{2(t-t_0)} } 
\right]
\nonumber \\
&& =  e^{ \mu (\vert x_0 \vert - \vert x \vert) }
 \frac{e^{- \frac{\mu^2}{2}  (t-t_0) - \frac{ x^2+x_0^2}{2(t-t_0)} }} { \sqrt{2 \pi (t-t_0)}} 
\left[ e^{ \frac{ x x_0}{(t-t_0)} } 
- e^{- \frac{\vert x x_0 \vert}{(t-t_0)} } 
\right]
\label{gabssgn}
\end{eqnarray}
in agreement with the method of images.

The Laplace transform $ {\hat \gamma}^{abs}_s ( x_0) $ of Eq. \ref{laplacegammaabs} reads
\begin{eqnarray}
{\hat \gamma}^{abs}_s ( x_0) 
  = \frac{ {\hat G}_s (0 \vert x_0) }{ {\hat G}_s (0 \vert 0) } = 
  e^{ (\mu -\sqrt{\mu^2+2 s } )\vert x_0 \vert   }
\label{laplacegammaabssgn}
\end{eqnarray}
Its Laplace inversion yields the absorption rate $\gamma^{abs}(t \vert x_0,t_0) $ 
\begin{eqnarray}
\gamma^{abs}(t \vert x_0,t_0)  = 
\frac{ \vert x_0 \vert  } { \sqrt{2 \pi } (t-t_0)^{\frac{3}{2}}} 
  e^{ \mu \vert x_0 \vert - \frac{\mu^2}{2} (t-t_0) - \frac{ x_0^2}{2(t-t_0)} } 
\label{gammaabssgn}
\end{eqnarray}
One can check the normalization to unity for any starting point $x_0$
\begin{eqnarray}
\int_{t_0}^{+\infty} dt \gamma^{abs}(t \vert x_0,t_0)  = {\hat \gamma}^{abs}_{s=0} ( x_0) = 1
\label{normagammaabssgn}
\end{eqnarray}

The survival probability $S^{abs} (t \vert x_0,t_0) $ of Eq. \ref{survival}
can be obtained from the 
integral over the final position $x$ of the propagator $G^{abs}(x,t \vert x_0,t_0) $ of Eq. \ref{gabssgn}
\begin{eqnarray}
S^{abs} (t \vert x_0,t_0) \equiv  \int_{-\infty}^{+\infty} dx G^{abs}(x,t \vert x_0,t_0)
=  \frac{e^{- \frac{\mu^2}{2}  (t-t_0) + \mu \vert x_0 \vert - \frac{ x_0^2}{2(t-t_0)}
}  } { \sqrt{2 \pi (t-t_0)}} 
\int_{-\infty}^{+\infty} dx
 e^{ - \mu  \vert x \vert - \frac{ x^2}{2(t-t_0)}}
\left[ e^{ \frac{ x x_0}{(t-t_0)} } 
- e^{- \frac{\vert x  x_0 \vert}{(t-t_0)} } 
\right] \ \ 
\label{survivalsgn}
\end{eqnarray}
Its asymptotic decay for large time $(t-t_0)$ is given by
\begin{eqnarray}
S^{abs} (t \vert x_0,t_0)  \opsimeq_{(t-t_0) \to + \infty} 
 \frac{e^{- \frac{\mu^2}{2}  (t-t_0) + \mu \vert x_0 \vert 
}  } { \sqrt{2 \pi (t-t_0)}} 
\int_{-\infty}^{+\infty} dx
 e^{ - \mu  \vert x \vert }
\left[ \frac{ x x_0 +\vert x  x_0 \vert }{t-t_0}   
\right] 
= \sqrt{\frac{2}{\pi} } \frac{  \vert   x_0 \vert e^{ \mu \vert x_0 \vert- \frac{\mu^2}{2}  (t-t_0)  }} { \mu^2   (t-t_0)^{\frac{3}{2}}}  
\label{survivalsgnlarget}
\end{eqnarray}


\subsection{ Joint propagator $  P(  x,A,t \vert   x_0,A_0,t_0) $ for the unconditioned joint process $[X(t),A(t)]$  }

The singular contribution of Eq. \ref{hatsSingularAbsinv}
involves the propagator $G^{abs}(x,t \vert x_0,t_0)  $ of Eq. \ref{gabssgn}
\begin{eqnarray}
 P^{Singular} (x,A,t  \vert x_0,A_0,t_0) 
 &&  =  \delta(A-A_0) G^{abs}(x,t \vert x_0,t_0)
 \nonumber \\
 && =    \delta(A-A_0)
e^{ \mu (\vert x_0 \vert - \vert x \vert) }
 \frac{e^{- \frac{\mu^2}{2}  (t-t_0) }  } { \sqrt{2 \pi (t-t_0)}} 
\left[ e^{- \frac{ ( x-x_0 )^2}{2(t-t_0)} } 
- e^{- \frac{(\vert x \vert +  \vert x_0 \vert)^2}{2(t-t_0)} } 
\right]
\label{hatsSingularAbsinvsgn}
\end{eqnarray}

The Laplace transform $ {\hat P}^{Regular}_{s} (x,A  \vert x_0,A_0)  $ of Eq. \ref{hatsRegular}
reads using Eqs \ref{laplacesgnxeq0}, \ref{laplacesgnx0eq0} and \ref{laplacesgn00}
\begin{eqnarray}
 {\hat P}^{Regular}_{s} (x,A  \vert x_0,A_0)  && = 
 \theta(A>A_0) \left[ \frac{ {\hat G}_s (x \vert 0){\hat G}_s (0 \vert x_0) }{ {\hat G}^2_s (0 \vert 0) } \right]
 e^{-  \frac{(A-A_0)}{{\hat G}_s (0 \vert 0)} }  
 \nonumber \\
 && = \theta(A>A_0)   e^{ \mu (\vert x_0 \vert - \vert x \vert +A-A_0) }
 e^{ -\sqrt{\mu^2+2 s } (\vert x_0 \vert +\vert x \vert  + A-A_0  ) }  
\label{hatsRegularsgn}
\end{eqnarray}
Equation \ref{laplacederi} allows to compute the Laplace inversion
\begin{eqnarray}
  P^{Regular}(  x,A,t \vert   x_0,A_0,t_0)
=  \theta(A>A_0) 
 e^{ \mu (\vert x_0 \vert - \vert x \vert +A-A_0) - \frac{\mu^2}{2}  (t-t_0) }
\left( \frac{\vert x_0 \vert +\vert x \vert  + A-A_0  }{ \sqrt{2 \pi }(t-t_0)^{\frac{3}{2}}} \right)  
e^{- \frac{(\vert x_0 \vert +\vert x \vert  + A-A_0  )^2}{2(t-t_0)} } 
\label{propagatorRegularinversessgn}
\end{eqnarray}
So the joint propagator $ P(  x,A,t \vert   x_0,A_0,t_0) $
involving the two contributions of Eq. \ref{hatsSingularAbsinvsgn}
and Eq. \ref{propagatorRegularinversessgn}
 reads
\begin{eqnarray}
  P(  x,A,t \vert   x_0,A_0,t_0)
&&  = 
 \delta(A-A_0)
e^{ \mu (\vert x_0 \vert - \vert x \vert) }
 \frac{e^{- \frac{\mu^2}{2}  (t-t_0) }  } { \sqrt{2 \pi (t-t_0)}} 
\left[ e^{- \frac{ ( x-x_0 )^2}{2(t-t_0)} } 
- e^{- \frac{(\vert x \vert +  \vert x_0 \vert)^2}{2(t-t_0)} } 
\right]
\nonumber \\
&& + \theta(A>A_0) 
 e^{ \mu (\vert x_0 \vert - \vert x \vert +A-A_0) - \frac{\mu^2}{2}  (t-t_0) }
\left( \frac{\vert x_0 \vert +\vert x \vert  + A-A_0  }{ \sqrt{2 \pi }(t-t_0)^{\frac{3}{2}}} \right)  
e^{- \frac{(\vert x_0 \vert +\vert x \vert  + A-A_0  )^2}{2(t-t_0)} } 
\label{propagatorinversessgn}
\end{eqnarray}
The integration of this joint propagator $P(  x,A,t \vert   x_0,A_0,t_0) $ of Eq. \ref{propagatorinversessgn} over the local time $A$
\begin{eqnarray}
 \int_{0}^{+\infty} dA \, P(  x,A,t \vert   x_0,A_0,t_0)
&&  = 
e^{ \mu (\vert x_0 \vert - \vert x \vert) }
 \frac{e^{- \frac{\mu^2}{2}  (t-t_0) }  } { \sqrt{2 \pi (t-t_0)}} 
\left[ e^{- \frac{ ( x-x_0 )^2}{2(t-t_0)} } 
- e^{- \frac{(\vert x \vert +  \vert x_0 \vert)^2}{2(t-t_0)} } 
\right]
 \\
&& + \int_{A_0}^{+\infty} dA \,
 e^{ \mu (\vert x_0 \vert - \vert x \vert +A-A_0) - \frac{\mu^2}{2}  (t-t_0) }
\left( \frac{\vert x_0 \vert +\vert x \vert  + A-A_0 }{ \sqrt{2 \pi }(t-t_0)^{\frac{3}{2}}} \right)  
e^{- \frac{(\vert x_0 \vert +\vert x \vert  + A-A_0  )^2}{2(t-t_0)} } 
\nonumber \\
&& = e^{-2 \mu \vert x \vert} \left[ \frac{1}{\sqrt{2 \pi (t-t_0)}} e^{ \mu (\vert x_0 \vert + \vert x \vert)- \frac{\mu^2}{2}(t-t_0) - \frac{(x-x_0)^2}{2(t-t_0)} } + \frac{\mu}{2} \erfc \left( \frac{\vert x_0\vert + \vert x \vert- \mu(t-t_0)}{\sqrt{2 (t-t_0)}}  \right)  \right] \nonumber
\label{propagatorinversessgnintegrated}
\end{eqnarray}
allows to recover the free propagator of the Brownian motion with alternating drift \cite{us_DoobKilling}
as it should.


\subsection{  Probability $\Pi (A,t  \vert x_0,A_0,t_0) $ to see the local time $A$ at time $t$ }

The probability $\Pi(A,t \vert x_0,A_0,t_0)  $ of Eq. \ref{propagAalone}
can be obtained via the integration of the joint propagator $ P(  x,A,t \vert   x_0,A_0,t_0) $ of Eq. \ref{propagatorinversessgn}
over the final position $x$
\begin{eqnarray}
 \Pi(A,t \vert x_0,A_0,t_0) \equiv \int_{-\infty}^{+\infty} dx  P(A,t \vert x_0,A_0,t_0)
\label{propagAalonesgn}
\end{eqnarray}

Its singular contribution of Eq. \ref{hatsintegratesingularinv} involves 
the survival probability $S^{abs} ( t \vert x_0, t_0)   $ of Eq. \ref{survivalsgn}
with the asymptotic behavior of Eq. \ref{survivalsgnlarget}
\begin{eqnarray}
 \Pi^{Singular} (A ,t \vert x_0,A_0,t_0) && =  \delta(A-A_0) S^{abs} ( t \vert x_0, t_0)  
 \nonumber \\
 && 
  \opsimeq_{(t-t_0) \to + \infty}  \delta(A-A_0)
\sqrt{\frac{2}{\pi} } \frac{  \vert   x_0 \vert e^{ \mu \vert x_0 \vert- \frac{\mu^2}{2}  (t-t_0)  }} { \mu^2   (t-t_0)^{\frac{3}{2}}}  
\label{hatsintegratesingularinvsgn}
\end{eqnarray}

Its regular contribution can be obtained from the integration over $x$ of Eq. \ref{propagatorRegularinversessgn}
\begin{eqnarray}
&& \Pi^{Regular}(A,t \vert x_0,A_0,t_0) \equiv \int_{-\infty}^{+\infty} dx  P^{Regular}(A,t \vert x_0,A_0,t_0)
\nonumber \\
&&  =  \theta(A>A_0) 
 \frac{ e^{- \frac{\mu^2}{2}  (t-t_0)+  \mu (\vert x_0 \vert  +A-A_0)  }}{ \sqrt{2 \pi }(t-t_0)^{\frac{3}{2}}}  
 \int_{-\infty}^{+\infty} dx
 e^{ - \mu \vert x \vert }
\left( \vert x \vert +\vert x_0 \vert  + A-A_0   \right)  
e^{- \frac{(\vert x \vert +\vert x_0 \vert  + A-A_0  )^2}{2(t-t_0)} } 
\nonumber \\
&&   =  \theta(A>A_0) 
 \frac{ e^{- \frac{\mu^2}{2}  (t-t_0)+  \mu (\vert x_0 \vert  +A-A_0)  }}{ \sqrt{2 \pi (t-t_0)} }  
 2 \int_{0}^{+\infty} dx
 e^{ - \mu x }
\frac{ \left( x +\vert x_0 \vert  + A-A_0   \right)  }{t-t_0}
e^{- \frac{(x +\vert x_0 \vert  + A-A_0  )^2}{2(t-t_0)} } 
\nonumber \\
 && = -  \theta(A>A_0) \sqrt{\frac{2}{\pi (t-t_0) } }
  e^{- \frac{\mu^2}{2}  (t-t_0)+  \mu (\vert x_0 \vert  +A-A_0)  }
 \left(  \left[ 
  e^{ - \mu x }  e^{- \frac{(x +\vert x_0 \vert  + A-A_0  )^2}{2(t-t_0)} } \right]_{x=0}^{x=+\infty}
 + \mu \int_{0}^{+\infty} dx e^{ - \mu x- \frac{(x +\vert x_0 \vert  + A-A_0  )^2}{2(t-t_0)} }
 \right)
 \nonumber \\
 && =   \theta(A>A_0) \sqrt{\frac{2}{\pi (t-t_0) } }
  e^{- \frac{\mu^2}{2}  (t-t_0)+  \mu (\vert x_0 \vert  +A-A_0)  }
 \left(  
  e^{- \frac{(\vert x_0 \vert  + A-A_0  )^2}{2(t-t_0)} } 
 - \mu e^{- \frac{(\vert x_0 \vert  + A-A_0  )^2}{2(t-t_0)}}
 \int_{0}^{+\infty} dx e^{ - \frac{x^2}{2(t-t_0)} 
 -   x \left[ \mu+\frac{ (\vert x_0 \vert  + A-A_0  )}{(t-t_0)}\right]}
 \right)
 \nonumber \\
 && =   \theta(A>A_0) \sqrt{\frac{2}{\pi (t-t_0) } }
  e^{- \frac{(\vert x_0 \vert  + A-A_0 - \mu (t-t_0) )^2}{2(t-t_0)} }
 \left(  
 1
 - \mu 
 \int_{0}^{+\infty} dx e^{ - \frac{x^2}{2(t-t_0)} 
 -   x \left[ \mu+\frac{ (\vert x_0 \vert  + A-A_0  )}{(t-t_0)}\right]}
 \right)
 \nonumber \\
  && =   \theta(A>A_0) e^{ - \frac{(\vert x_0 \vert + A-A_0 - \mu (t-t_0) )^2}{2(t-t_0)} } \left[\sqrt{\frac{2}{\pi (t-t_0) } } - \mu e^{ \frac{(\vert x_0 \vert + A-A_0 + \mu (t-t_0) )^2}{2(t-t_0)} }\erfc \left( \frac{\vert x_0\vert+\mu(t-t_0) + A-A_0 }{\sqrt{2 (t-t_0)}}  \right) \right]
\label{propagAalonesgnreg}
\end{eqnarray}

For large time interval $(t-t_0)$,
the leading behavior is given by
\begin{eqnarray}
 && \Pi^{Regular}(A,t \vert x_0,A_0,t_0)
  \opsimeq_{(t-t_0) \to + \infty}
   \theta(A>A_0) \sqrt{\frac{2}{\pi (t-t_0) } }
  e^{- \frac{(\vert x_0 \vert  + A-A_0 - \mu (t-t_0) )^2}{2(t-t_0)} }
 \left(  1 - \mu 
 \int_{0}^{+\infty} dx e^{ -   x \left[ \mu+\frac{ (\vert x_0 \vert  + A-A_0  )}{(t-t_0)}\right]}
 \right)
 \nonumber \\
 && \opsimeq_{(t-t_0) \to + \infty} 
   \theta(A>A_0) \sqrt{\frac{2}{\pi (t-t_0) } }
  e^{- \frac{(\vert x_0 \vert  + A-A_0 - \mu (t-t_0) )^2}{2(t-t_0)} }
 \left(  1 - \frac{ \mu }{ \mu+\frac{ (\vert x_0 \vert  + A-A_0  )}{(t-t_0)} }
 \right)
 \nonumber \\
 && \opsimeq_{(t-t_0) \to + \infty} 
   \theta(A>A_0) \sqrt{\frac{2}{\pi (t-t_0) } }
  e^{- \frac{(\vert x_0 \vert  + A-A_0 - \mu (t-t_0) )^2}{2(t-t_0)} }
 \left(  \frac{  \vert x_0 \vert  + A-A_0   }{ \mu(t-t_0) + \vert x_0 \vert  + A-A_0  }
 \right)
 \label{propagAalonesgnregtlarge}
 \end{eqnarray}
Note that for $\mu=0$, we recover the expression $\Pi^{Regular}(A,t \vert x_0,A_0,t_0)$ of the standard Brownian motion Eq. \ref{hatsintegrateregularfreemuzerosinv}, as expected.


\subsection{ Large deviations properties of the intensive local time $a= \frac{A-A_0}{t-t_0} \in [0,+\infty[$ }

The probability to see $A=A_0+(t-t_0)a$ in Eq. \ref{propagatorinversessgn} reads
\begin{eqnarray}
&&  P(  x,A=A_0+(t-t_0)a,t \vert   x_0,A_0,t_0)
  = 
 \delta((t-t_0)a)
e^{ \mu (\vert x_0 \vert - \vert x \vert) }
 \frac{e^{- \frac{\mu^2}{2}  (t-t_0) }  } { \sqrt{2 \pi (t-t_0)}} 
\left[ e^{- \frac{ ( x-x_0 )^2}{2(t-t_0)} } 
- e^{- \frac{(\vert x \vert +  \vert x_0 \vert)^2}{2(t-t_0)} } 
\right]
\nonumber \\
&& + \theta(a>0) 
 e^{ \mu (\vert x_0 \vert - \vert x \vert +(t-t_0)a) - \frac{\mu^2}{2}  (t-t_0) }
\left( \frac{\vert x_0 \vert +\vert x \vert  + (t-t_0)a  }{ \sqrt{2 \pi }(t-t_0)^{\frac{3}{2}}} \right)  
e^{- \frac{(\vert x_0 \vert +\vert x \vert  + (t-t_0)a  )^2}{2(t-t_0)} } 
\label{propagatorinversessgnintensive}
\end{eqnarray}
The large deviations of the intensive local time $a= \frac{A-A_0}{t-t_0} \in [0,+\infty[ $
 \begin{eqnarray}
 P(  x,A=A_0+(t-t_0)a,t \vert   x_0,A_0,t_0) \oppropto_{(t-t_0) \to +\infty} e^{- (t-t_0) I ( a )}
\label{Plevel1sgn}
\end{eqnarray} 
thus involve the rate function \cite{occupationsinai}
 \begin{eqnarray}
 I ( a ) =  -  \mu a +  \frac{\mu^2}{2}  + \frac{a^2}{2} = \frac{ (a- \mu)^2 }{2} \ \ \ {\rm for } \ \ a \in [0,+\infty[
\label{Iasgn}
\end{eqnarray} 
 The boundary value $ I ( a=0 ) $ at the origin $a=0$ 
 \begin{eqnarray}
 I ( a=0 ) =  \frac{\mu^2}{2}   
\label{Iasgn0}
\end{eqnarray} 
governs the decay of the survival probability of Eq. \ref{sabsI0}.

The equilibrium value $a_{eq}$ of Eq. \ref{iaeqvanish}
where the rate function $I(a)$ of Eq. \ref{Iasgn}
and its first derivative $I'(a)$ vanish (Eq. \ref{iaeqvanish}) is simply $a_{eq}=\mu$.
It coincides with the value $G_{eq}(0)$ of the equilibrium distribution of $G_{eq}(x)$ 
of Eq. \ref{boltzmannsgn}
at the origin $x=0$
as it should
\begin{eqnarray}
 a_{eq} = \mu = G_{eq}(  0) 
\label{g1sgn}
\end{eqnarray}

If one includes the prefactors, the leading order of the regular contribution of Eq. \ref{propagatorinversessgnintensive} reads
\begin{eqnarray}
  P^{Regular}(  x,A=A_0+(t-t_0)a,t \vert   x_0,A_0,t_0)
  \opsimeq_{(t-t_0) \to +\infty}
 \frac{a  }{ \sqrt{2 \pi (t-t_0) }} 
 e^{- (\mu+a) \vert x \vert  + (\mu-a)  \vert x_0 \vert   } 
e^{- (t-t_0) I ( a )}
\label{propagatorinversessgnintensiveReg}
\end{eqnarray}
The agreement with the general formula of Eq. \ref{laplaceinverseasaddleP}
can be checked using Eqs \ref{laplacesgnxeq0},
\ref{laplacesgnx0eq0},
 \ref{laplacesgn00}
as well as Eq. \ref{saddlesa}
\begin{eqnarray}
  0  =    a   \partial_s \left[  \sqrt{\mu^2+2 s }-  \mu     \right] -1 = \frac{a }{\sqrt{\mu^2+2 s }} -1
\label{saddlesasgn}
\end{eqnarray}
that leads to the saddle-point
\begin{eqnarray}
  s_a  =  \frac{a^2-\mu^2}{2}
\label{saddlesasgnsol}
\end{eqnarray}


\subsection{  Conditioning towards the position $x_T^*$ and the local time $A_T^*$ at the finite time horizon $T$}

\label{subsec_bridgexasgn}

Let us now apply the framework described in the subsection \ref{subsec_bridgexa}.
Using the explicit joint propagator of Eq. \ref{propagatorinversessgn}
\begin{eqnarray}
 && \ln  P(  x_T,A_T,T \vert   x,A,t) 
 = \mu (\vert x \vert - \vert x_T \vert) - \frac{\mu^2}{2}  (T-t) - \ln (\sqrt{2 \pi (T-t)} ) 
 \nonumber \\
 &&
 + \ln \left( \delta(A_T-A)
\left[ e^{- \frac{ ( x_T-x )^2}{2(T-t)} } 
- e^{- \frac{(\vert x_T \vert +  \vert x \vert)^2}{2(T-t)} } 
\right]
 + \theta(A_T>A) 
\left( \frac{\vert x \vert +\vert x_T \vert  + A_T-A  }{ T-t} \right)  
e^{ \mu (A_T-A) - \frac{(\vert x \vert +\vert x_T \vert  + A_T-A  )^2}{2(T-t)} } 
 \right)
  \nonumber \\
 \label{propagatorendsgn}
\end{eqnarray}
one obtains the conditioned drift of Eq. \ref{mustarbridge}
\begin{eqnarray} 
&&  \mu^{[x_T^*,A_T^*]}_T( x,A , t )  = - \mu \, {\rm sgn}(x)  +  \partial_x    \ln P( x_T^*,A_T^*,T \vert  x,A,t)
  \nonumber \\
  && =  \partial_x \ln \left( \delta(A_T^*-A)
\left[ e^{- \frac{ ( x_T^*-x )^2}{2(T-t)} } 
- e^{- \frac{(\vert x_T^* \vert +  \vert x \vert)^2}{2(T-t)} } 
\right]
 + \theta(A_T^*>A) 
\left( \frac{\vert x \vert +\vert x_T^* \vert  + A_T^*-A  }{ T-t} \right)  
e^{ \mu (A_T^*-A) - \frac{(\vert x \vert +\vert x_T^* \vert  + A_T^*-A  )^2}{2(T-t)} } 
 \right)
   \nonumber \\
\label{mustarbridgederivative}
\end{eqnarray}
It actually coincides with the conditioned drift of Eq. \ref{mustarbridgebrown}
with its two regions of Eqs \ref{mustarbridgebrownbelow} and
\ref{mustarbridgebrownabove} 
\begin{eqnarray}
  \mu^{[x_T^*,A_T^*]}_T( x,A<A_T^* , t ) &&  = 
  {\rm sgn}(x)
\left[ \frac{1}{\vert x_T^* \vert +\vert x \vert +A_T^*-A} -
 \frac{\vert x_T^* \vert +\vert x \vert +A_T^*-A}{T-t} 
\right] 
\nonumber \\
  \mu^{[x_T^*,A_T^*]}_T( x , A=A_T^*, t ) &&  = 
  \frac{ \frac{x_T^*-x}{T-t} e^{- \frac{(x_T^*-x)^2}{2(T-t)} }
  + \frac{x + {\rm sgn}(x) \vert x_T^* \vert}{T-t} e^{ - \frac{(\vert x_T^* \vert +\vert x \vert )^2}{2(T-t)} }    
      }
{ e^{- \frac{(x_T^*-x)^2}{2(T-t)} }- e^{ - \frac{(\vert x_T^* \vert +\vert x \vert )^2}{2(T-t)} }    
}
\label{mustarbridgesgn}
\end{eqnarray}
Corresponding stochastic trajectories have already been shown on figure \ref{fig2}.


\subsection{  Conditioning towards the local time $A_T^*$ at the finite time horizon $T$}

\label{subsec_bridgeasgn}

Let us now apply the framework described in the subsection \ref{subsec_bridgea}.

(i) In the region $A_0=0 \leq A<A_T^*$ where the local time $A$ has not yet reached its conditioned final value $A_T^*$, the conditioned drift of Eq. \ref{mustarbridgepibelow} involves the regular contribution
$\Pi^{Regular}( A_T^*,T \vert  x,A,t) $ of Eq. \ref{propagAalonesgnreg}
\begin{eqnarray}
&&  \mu^{[A_T^*]}_T( x,A<A_T^* , t )   =  - \mu \, {\rm sgn}(x) +  \partial_x    \ln \Pi^{Regular}( A_T^*,T \vert  x,A,t)
\nonumber \\
&&  =  - \mu \, {\rm sgn}(x) +  \partial_x    \ln 
\left[
e^{ - \frac{(\vert x \vert + A_T^*-A - \mu (T-t) )^2}{2(T-t)} } \left( \sqrt{\frac{2}{\pi (T-t) } } - \mu e^{ \frac{(\vert x \vert + A_T^*-A + \mu (T-t) )^2}{2(T-t)} }\erfc \left( \frac{\vert x \vert+\mu(T-t) + A_T^*-A }{\sqrt{2 (T-t)}}  \right) \right)
\right]
\nonumber \\
&&  = - {\rm sgn}(x) \frac{\vert x \vert  + A_T^*-A }{T-t} 
 +  \partial_x    \ln 
\left(  
1
- \mu 
\int_{0}^{+\infty} dy e^{ - \frac{y^2}{2(T-t)} 
-   y \left[ \mu+\frac{ (\vert x \vert  + A_T^*-A  )}{(T-t)}\right]}
\right)
\nonumber \\
&&  =   \mu \, {\rm sgn}(x) - \frac{2}{(T-t)}\frac{(\vert x \vert  + A_T^*-A  )}{2 - \mu \sqrt{2 \pi (T-t)} e^{ \frac{(\vert x \vert  + \mu (T-t) + A_T^*-A)^2}{2(T-t)} } \erfc \left( \frac{\vert x \vert +\mu(T-t) + A_T^*-A }{\sqrt{2 (T-t)}}  \right)  } {\rm sgn}(x) 
\label{mustarbridgepibangbangbelow}
\end{eqnarray}

(ii) In the region $A=A_T^*$ where the local time $A$ has already reached its conditioned final value $A_T^*$,
and where the position $x$ cannot visit the origin $x=0$ anymore,
the conditioned drift of Eq. \ref{mustarbridgepiabove} involves the survival probability $S^{abs}(T \vert x,t)  $ 
of Eq. \ref{survivalsgn} 
\begin{eqnarray}
\label{mustarbridgepibangbangabove}
 \mu^{[A_T^*]}_T( x , A=A_T^*, t ) &&  =  - \mu \, {\rm sgn}(x)+  \partial_x    \ln S^{abs}(T \vert x,t) 
\nonumber \\
&& = - \mu \, {\rm sgn}(x)+  \partial_x    \ln \left(
 \frac{e^{- \frac{\mu^2}{2}  (T-t) + \mu \vert x \vert - \frac{ x^2}{2(T-t)}
}  } { \sqrt{2 \pi (T-t)}} 
\int_{-\infty}^{+\infty} dy
 e^{ - \mu  \vert y \vert - \frac{ y^2}{2(T-t)}}
\left[ e^{ \frac{ y x}{(T-t)} } 
- e^{- \frac{\vert y  x \vert}{(T-t)} } 
\right] \right)
\nonumber \\
&& =   \partial_x    \ln \left(
\int_{-\infty}^{+\infty} dy
 e^{ - \mu  \vert y \vert - \frac{ y^2}{2(T-t)}}
\left[ e^{ \frac{ y x}{(T-t)} } 
- e^{- \frac{\vert y  x \vert}{(T-t)} } 
\right] \right)
  \nonumber \\
 && 
= 2 \sqrt{\frac{2}{\tau}} 
\frac{
\, {\rm sgn}(x) \left( \frac{1}{\sqrt{\pi }}-  e^{\frac{(\vert x \vert + \mu \tau)^2}{2\tau}} \mathcal{F} \left( \frac{\vert x \vert + \mu \tau}{\sqrt{2 \tau}}  \right) \right) +  e^{\frac{(x - \mu \tau)^2}{2\tau}} \left(e^{2 \mu x} \mathcal{F} \left( \frac{ x + \mu \tau}{\sqrt{2 \tau}}  \right)   -\mathcal{F} \left( \frac{ \mu \tau -x}{\sqrt{2 \tau}}  \right)   \right) 
     }
     {
   e^{\frac{(x + \mu \tau)^2}{2 \tau} } \erfc \left( \frac{x + \mu \tau}{\sqrt{2 \tau}}  \right) 
  + e^{\frac{(x - \mu \tau)^2}{2 \tau} } \erfc \left( \frac{\mu \tau - x }{\sqrt{2 \tau}}  \right)
  -2 e^{\frac{(\vert x \vert + \mu \tau)^2}{2 \tau}} \erfc \left( \frac{\vert x \vert + \mu \tau }{\sqrt{2 \tau}} 
   \right)
     } 
\end{eqnarray} 
 where $\mathcal{F}(x) = x \erfc(x)$ and $\tau = T-t$.

In the second region, the asymptotic behavior near the origin $x \to 0$ is
\begin{eqnarray}
   \mu^{[A_T^*]}_T( x, A=A_T^*, t ) \opsimeq_{x \to 0} \frac{1}{x} + x \left( \frac{1}{T-t} + \frac {\mu^2}{3} + \frac{1}{-3 (T-t) + 3 \mu \sqrt{\frac{\pi (T-t)^3}{2} } e^{\frac{(T-t) \mu^2}{2}} \erfc \left( \mu \sqrt{\frac{T-t}{2}}  \right)   
}  \right)
\label{mustarbridgepibangbangabovexto0}
\end{eqnarray}
Again, the $1/x$ term prevents the process from crossing the origin and the local time cannot increase anymore.

\begin{figure}[h]
\centering
\includegraphics[width=5.2in,height=4.5in]{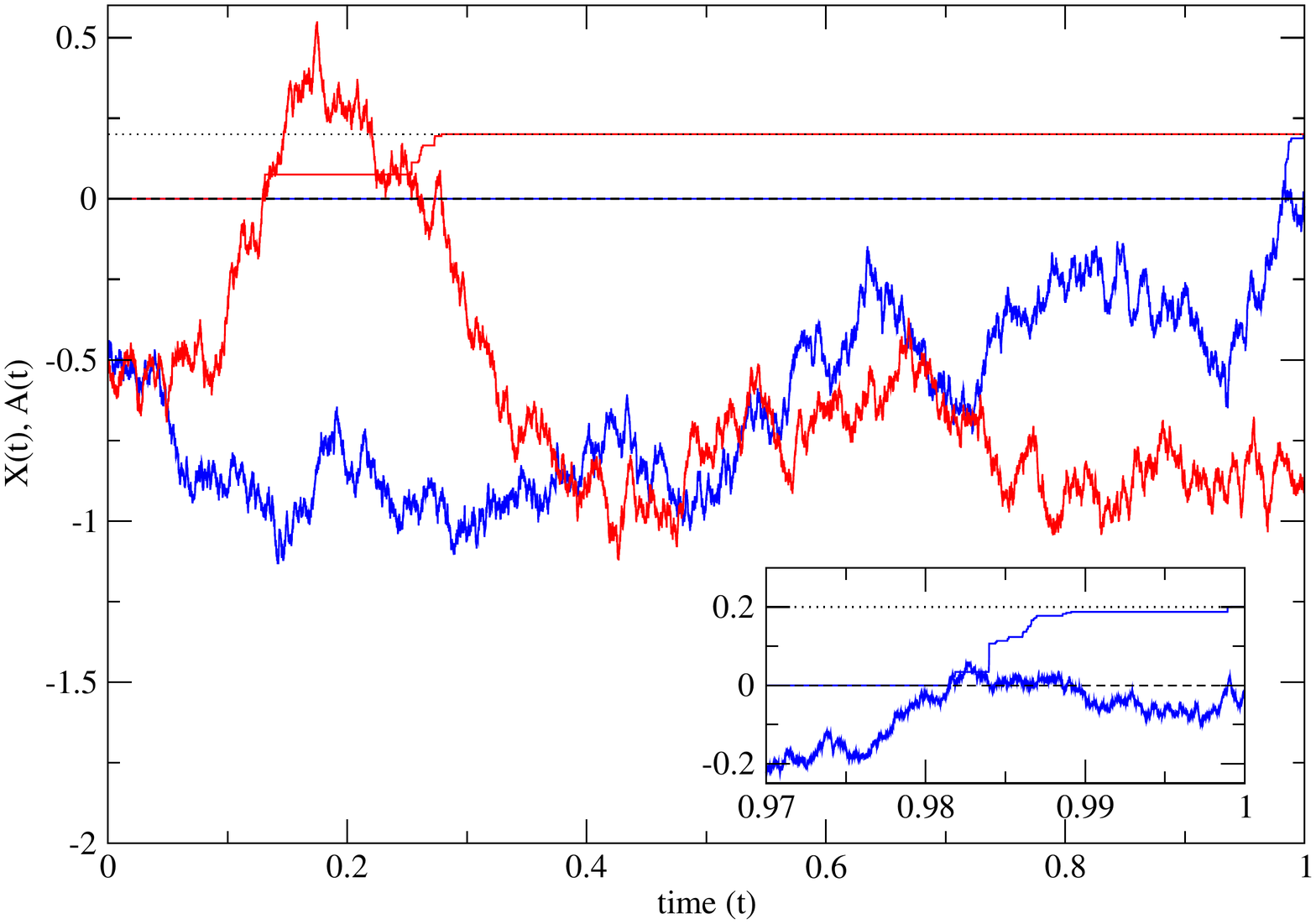}
\setlength{\abovecaptionskip}{15pt}  
\caption{Examples of realization of the Brownian process with alternating drift conditioned to have the final local time $A_T^*= 0.2$ at the finite time horizon $T=1$ (see the conditioned drift of Eqs. \ref{mustarbridgepibangbangbelow} and \ref{mustarbridgepibangbangabove}). For each trajectory, the associated local time $A(t)$ is shown as a function of the time $t \in [0,T]$. The process can end at any final position $x_T$, while the initial position is $x_0=-0.5$ here . Observe that the blue realization reaches the desired local time value only at the very end of the time-window (as shown in the encapsulated plot). The time step used in the discretization is $dt = 10^{-5}$.}
\label{fig5}
\end{figure}


\subsection{ Case $\mu=0$ : Conditioning towards the intensive value $a^*=\frac{A_T^*}{T}$ in the limit $T\to +\infty$ }

\label{subsec_sgnintensive}

A direct consequence of Eq. \ref{mustarbridgesgn} is that 
the conditioning towards the intensive value $a^*=\frac{A_T^*}{T}$ in the limit $T\to +\infty$
will give exactly the same conditioned drift of Eq. \ref{mustarbridgepibelowmu0intensive}
\begin{eqnarray}
\mu_T^{[T a^*]}( x,A<T a^* , t )  
  \opsimeq_{T \to +\infty } - {\rm sgn} (x) a^* \equiv  \mu_{\infty}^{[a^*]}( x )
\label{mustarbridgepibelowsgnintensive}
\end{eqnarray}

The agreement with the general formula of Eq. \ref{mustarbridgepibelowintensiveasympto}
for the conditioned drift $\mu_{\infty}^{[a^*]}( x ) $
can be checked by Eq. \ref{laplacesgnxeq0}
\begin{eqnarray}
{\tilde G}_s (0 \vert x)  
 = \frac{ e^{ (\mu -\sqrt{\mu^2+2 s } )\vert x \vert   } } { \sqrt{\mu^2+2 s }-  \mu}  
\label{laplacesgnxeq0intense}
\end{eqnarray}
and the saddle-point value $s_{a^*}=\frac{(a^*)^2-\mu^2}{2} $ of Eq. \ref{saddlesasgnsol}
to obtain
\begin{eqnarray}
\mu_{Bridge[a^*]}^*( x ) && = - \mu \, {\rm sgn} (x) +   \partial_x    \ln {\hat G}_{s_{a^*}} (0 \vert x)  
=- \mu \,  {\rm sgn} (x) +  \partial_x    \left( (\mu -\sqrt{\mu^2+2 s_{a^*} } )\vert x \vert -\ln( \sqrt{\mu^2+2 s_{a^*} }-  \mu)  \right) 
\nonumber \\
&& = - \sqrt{\mu^2+2 s_{a^*} }{\rm sgn} (x)
= -  {\rm sgn} (x) a^*
\label{mustarbridgepibelowintensiveasymptosgn}
\end{eqnarray}
As explained in the Appendices,
this result can be also recovered via the appropriate canonical conditioning leading to Eq. \ref{mustarpinteriorgssgntrad}.


\section{ Conclusion }

\label{sec_conclusion}

In the present paper, we have analyzed the conditioning of a diffusion process $X(t)$ of drift $\mu(x)$
and of diffusion coefficient $D=1/2$ with respect to its local time $A_{x=0}(t) =A(t)$ at the origin $x=0$.
Our goal was to construct various conditioned joint processes $[X^*(t),A^*(t)] $
satisfying certain conditions involving the local time $A^*(T)$ at the finite time horizon $T$
or in the limit of the infinite time horizon $T \to +\infty$. \\
For the case of the finite time horizon $T$,
we have studied :

(a) the conditioning towards the final position $X^*(T)$ and towards the final local time $A^*(T)$. In other words, this case corresponds to conditioning a generalized Brownian bridge with respect to its local time at the final time.

(b) the conditioning towards the final local time $A^*(T)$ alone 
without any condition on the final position $X^*(T)$. \\
In the limit of the infinite time horizon $T \to +\infty$, we have analyzed :

(1) the conditioning towards the finite asymptotic local time $A_{\infty}^*<+\infty$

(2)  the conditioning towards the intensive local time $a^* $ corresponding to the extensive behavior $A_T \simeq T a^*$, that we have compared in the Appendices to the appropriate 'canonical conditioning' based on the generating function of the local time in the regime of large deviations. \\
This general construction has been applied to generate various constrained stochastic trajectories for three unconditioned diffusions with different recurrence/transience properties : 

(i) as simplest example of a transient diffusion, we have considered the uniform strictly positive drift $\mu(x)=\mu>0$

(ii) as simplest example of a diffusion converging towards an equilibrium, we have chosen the drift $\mu(x)=- \mu \, {\rm sgn}( x)$ of parameter $\mu>0$
 
(iii) as simplest example of a recurrent diffusion that does not converge towards an equilibrium, we have focused on the Brownian motion without drift $\mu=0$.

The generalization of the present work to analyze the conditioning with respect to two local times
is described in \cite{us_TwoLocalTimes}.


\appendix

\section{ Notion of canonical conditioned process $X_p^*(t)$ of parameter $p$ conjugated to the local time}

\label{app_canonical}

As recalled in the Introduction, the 'canonical conditioning' based on generating functions of time-additive observables for Markov processes over a large time-window $T$ has recently been used extensively in the field of non-equilibrium statistical physics \cite{peliti,derrida-lecture,tailleur,sollich_review,lazarescu_companion,lazarescu_generic,jack_review,vivien_thesis,lecomte_chaotic,lecomte_thermo,lecomte_formalism,lecomte_glass,kristina1,kristina2,jack_ensemble,simon1,simon2,simon3,Gunter1,Gunter2,Gunter3,Gunter4,chetrite_canonical,chetrite_conditioned,chetrite_optimal,chetrite_HDR,touchette_circle,touchette_langevin,touchette_occ,touchette_occupation,derrida-conditioned,derrida-ring,bertin-conditioned,garrahan_lecture,Vivo,chemical,touchette-reflected,touchette-reflectedbis,c_lyapunov,previousquantum2.5doob,quantum2.5doob,quantum2.5dooblong,c_ruelle,lapolla,chabane,chabane_thesis}. Its physical meaning comes 
from the equivalence at the level of the large deviations for large time $T$
 between the 'canonical conditioning'
and the 'microcanonical conditioning' (see the two detailed papers \cite{chetrite_conditioned,chetrite_optimal}
and the HDR thesis \cite{chetrite_HDR} with references therein).
In this Appendix, it is thus interesting to analyze the 'canonical conditioning' 
 of parameter $p$ conjugated to the local time increment $[A(t)-A(t_0)]$
 and to compare with the 'microcanonical conditioning' described in the main text.

\subsection{ Canonical conditioned process $X_p^*(t)$ of parameter $p$ based on the Laplace transform $ {\tilde P}_{p} (x,t  \vert x_0,t_0) $ }

The canonical conditioning is based 
on the Laplace transform $ {\tilde P}_{p} (x,t  \vert x_0,t_0) $ of Eq. \ref{laplaceA}
with respect to the local time increment $(A-A_0)$, where the Laplace parameter $p$ 
conjugated to the local time increment $(A-A_0)$ is fixed.

For the bridge conditioned to end at the position $x_T^*$ at the time horizon $T$,
the conditioned probability for the position $x$ at an interior time $t \in [0,T]$ reads
\begin{eqnarray}
P^{[x_T^*;p]}_T(x,t) =  \frac{{\tilde P}_{p}(x_T ,T\vert x,t) {\tilde P}_{p}(x,t \vert x_0,0)}{ {\tilde P}_{p}(x_T ,T\vert x_0,0)}
\label{markovcondk}
\end{eqnarray}
The corresponding Ito dynamics for the conditioned process $X_p^*(t) $ of parameter $p$
\begin{eqnarray}
dX_p^*(t)  =  \mu_p^*( X_p^*(t),t ) dt + dB(t)
\label{Doobp}
\end{eqnarray}
involves the conditioned drift
\begin{eqnarray}
  \mu^{[x_T^*;p]}_T( x,t )  = \mu(x) + \partial_x \ln {\tilde P}_p(x_T ,T\vert x,t)
\label{mustarp}
\end{eqnarray}


\subsection{ Properties of the $p$-deformed propagator $ {\tilde P}_{p} (x,t  \vert x_0,t_0) $ }

The forward dynamics of the propagator $  {\tilde P}_{p} (x,t  \vert x_0,t_0)$
given by 
the Feynman-Kac formula of Eq. \ref{feynmankac}
\begin{eqnarray}
\partial_t  {\tilde P}_{p} (x,t  \vert x_0,t_0)
= -  p \delta(x)  {\tilde P}_{p} (x,t  \vert x_0,t_0)
 -    \partial_x \left[ \mu(x)  {\tilde P}_{p} (x,t  \vert x_0,t_0) \right] +  \frac{1}{2} \partial_x^2   {\tilde P}_{p} (x,t  \vert x_0,t_0) \equiv {\cal F}_p {\tilde P}_{p} (x,t  \vert x_0,t_0)
\label{forward}
\end{eqnarray}
involves the generator 
\begin{eqnarray}
{\cal F}_p \equiv -  p \delta(x)   -    \partial_x  \mu(x)  +  \frac{1}{2} \partial_x^2  
 \label{generator}
\end{eqnarray}
Its adjoint
\begin{eqnarray}
{\cal F}_p^{\dagger}  = -  p \delta(x)   + \mu(x)     \partial_x  +  \frac{1}{2} \partial_x^2  
 \label{adjoint}
\end{eqnarray}
governs the backward dynamics of the propagator ${\tilde P}_{p}(x_T ,T\vert x,t)$
\begin{eqnarray}
\partial_t  {\tilde P}_{p}(x_T ,T\vert x,t)
=  {\cal F}_p^{\dagger} {\tilde P}_{p}(x_T ,T\vert x,t)
=  -  p \delta(x) {\tilde P}_{p}(x_T ,T\vert x,t)  
+ \mu(x)     \partial_x  {\tilde P}_{p}(x_T ,T\vert x,t)
+  \frac{1}{2} \partial_x^2  {\tilde P}_{p}(x_T ,T\vert x,t)
\label{backward}
\end{eqnarray}

\subsubsection{Physical meaning of the $p$-deformed Fokker-Planck dynamics} 

With respect to the dynamics of Eq. \ref{forward1d}
corresponding to $p=0$, 
the additional term in the forward dynamics of Eq. \ref{forward}
corresponds to the killing rate of amplitude $p>0$ localized at the origin $x=0$
\begin{eqnarray}
k(x)= p \delta(x) 
\label{killingrate}
\end{eqnarray}
It is however also interesting to consider the case $p<0$ in the Laplace transform of Eq. \ref{laplaceA} : 
then the additional term in the Feynman-Kac formula of Eq. \ref{feynmankac}
corresponds instead to the reproducing rate of amplitude $(-p)>0$ localized at the origin $x=0$
\begin{eqnarray}
r(x)= (-p) \delta(x) 
\label{cloningrate}
\end{eqnarray}


\subsubsection{Physical meaning of the associated $p$-deformed quantum Hamiltonian $H_p$}

Via the similarity transformation analogous to Eq. \ref{defpsi}
that involves the potential $U(x)$ of Eq. \ref{potentialU}
\begin{eqnarray}
{\tilde P}_{p} (x,t  \vert x_0,t_0) =e^{- \frac{  U(x) }{2}} \psi(x,t \vert x_0,t_0)e^{ \frac{  U(x_0) }{2}}
= e^{ \int_{x_0}^{x} dy \mu(y)} \psi_p(x,t \vert x_0,t_0)
\label{defpsip}
\end{eqnarray}
the forward dynamics of Eq. \ref{forward}
for ${\tilde P}_{p} (x,t  \vert x_0,t_0)$ translates into the Euclidean Schr\"odinger Equation for 
$\psi_p(x,t \vert x_0,t_0) $
\begin{eqnarray}
- \partial_t \psi_p (  x,t \vert   x_0,t_0) = H_p \psi_p(  x,t \vert   x_0,t_0) 
\label{schrodingerp}
\end{eqnarray}
With respect to the Hamiltonian $H$ of Eq. \ref{hamiltonian} involving the potential $V(x)$ of Eq. \ref{susy},
the quantum Hamiltonian $H_p$
contains an additional delta potential of amplitude $p$ localized at the origin $ x=0$
\begin{eqnarray}
 H_p = H + p \delta(x) = -  \frac{1}{2} \partial_x^2 +V(x) + p \delta(x) 
 = -  \frac{1}{2} \partial_x^2 + \frac{ \mu^2(x)}{2} + \frac{\mu'(x)}{2}  + p \delta(x) 
\label{hamiltonianp}
\end{eqnarray}
So the case $p>0$ associated to the killing rate of Eq. \ref{killingrate} corresponds to an
additional repulsive delta potential, while the case $p<0$ associated to the reproducing rate of Eq. \ref{cloningrate} corresponds to an additional attractive delta potential.


\subsection{ Canonical conditioning for large horizon $T $
when the Hamiltonian $H_p$ has a normalizable ground-state }


\subsubsection{ Propagator $ {\tilde P}_{p} (x,t  \vert x_0,t_0) $ for large time $(t-t_0)$ when the $p$-deformed Hamiltonian $H_p$ has a normalizable ground-state }

When the $p$-deformed quantum Hamiltonian $H_p$ has 
a normalizable ground-state $\phi_p^{GS}(  x) $ of energy $E_p$
\begin{eqnarray}
 H_p \phi_p^{GS}(  x) =E_p \phi_p^{GS}(  x) 
\label{eigenhamiltonianp}
\end{eqnarray}
the ground state can be chosen real and positive $\phi_p^{GS}(  x) \geq 0$
with the normalization
\begin{eqnarray}
\langle  \phi_p^{GS} \vert  \phi_p^{GS} \rangle = \int_{-\infty}^{+\infty} dx \left[  \phi_p^{GS}(  x) \right]^2=1 
\label{quantumnorma}
\end{eqnarray}
This ground-state $\phi_p^{GS}(  x) $ and its energy $E_p$
determine the leading asymptotic behavior
of the quantum propagator 
\begin{eqnarray}
 \psi_p (  x,t \vert   x_0,t_0) \opsimeq_{(t-t_0) \to + \infty}  e^{- (t-t_0) E_p}
 \phi_p^{GS}(  x) \phi_p^{GS}(  x_0) 
\label{psipGS}
\end{eqnarray}
The corresponding asymptotic behavior of the propagator ${\tilde P}_{p} (x,t  \vert x_0,t_0) $ given by 
the similarity transformation of Eq. \ref{defpsip} reads
\begin{eqnarray}
{\tilde P}_{p} (x,t  \vert x_0,t_0) && =e^{- \frac{  U(x) }{2}} \psi_p(x,t \vert x_0,t_0)e^{ \frac{  U(x_0) }{2}}
\nonumber \\
&& \opsimeq_{(t-t_0) \to + \infty}  e^{- (t-t_0) E_p}
\left[ e^{- \frac{  U(x) }{2}} \phi_p^{GS}(  x) \right] \left[e^{ \frac{  U(x_0) }{2}} \phi_p^{GS}(  x_0) \right]
\equiv e^{- (t-t_0) E_p} r_p(x) l_p(x_0)
\label{ppGS}
\end{eqnarray}
where
\begin{eqnarray}
  r_p(x) && \equiv e^{- \frac{  U(x) }{2}} \phi_p^{GS}(  x) 
  = e^{ \int_{0}^{x} dy \mu(y)} \phi_p^{GS}(  x)
\nonumber \\
  l_p(x_0) && \equiv e^{ \frac{  U(x_0) }{2}} \phi_p^{GS}(  x_0) 
  = e^{ - \int_{0}^{x_0} dy \mu(y)} \phi_p^{GS}(  x_0)
\label{rightleft}
\end{eqnarray}
correspond to the positive right and left eigenvectors of the generator of Eq. \ref{generator}
associated to the eigenvalue $(-E_p)$
\begin{eqnarray}
- E_p r_p(x) && = {\cal F}_p r_p(x)  = -  p \delta(x) r_p(x)   -    \partial_x \left[ \mu(x) r_p(x) 
\right] +  \frac{1}{2} \partial_x^2  r_p(x) 
\nonumber \\
- E_p l_p(x) && = {\cal F}_p^{\dagger} l_p(x) = -  p \delta(x) l_p(x)  + \mu(x)     \partial_x l_p(x) +  \frac{1}{2} \partial_x^2  l_p(x)
 \label{eigen}
\end{eqnarray}
The normalization inherited from Eq. \ref{quantumnorma} reads
\begin{eqnarray}
\langle l_p \vert r_p \rangle = \int_{-\infty}^{+\infty} dx r_k(x) l_k(x) = \int_{-\infty}^{+\infty} dx \left[  \phi_p^{GS}(  x) \right]^2=1 
\label{FPnorma}
\end{eqnarray}

For the double Laplace transform $ {\hat {\tilde P}}_{s,p} (x \vert x_0)$
of Eq. \ref{laplacedouble}, the asymptotic behavior of Eq. \ref{ppGS} for large $(t-t_0)$
means that  $ {\hat {\tilde P}}_{s,p} (x \vert x_0)$ exists for $s \in ]-E_p,+\infty[$
with the following pole singularity for $s \to (-E_p)^+ $
\begin{eqnarray}
{\hat {\tilde P}}_{s,p} (x \vert x_0) 
\opsimeq_{s \to (-E_p)^+}
 \int_{t_0}^{+\infty} dt e^{- (s+E_p) (t-t_0) }  r_p(x) l_p(x_0)
= \frac{ r_p(x) l_p(x_0) } {s+E_p } 
\label{laplacedoublepole}
\end{eqnarray}


\subsubsection{ Simplifications for the canonical conditioning for large horizon $T $ when $H_p$ has a normalizable ground-state} 

\label{sec_canonicalGS}

When the $p$-deformed quantum Hamiltonian $H_p$ has 
a normalizable ground-state $\phi_p^{GS}(  x) $,
the asymptotic behavior of Eq. \ref{ppGS}
can be plugged into the three propagators of Eq. \ref{markovcondk}
to obtain that the conditioned density at any interior time $0 \ll t \ll T$
\begin{eqnarray}
P^{[x_T^*;p]}_T(x,t) \opsimeq_{ 0 \ll t \ll T}  
 \frac{ e^{-E_p (T-t) } r_p(x_T) l_p(x) e^{-E_pt} r_p(x) l_p(x_0)}{e^{-E_pT} r_p(x_T) l_p(x_0) }
 = l_p(x) r_p(x) \equiv P^*_p(x) 
\label{pkinterior}
\end{eqnarray}
does not depend on the interior time $t$ anymore.
This steady conditioned density $P^*_p(x) $ only involves the product 
of the left and right eigenvectors of Eq. \ref{rightleft}
and can be thus rewritten as the square of the ground-state $\phi_p^{GS}(  x) $ of the quantum Hamiltonian $H_p$ alone
\begin{eqnarray}
P^*_p(x) =  l_p(x)  r_p(x) = \left[ \phi_p^{GS}(  x) \right]^2
\label{pkgs}
\end{eqnarray}

The corresponding conditioned drift of Eq. \ref{mustarp} is also independent of the interior time $t$
\begin{eqnarray}
  \mu^{[x_T^*;p]}_T( x,t )  \opsimeq_{ 0 \ll t \ll T} 
    \mu(x) + \partial_x \ln \left[ e^{-E_p (T-t) } r_p(x_T) l_p(x)\right]
    =  \mu(x) + \partial_x \ln \left[  l_p(x)\right] \equiv  \mu_p^*( x )
\label{mustarpinterior}
\end{eqnarray}
Since $\mu_p^*( x ) $ involves the initial drift $ \mu(x)$ 
and the logarithmic derivative of the left eigenvector $ l_p(x) $
of Eq. \ref{rightleft}, it 
can be rewritten in terms of the logarithmic derivative of the ground-state $\phi_p^{GS}(  x) $
of the quantum Hamiltonian $H_p$ alone
\begin{eqnarray}
  \mu_p^*( x )    =  \mu(x) + \partial_x \ln \left[  e^{ - \int_{0}^{x} dy \mu(y)} \phi_p^{GS}(  x)\right] 
  = \partial_x \ln \left[  \phi_p^{GS}(  x)\right] 
\label{mustarpinteriorgs}
\end{eqnarray}


\subsubsection{ Conclusion }

In summary, when the $p$-deformed quantum Hamiltonian $H_p$ has 
a normalizable ground-state $\phi_p^{GS}(  x) $, 
then the canonical conditioned process $X_p^*(t)$ 
becomes simple for large time horizon $T\to +\infty$ in the region of interior times $0 \ll t \ll T$: 
its steady density $P^*_p(x) $ of Eq. \ref{pkgs} and the conditioned drift $ \mu_p^*( x ) $ of Eq. \ref{mustarpinteriorgs} are time-independent
and involve only the normalizable ground-state $\phi_p^{GS}(  x) $ of $H_p$.

The physical meaning of this conditioned process $X_p^*(t)$ 
depends on whether $H=H_{p=0}$ has also a normalizable ground-state or not :

(i) the case where $H=H_{p=0}$ has also a normalizable ground-state $\phi_{p=0}^{GS}(  x) $,
i.e. where the unconditioned process $X(t)$ converges towards an equilibrium state $G_{eq}(x)$ 
is discussed in Appendix \ref{app_canonicalEq}.

(ii) the case where $H=H_{p=0}$ has no normalizable ground-state is discussed
 in Appendix \ref{app_canonicalnoEq}.


\section{ Canonical conditioning when the unconditioned process $X(t)$ has an equilibrium state }

\label{app_canonicalEq}

In this Appendix, we consider the case where the unconditioned process $X(t)$ 
converges towards the equilibrium state $G_{eq}(x)$ of Eq. \ref{boltzmann},
so that the quantum Hamiltonian $H$ of Eq. \ref{hamiltonian}
has a normalizable ground-state $ \phi^{GS}(  x) $ given by Eq. \ref{gs}.
Then the ground-state $ \phi_{p}^{GS}(  x)$ of the Hamiltonian $H_p$ of Eq. \ref{hamiltonianp}
 can be interpreted as a deformation of this unperturbed
ground-state $\phi^{GS}(  x) = \phi_{0}^{GS}(  x) $,
with the following consequences for the physical meaning of the canonical conditioning of parameter $p$


\subsection{ Link between  $E_p$ and the rate function $I(a)$ governing the large deviations of the intensive local time $a$}

The ground-state energy $E_p$ of the $p$-deformed Hamiltonian $H_p$
governs the asymptotic behavior of Eq. \ref{ppGS}
of the Laplace transform $ {\tilde P}_{p} (x,T  \vert x_0,0) $ of the local time $A_T$
introduced in Eq. \ref{laplaceA}
\begin{eqnarray}
 {\tilde P}_{p} (x,T  \vert x_0,0) && \equiv 
\int_{0}^{+\infty} dA e^{-p A }P(x,A,T \vert x_0,A_0,0)
\nonumber \\
&& \opsimeq_{T \to + \infty}  e^{- T E_p}
\left[ e^{- \frac{  U(x) }{2}} \phi_p^{GS}(  x) \right] \left[e^{ \frac{  U(x_0) }{2}} \phi_p^{GS}(  x_0) \right]
\equiv e^{- T E_p} r_p(x) l_p(x_0)
\label{laplaceAtlarge}
\end{eqnarray}

The large deviations properties of Eq. \ref{level1def}
can also be used to evaluate 
the generating function of $(A_T-A_0)=T a$
via the saddle-point method for large $T$
\begin{eqnarray}
 \langle e^{-p (A_T-A_0) }  \rangle 
 = \langle e^{-p T a } \rangle =\int_0^{+\infty} da  e^{-p T a} P_T( a ) 
 \opsimeq_{T \to +\infty} 
 \int_0^{+\infty} da e^{ -T \left[ p a + I ( a ) \right] }\opsimeq_{T \to +\infty} e^{ -T E_p }
\label{level1gen}
\end{eqnarray} 
So the energy $E_p$ governing the asymptotic behavior of Eq. \ref{laplaceAtlarge}
for the propagator ${\tilde P}_{p} (x,t  \vert x_0,t_0) $ 
 is the Legendre transform of the rate function $ I ( a ) $
 \begin{eqnarray}
 p a + I ( a ) && =  E_p
 \nonumber \\
 p+ I'(a) && =0
\label{legendre}
\end{eqnarray} 
while the reciprocal Legendre transform reads
 \begin{eqnarray}
I(a) && =E_p-pa 
 \nonumber \\
a && = \frac{dE_p}{dp}
\label{legendrereci}
\end{eqnarray} 
As a consequence, the canonical conditioning of parameter $p$ 
discussed in section \ref{sec_canonicalGS}
can be considered as asymptotically equivalent for large $T$
to the microcanonical conditioning of subsection \ref{subsec_doobintensive}
towards 
the intensive local time $a^*_p=\frac{dE_p}{dp}$ corresponding to the Legendre value of Eq. \ref{legendrereci}.
Note that this relation $a^*_p=\frac{dE_p}{dp} $ has a very simple interpretation via
the first-order perturbation theory for the energy $E_p$ of the ground state $\phi^{GS}_p(x)$
 in quantum mechanics when the parameter $p$ is changed into $(p+\epsilon)$
\begin{eqnarray}
a^*_p && = \frac{dE_p}{dp} = \lim_{\epsilon \to 0} \left(\frac{E_{p+\epsilon}-E_p}{\epsilon} \right)
= \langle \phi^{GS}_p \vert \delta(x) \vert \phi^{GS}_p \rangle = 
\left[\phi^{GS}_p(x=0) \right]^2 = P^*_p(x=0) 
\label{quantumperturbationfirstorder}
\end{eqnarray} 
that corresponds to the conditioned steady state $P^*_p(x) $ of Eq. \ref{pkgs} at the origin $x=0$.


\subsubsection*{ Relation between the Laplace parameter $p$ and
the time-Laplace parameter $s$ in the large deviations analysis     }

The comparison between 

(i) the Legendre transform of Eqs. \ref{legendre}, \ref{legendrereci} between $I(a)$ and $E_p$

(ii) the quasi-Legendre transform of Eqs. \ref{regularlargedevcompatibilityasaddle},
\ref{saddlesa} and \ref{iafromsaddle}
between $I(a)$ and $ \frac{ 1 }{{\hat G}_s (0 \vert 0)} $

allows to eliminate the variable $a$ to obtain the following relations between the 
Laplace parameter $p$ and the time-Laplace parameter $s$ 
 \begin{eqnarray}
  s && = - E_p
  \nonumber \\
  p && = - \frac{ 1 }{{\hat G}_s (0 \vert 0)}
\label{legendreps}
\end{eqnarray} 


\subsection{ Example of the drift $\mu(x)=-\mu \, {\rm sgn}(x)$ with $\mu>0$  }


\subsubsection{ Computation of the energy $E_p$ via the Legendre transform of the rate function $I(a)$   }

For the explicit rate function $I(a)$ of Eq. \ref{Iasgn},
the Legendre transform of Eq. \ref{legendre}  yields the following properties.

(i) The microcanonical conditioning to the intensive local time $a^*$ is 
asymptotically equivalent in the thermodynamic limit $T\to + \infty$
to the canonical conditioning of parameter
 \begin{eqnarray}
  p  = - I'(a^*) = \mu-a^*
\label{legendresgn1}
\end{eqnarray} 
so that the domain $a^* \in [0,+\infty[$ of definition for the intensive local time 
corresponds to the following domain for the Laplace variable $p$
 \begin{eqnarray}
p \in ]-\infty,\mu]
\label{domainp}
\end{eqnarray} 
Reciprocally, the canonical conditioning of parameter $p$ 
is 
asymptotically equivalent in the thermodynamic limit $T\to + \infty$
to the microcanonical conditioning of the intensive local time 
 \begin{eqnarray}
  a^*_p  = \mu-p
\label{apstarsign}
\end{eqnarray} 

(ii) The energy $E_p$ of Eq. \ref{legendre}
 reads using Eq. \ref{Iasgn} and Eq. \ref{legendresgn1}
 \begin{eqnarray}
 E_p && = p a + I ( a ) =  p a + \frac{ (a- \mu)^2 }{2} = p (\mu-p) + \frac{ p^2 }{2} = p \mu - \frac{ p^2 }{2}
\label{legendresgn2}
\end{eqnarray}


\subsubsection{ Direct analysis of the ground-state of the $p$-deformed Hamiltonian $H_p$   }

For the drift $ \mu(x)=-\mu \, {\rm sgn}(x)$ with $\mu>0$, 
the Hamiltonian $H=H^{[\mu]}$ of Eq. \ref{Hsgn} of parameter $\mu>0$
\begin{eqnarray}
H^{[\mu]}  =  -  \frac{1}{2} \partial_x^2 + \frac{ \mu^2}{2}   - \mu \delta(x)
\label{Hsgnbis}
\end{eqnarray}
has the zero-energy normalized ground-state of Eq. \ref{gs} using Eq. \ref{boltzmannsgn}
\begin{eqnarray}
 \phi^{GS[\mu]}(  x) =  \sqrt{G_{eq}(  x) } = \sqrt{\mu} e^{- \mu \vert x \vert }
\label{gssgn}
\end{eqnarray}

The $p$-deformed Hamiltonian $H_p= H_p^{[\mu]}$ of Eq. \ref{hamiltonianp}
\begin{eqnarray}
 H_p^{[\mu]} && = H^{[\mu]} + p \delta(x) =  -  \frac{1}{2} \partial_x^2 + \frac{ \mu^2}{2}   - ( \mu-p)  \delta(x)
 \nonumber \\
 && = -  \frac{1}{2} \partial_x^2 + \frac{ (\mu-p)^2}{2}   - ( \mu-p)  \delta(x) + \frac{\mu^2-(\mu-p)^2}{2}
 \nonumber \\
 && \equiv H^{[\mu-p]}  + E_p
\label{hamiltonianpsgn}
\end{eqnarray}
can be thus interpreted in the domain $p \in ]-\infty,\mu [ $ of Eq. \ref{domainp} as the sum of :

(i)  the Hamiltonian $H^{[\mu-p]} $ of effective drift $(\mu-p)>0$ in Eq. \ref{Hsgnbis},
with its zero-energy normalized ground-state of Eq. \ref{gssgn}
\begin{eqnarray}
 \phi^{GS[\mu-p]}(  x) =   \sqrt{\mu-p} e^{- (\mu-p) \vert x \vert }
\label{gssgnp}
\end{eqnarray}

(ii) the remaining constant in Eq. \ref{hamiltonianpsgn}
 \begin{eqnarray}
 E_p  = \frac{\mu^2-(\mu-p)^2}{2} = p \mu - \frac{ p^2 }{2}
\label{legendresgn}
\end{eqnarray} 
that directly represents the ground-state energy of $H_p^{[\mu]} $ and that 
coincides with the alternative analysis of Eq. \ref{legendresgn} as it should.


\subsubsection{ Canonical conditioned process $X_p^*(t)$ of parameter $p$   }

Since Eq. \ref{gssgnp} is the ground-state of the $p$-deformed Hamiltonian $ H_p^{[\mu]}  $
\begin{eqnarray}
\phi_p^{GS[\mu]}(  x) =  \phi^{GS[\mu-p]}(  x) =   \sqrt{\mu-p} e^{- (\mu-p) \vert x \vert }
\label{gssgnpp}
\end{eqnarray}
one obtains that the conditioned drift of Eq. \ref{mustarpinteriorgs} reduces to
\begin{eqnarray}
  \mu_p^*( x )    = \partial_x \ln \left[  \phi_p^{GS[\mu]}(  x)\right] = -  (\mu-p) \, {\rm sgn} (x)
\label{mustarpinteriorgssgn}
\end{eqnarray}
So the canonical conditioning of parameter $p \in ]-\infty,\mu [$
simply amounts to change the amplitude $\mu$ of the unconditioned drift $\mu(x)=-\mu \, {\rm sgn}(x) $
into the amplitude $(\mu-p)$.
As a consequence, the corresponding conditioned equilibrium state 
\begin{eqnarray}
G^{eq[\mu]}_p(x) = G^{eq[\mu-p]} (x)=   (\mu-p) e^{- 2 (\mu-p) \vert x \vert }
\label{geqp}
\end{eqnarray}
will produce the following equilibrium value for the intensive local time
\begin{eqnarray}
a^{eq[\mu]}_p =G^{eq[\mu]}_p(x=0) =   \mu-p 
\label{aeqp}
\end{eqnarray}
in agreement with the Legendre correspondence of Eq. \ref{legendresgn1}.

Finally, the conditioned drift of Eq. \ref{mustarpinteriorgssgn} can be rewritten 
in terms of $a^*_p  = \mu-p $ of Eq. \ref{apstarsign}
as
\begin{eqnarray}
  \mu_p^*( x )  = - a^*_p  \, {\rm sgn} (x)
\label{mustarpinteriorgssgntrad}
\end{eqnarray}
in agreement with the microcanonical conditioning of Eq. \ref{mustarbridgepibelowintensiveasymptosgn} in the main text.


\section{ Canonical conditioning when the unconditioned process $X(t)$ has no equilibrium state }

\label{app_canonicalnoEq}

In this Appendix, we consider the case where the unconditioned process $X(t)$ 
has no equilibrium state,
so that the quantum Hamiltonian $H$ of Eq. \ref{hamiltonian} has no bound state.
For the Hamiltonian $H_p$ of Eq. \ref{hamiltonianp}, the presence of a bound state depends on the sign of $p$ as follows.

(i) The case $p>0$ corresponds to an additional repulsive delta potential at the origin $x=0$
and will not change the range $]V_{\infty},+\infty[$
of the continuous spectrum of $H$, since $H$ and $H_p$ have the same potential at $x \to \pm \infty$
in Eq. \ref{vinfty}.

(ii)  The case $p<0$ corresponds to an additional attractive delta potential at the origin $x=0$
that produces 
a normalizable ground state for $H_p$ as we now describe.


\subsection{ Emergence of a bound state in the attractive case $p<0$}

For the double Laplace transform $ {\hat {\tilde P}}_{s,p} (x \vert x_0)$ of Eq. \ref{laplacedouble},
the result of Eq. \ref{pole}
\begin{eqnarray}
{\hat {\tilde P}}_{s,p}  (x \vert x_0)
 = \left[ {\hat G}_s (x \vert x_0) - \frac{ {\hat G}_s (x \vert 0){\hat G}_s (0 \vert x_0) }{ {\hat G}_s (0 \vert 0) } \right]
 + \left[ \frac{ {\hat G}_s (x \vert 0){\hat G}_s (0 \vert x_0) }{ {\hat G}^2_s (0 \vert 0) } \right]
 \frac{ 1   }{  p + \frac{1}{{\hat G}_s (0 \vert 0)} }  
\label{polep}
\end{eqnarray}
shows that for $p<0$, a new singularity will appear in ${\hat {\tilde P}}_{s,p}  (x \vert x_0)$
with respect to ${\hat G}_s (x \vert x_0) $ when the variable $s$ make the denominator vanish
\begin{eqnarray}
0  =p + \frac{1}{{\hat G}_s (0 \vert 0)}  
\label{spole}
\end{eqnarray}
The comparison with the pole in Eq. \ref{laplacedoublepole} 
shows that the value of $s$ satisfying Eq. \ref{spole} is directly related to the ground-state energy $E_p$ of $H_p$
\begin{eqnarray}
s= - E_p
\label{spole2}
\end{eqnarray}
i.e. the relations between $p$ and $s$ of Eqs \ref{spole} and \ref{spole2} 
are the same as in Eq. \ref{legendreps}.


\subsection{Example of the uniform drift $\mu \geq 0 $  }


\subsubsection{ Computation of the energy $E_p$ via the pole analysis of Eqs \ref{spole} \ref{spole2}  }

For the case of the uniform drift $\mu \geq 0 $, the Laplace transform ${\hat G}_s (0 \vert 0) $ of Eq. \ref{laplacefree}
yields that Eq. \ref{spole} reads
\begin{eqnarray}
0 = p +\sqrt{\mu^2+2 s }  
\label{spolemu}
\end{eqnarray}
So for any $p<0$, its solution $s=-E_p$ of Eq. \ref{spole2} leads to the energy
\begin{eqnarray}
E_{p<0} = \frac{\mu^2-p^2}{2} 
\label{spolemuep}
\end{eqnarray}
for the ground state of $H_p$ that emerges below the continuous spectrum $]\frac{\mu^2}{2}  ,+\infty[ $.


\subsubsection{ Direct analysis of the ground-state of the $p$-deformed Hamiltonian $H_p$ for $p<0$  }

For the drift $ \mu(x)=\mu $, 
the Hamiltonian $H$ of Eq. \ref{hamiltonian}
\begin{eqnarray}
H  =  -  \frac{1}{2} \partial_x^2 + \frac{ \mu^2}{2} 
\label{Hfree}
\end{eqnarray}
has no bound-state, but only a continuous spectrum $]\frac{ \mu^2}{2},+\infty[$.

The $p$-deformed Hamiltonian $H_p$ of Eq. \ref{hamiltonianp} reads
\begin{eqnarray}
 H_p^{[\mu]} && = H^{[\mu]} + p \delta(x) =  -  \frac{1}{2} \partial_x^2 + \frac{ \mu^2}{2}   +p \delta(x)
\label{hamiltonianpmu}
\end{eqnarray}

For the repulsive case $p>0$, $H_p$ keeps the continuous spectrum $]\frac{ \mu^2}{2},+\infty[$.

However, for the attractive case $p<0$, 
a bound state emerges below the continuous spectrum $]\frac{ \mu^2}{2},+\infty[$.
 It is exponentially localized around the origin
\begin{eqnarray}
 \phi_{p<0}^{GS}(  x) =  \sqrt{ (-p) }e^{ - (-p)  \vert x \vert } 
\label{gspneg}
\end{eqnarray}
Its energy
\begin{eqnarray}
 E_{p<0} =  \frac{\mu^2-p^2}{2}
\label{egspneg}
\end{eqnarray}
coincides with the other derivation of Eq. \ref{spolemuep} as it should.


\subsubsection{ Canonical conditioned process $X_p^*(t)$ of parameter $p<0$   }

The conditioned drift of Eq. \ref{mustarpinteriorgs} reads
\begin{eqnarray}
  \mu_{p<0}^*( x )    = \partial_x \ln \left[  \phi_p^{GS}(  x)\right] = - (-p)  \, {\rm sgn} (x)
\label{mustarpneg}
\end{eqnarray}
The corresponding conditioned equilibrium state reads
\begin{eqnarray}
G^{eq}_p(x) = \left[ \phi_{p<0}^{GS}(  x) \right]^2 =   (-p) e^{- 2 (-p) \vert x \vert }
\label{gpneg}
\end{eqnarray}

So the canonical conditioning of parameter $p<0$ is asymptotically equivalent to
 the microcanonical conditioning towards the intensive local time  
\begin{eqnarray}
a^*_p= G^{eq}_p(x=0)=-p >0
\label{apneg}
\end{eqnarray}
So the conditioned drift of Eq. \ref{mustarpneg} can be rewritten as
\begin{eqnarray}
  \mu_{p<0}^*( x )    = - a_p^*  \, {\rm sgn} (x)
\label{mustarpnegtrada}
\end{eqnarray}
in agreement with the microcanonical conditioning of Eq. \ref{mustarbridgepibelowintensiveasymptobrown} in the main text.


\end{document}